\documentclass[fleqn,compsoc,conference,a4paper,10pt,times]{IEEEtran}
\IEEEoverridecommandlockouts
\usepackage{cite}
\usepackage{authblk}
\usepackage{amsmath,amssymb,amsfonts}
\usepackage{algorithm}
\usepackage{amssymb}
\usepackage{algpseudocode}
\usepackage{enumitem}
\usepackage{graphicx}
\usepackage{textcomp}
\usepackage{bmpsize}
\usepackage{xcolor}
\usepackage{lipsum}
\usepackage[utf8]{inputenc}
\usepackage[T1]{fontenc}
\usepackage[colorlinks=true,urlcolor=black]{hyperref}
\usepackage{subcaption}
\usepackage{url}
\DeclareUnicodeCharacter{0430}{\unicodea}
\DeclareUnicodeCharacter{1EF5}{\unicodey}
\def\BibTeX{{\rm B\kern-.05em{\sc i\kern-.025em b}\kern-.08em
    T\kern-.1667em\lower.7ex\hbox{E}\kern-.125emX}}

\begin{document}
\urlstyle{tt}
\title{PhishMatch: A Layered Approach for Effective Detection of Phishing URLs}
\author{Harshal Tupsamudre}
\author{Sparsh Jain}
\author{Sachin Lodha}
\affil{TCS Research, India}

\maketitle
\thispagestyle{plain}
\pagestyle{plain}

\begin{abstract}
Phishing attacks continue to be a significant threat on the Internet. Prior studies show that it is possible to determine whether a website is phishing or not just by analyzing its URL more carefully. 
A major advantage of the URL based approach is that it can identify a phishing website even before the web page is rendered in the browser, thus avoiding other potential problems such as cryptojacking and drive-by downloads. However, traditional URL based approaches have their limitations. Blacklist based approaches are prone to zero-hour phishing attacks, advanced machine learning based approaches consume high resources, and other approaches send the URL to a remote server which compromises user's privacy. In this paper, we present a layered anti-phishing defense, PhishMatch, which is robust, accurate, inexpensive, and client-side. 
We design a space-time efficient Aho-Corasick algorithm for exact string matching and $n$-gram based indexing technique for approximate string matching to detect various cybersquatting techniques in the phishing URL.
To reduce false positives, we use a global whitelist and personalized user whitelists. We also determine the context in which the URL is visited and use that information to classify the input URL more accurately. 
The last component of PhishMatch involves a machine learning model and controlled search engine queries to classify the URL.
A prototype plugin of PhishMatch, developed for the Chrome browser, was found to be fast and lightweight.
Our evaluation shows that PhishMatch is both efficient and effective.
\end{abstract}

\begin{IEEEkeywords}
Phishing, Pattern matching, Cybersquatting
\end{IEEEkeywords}

\section{Introduction}
The proliferation of the Internet has not only facilitated the revolution of electronic commerce, but also opened new avenues for committing cyber-crimes such as financial frauds and credential thefts. In phishing attack, one of the many cyber-crimes, unsuspecting users are lured ({\em e.g.,} via carefully crafted email) into visiting fraudulent websites that are look-alike of their legitimate counterparts and trick them into divulging sensitive information such as credit card details and passwords. Phishing has become a major threat to individuals as well as organizations. According to FBI~\cite{FBI}, phishing scams caused more than \$12.5 billion in damages between October 2013 and May 2018, involving 78,617 incidents in 150 countries. Further, as per Verizon's 2021 data breach investigation report~\cite{Verizon}, 36\% of the data breaches in organizations are caused due to phishing. Phishing attacks are also continuously evolving and becoming more sophisticated to evade detection. According to the phishing activity trends report (fourth quarter 2020) released by Anti-Phishing Work Group (APWG)~\cite{APWGReport}, more than 80\% of phishing websites were hosted on HTTPS infrastructure in order to deceive users into believing that they are visiting a safe website. 

Various factors can be considered while determining a website as phishy or benign such as URL, WHOIS lookup, web page features like visual similarity and DOM objects, etc. However, the main benefit of relying only on URL based features is that we can determine the label of a new URL before the page is loaded by the web browser, thus preventing potential dangers like drive-by download and cryptojacking attacks. Different URL obfuscation techniques (typosquatting, combosquatting, etc.) have been reported~\cite{Garera:2007} and their prevalence is measured~\cite{Wang:2006,banerjee:2008,agten2015seven,Kintis:2017,Holgers:2006,Quinkert:2019}.
Thus, in many cases (where URL obfuscation techniques are used) it is possible to determine whether a website is phishing or not just by inspecting its URL string more carefully. In particular, attackers employ two common strategies for constructing phishing URLs. 
\begin{enumerate}[labelindent=0em,noitemsep,nolistsep,leftmargin=*]
\item Attackers include the organization's name or target domain's name in the phishing URLs so that they appear more authentic to users. For instance, in what is known as combosquatting~\cite{Kintis:2017}, attackers register domains (\texttt{paypalsecure-verification.com}) that combine a popular brand name (\texttt{paypal}) with one or more phrases (\texttt{secure-verification}), whereas in subdomain spoofing (\texttt{www.paypal.com.login-safe.net}) attackers lure users by placing an authentic domain (\texttt{paypal.com}) in front of the fake domain (\texttt{login-safe.net}). 
\item Attackers register domains that are similar to popular domain names. For instance, in a form of domain squatting known as typosquatting~\cite{agten2015seven}, attackers register common misspellings of popular websites (\texttt{papyal.com}) to exploit typing errors users make when they enter URLs in the address bar of their browser. In a more advanced form of attacks known as IDN homograph attacks~\cite{Fu:2006}, attackers attempt to trick users by registering domains that look visually similar to the target domain, {\em e.g.}, \texttt{paypal.com} where Latin letter `a' (U+0061) is replaced by a similar looking Cyrillic letter `a' (U+0430).
\end{enumerate}

In this work, we develop a layered anti-phishing defense, called {\em PhishMatch}, which is robust, effective, fast and client-side. More specifically, it employs a list of 50,000 highly popular and trusted domains~\cite{pochat2019tranco}, and check the input URLs for different obfuscation techniques including subdomain spoofing, combosquatting, wrongTLDsquatting, typosquatting and IDN homograph. We resort to machine learning based approach if the URL does not contain any brand name or misspelled domain, and query search engines only if the machine learning model does not classify the URL with a high confidence score. The resulting system requires around 20 MB memory and can be deployed at client-side as a browser extension. Specifically our contributions are as follows:
\begin{itemize}[labelindent=0em,noitemsep,nolistsep,leftmargin=*]
	\item We exploit the structure of domains and construct a novel space-efficient Aho-Corasick pattern matching machine for storing popular domains which enables the detection of combosquatting, wrongTLDsquatting, subdomain spoofing and other obfuscating URLs. The resulting optimized pattern matching machine consumes around ~46\% less memory than the state-of-the-art~\cite{Tuck:2004}.
	\item We devise a fast $n$-gram based indexing technique to detect common misspellings of popular domains, also known as typosquatting domains. Our $n$-gram based technique allows efficient filtering of the probable target domains on which to calculate the similarity metric.
	\item To reduce the number of false positives, we employ four different whitelists, namely {\em session}, {\em local}, {\em community} and {\em global}. The local and community whitelists are created from the browsing history of users.
	\item To speed up the detection process, we determine the context in which the URL is being visited, {\em i.e.}, whether the URL is typed or clicked.
	\item We also employ a lightweight classifier trained on segmented bag-of-words (SBoW) features extracted from hostnames. 
	\item We query search engines only when the classifier predicts the class of URL with a low confidence score.
	\item We develop a prototype Chrome browser PhishMatch extension and compare its efficacy with the state-of-the-art deep learning based URLNet models~\cite{le2018urlnet}. Our results show that PhishMatch outperforms URLNet models with a reduced memory usage and faster prediction.
\end{itemize}

\subsection{Preliminaries}
\begin{figure*}[h]
\centering
\includegraphics[scale=0.5]{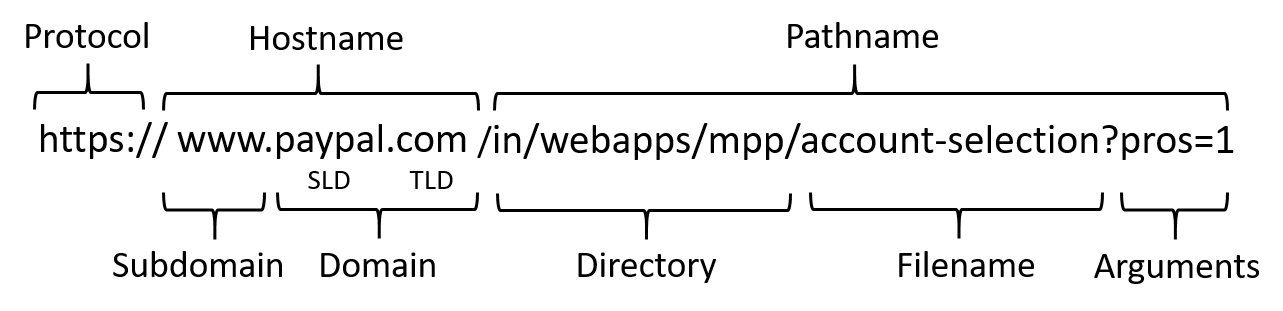}
\caption{The URL of PayPal signup page and its components.}
\label{fig:url_parts}
\end{figure*}
\begin{table*}[h]
\centering
\scriptsize
\caption{Commonly used cybersquatting techniques illustrated using PayPal brand.}\label{tab:spoofing}
\renewcommand{\arraystretch}{1.2}
\begin{tabular}{|l|l|}
\hline
\textbf{Description} & \textbf{Examples}\\
\hline
brand in subdomain & \texttt{http://paypal.com.elvalorsocial.com/}\\    
(subdomain spoofing)  & \texttt{https://auth-signpaypalalertcustomerapp.apacayus.com/myaccount/}\\
\hline
brand in domain   & \texttt{https://confirm-informations-paypal.com/}\\
(combosquatting)	 & \texttt{http://ssl-paypalupdate.com/}\\                      
\hline
brand in domain & \texttt{http://paypal.net/}\\
(wrongTLDsquatting) & \texttt{http://paypal.is/}\\
\hline
brand in pathname & \texttt{http://159.203.6.191/servicepaypal/}\\
			   & \texttt{https://hollywoodbytez.com/paypal.com/signin/}\\	            
\hline
misspelled brand &  \texttt{http://papyal.com}\\
(typosquatting) & \texttt{https://paytpal.com/}\\
\hline
similar brand & \texttt{http://paypa1.com}\\
(homograph) & \texttt{http://www.pay}$_{.}$\texttt{pal.com/}\\
\hline
unrelated domains &  \texttt{http://signin-ppl-users-542544558966554-com.umbler.net/f/ppl/}\\
	& \texttt{http://www.nwolb.co.uk.secureonlinelogin.s-secureuk.com/}\\
\hline
\end{tabular}
\end{table*}
In this section, we define the terms used throughout the paper. The term URL stands for {\em Uniform Resource Locator}, which is a unique address for identifying files and other resources on the World Wide Web. A URL comprises of three main parts: (i) protocol, (ii) hostname, and (iii) pathname. The hostname specifies the server on which the resource is located and the pathname specifies the location of the resource on the server. The hostname is further divided into two sub-parts: subdomain and domain. The domain is made up of two components: second-level domain (SLD) and top-level domain (TLD). SLD is also referred as brand name, trademark name or organization name. The pathname is divided into three sub-parts: directory, filename and arguments. An example URL (pointing to PayPal signup page) along with its different components are shown in Figure~\ref{fig:url_parts}.

We now describe different obfuscation techniques that have been shown prevalent in the literature~\cite{Garera:2007,Wang:2006,banerjee:2008,agten2015seven,Kintis:2017,Holgers:2006,Quinkert:2019}.
The real-world examples of each obfuscation type are shown in Table \ref{tab:spoofing}. We use the terms spoofing, obfuscation and cybersquatting interchangeably.

\begin{itemize}[labelindent=0em,noitemsep,nolistsep,leftmargin=*]
\item \textbf{Subdomain spoofing}: In this form of spoofing, the name of the organization being phished is present in the subdomain part of the URL. 
One such example is \texttt{paypal.com.elvalorsocial.com} where the legitimate domain \texttt{paypal.com} is placed in the subdomain. 

\item \textbf{Combosquatting}: In this form of cybersquatting, attackers register domains that combine popular brand names with one or more words. 
For example, \texttt{ssl-paypalupdate.com} is the combosquatting domain where the brand \texttt{paypal} is concatenated with the words \texttt{ssl} and \texttt{update}. 

\item \textbf{WrongTLDsquatting}: In this form of cybersquatting, the second-level domain (SLD) remains unchanged but a different top-level domain (TLD) is used. 
For example, \texttt{paypal.net} belongs to the wrongTLDsquatting category since the TLD \texttt{com} is replaced by the TLD \texttt{net}.

\item \textbf{Typosquatting}: In this form of cybersquatting, attackers register domains with common spelling mistakes users may make while typing the domain names, such as addition or deletion of a character, replacing a character, or re-ordering two consecutive characters. 
For example, \texttt{paytpal.com} is a typosquatting domain obtained by inserting letter `t' (while typing letter `y' on qwerty keyboard) in the target domain \texttt{paypal.com}.

\item \textbf{Homograph}: It is a form of visual spoofing in which attackers register domains that look similar to the target domain. 
The domain \texttt{paypa1.com} is an instance of homograph attack since digit `1' is used in place of the lowercase letter `l' to imitate the look of target domain \texttt{paypal.com}. 
More advanced homograph squatting employs unicode characters that are look-alike of ASCII characters. For example, \texttt{pay$_.$pal.ga} is an internationalized domain name (IDN) where letter `y' (U+0079) is replaced with character 
`y$_.$' (U+1EF5). 

\item \textbf{Unrelated domains}: In the remaining cases, phishing URLs do not contain popular brand names, however they might contain suspicious words such as \texttt{secure}, \texttt{signin} and \texttt{online} that are suggestive of phishing attacks.

\end{itemize}

The organization of this paper is as follows. First, we discuss the related work and its limitations that motivated us to develop PhishMatch. Then, we describe different components of the PhishMatch system. 
We developed a prototype of PhishMatch as a Chrome Browser extension, that we use to demonstrate the efficacy of PhishMatch by simulating it on the browsing history of 25 users and 18,000 phishing URLs from PhishTank~\cite{PhishTank}.
Finally, we conclude the paper and discuss some future research directions.

\section{Related Work}
Different URL obfuscation techniques have been reported in the literature. Garera {\em et al}.~\cite{Garera:2007} identified four distinct types of URL obfuscation techniques attackers commonly use to spoof legitimate websites. Most of the phishing URLs contain the name of the organization being phished in exact or approximate form. Many studies measured the prevalence of typosquatting~\cite{Wang:2006, banerjee:2008, agten2015seven}, combosquatting~\cite{Kintis:2017} and homograph attacks~\cite{Holgers:2006, Quinkert:2019}. Agten {\em et al}.~\cite{agten2015seven} used Alexa top 500 domains to generate a list of domains within edit distance 1. They found that 95\% of the popular domains are targeted by typosquatters, and only few brand owners proactively protect themselves by registering typosquatting domains. Kintis {\em et al}.~\cite{Kintis:2017} studied the extent of combosquatting domains, a form of cybersquatting, and found that more than 60\% of combosquatting domains live for more than 1,000 days. Holgers {\em et al}.~\cite{Holgers:2006} characterized the prevalence of homograph attacks in which a Latin letter is replaced with a similar looking international letter and found that most of such squatting domains tend to have a single confusable character. In the following subsections, we categorize the existing anti-phishing solutions on the basis of core approach used and describe their merits and demerits.

\subsection{Blacklists}
Blacklists are essentially a repository of URLs that have been confirmed to be phishing in the past. Blacklist based solutions are fast, but not effective against zero-hour phishing attacks~\cite{sheng:2009} and can be easily evaded by attackers~\cite{Garera:2007}. To overcome the problems associated with blacklists, researchers proposed a proactive approach PhishNet~\cite{prakash:2010}, which uses simple heuristics to generate new candidate phishing URLs from the already blacklisted URLs. One way to generate new candidate URLs is to replace a brand name in the blacklisted URLs with other brand names or to replace a TLD with other TLDs. PhishNet can also be used to identify new phishing URLs by performing approximate matching (based on regular expressions) with the existing blacklisted URLs. While URL blacklisting has been effective to some extent, it is rather easy for an attacker to deceive the system by slightly modifying one or more components of the URL string~\cite{Vanhoenshoven}.

\subsection{Machine Learning}
\label{sec:rel_ml}
To improve the detection of newly emerged phishing websites, researchers turned to machine learning (ML) techniques. McGrath {\em et al.}~\cite{McGrath:2008} found that length and character distributions of phishing URLs are very different from the non-phishing URLs. Consequently, subsequent studies trained various machine learning models using URL based features and predicted the phishing URLs with high accuracy. Ma {\em et al}.~\cite{Ma:2009} demonstrated a phishing detection approach that employs two types of URL features; lexical features extracted from URL names such as URL length and bag-of-words (BoW), and external features acquired by querying remote servers such as whois lookup. Le {\em et al}.~\cite{Le:2011} proposed PhishDef and showed that lexical features are as effective as using full features.
Sahingoz \emph{ et al}.~\cite{ml:Sahingoz2019} explored the use of NLP based features, word vectors, and hybrid features with different machine learning algorithms and found them to be effective.
Verma {\em et al}.~\cite{Verma:2017} explored the efficacy of unigrams, bigrams and trigrams features to learn the nature and construction of phishing URLs as opposed to benign URLs, and found that the classifiers trained on $n$-gram features achieve a higher classification accuracy.

A major challenge in using machine learning for phishing detection is obtaining representative URL samples to train a classifier. The most widely used URL sources in the literature~\cite{McGrath:2008, Ma:2009, Le:2011,tupsamudre:2019,ml:Sahingoz2019,Verma:2017} are PhishTank~\cite{PhishTank} which consists of human verified phishing URLs and DMOZ~\cite{DMOZ} which consists of human verified benign URLs. However, the URLs reported in PhishTank are biased (submitted by few users). Moreover, it contains phishing data pertaining to only few hundred brands, most of them US based~\cite{Moore:2008}. DMOZ on the other hand stopped functioning in 2017 and has not been updated since. Further, DMOZ dataset mostly contains hostnames of benign websites, whereas PhishTank dataset contains full URLs of phishing websites. 

Detection capabilities of the machine learning models could be further improved by including the web page features~\cite{Zhang:2007,Khonji:2013,chen:2010,ml:li2019,dom:2019} like title, text content, HTML tags, JavaScript content, etc. Zhang {\em et al}.~\cite{Zhang:2007} proposed CANTINA that identifies a phishing website using TF-IDF information extracted from the web page which was further imporved by Xiang \emph{et al}.~\cite{Xiang:2011}. Whittaker {\em et al}.~\cite{whittaker:2009} described the design of the Google's proprietary machine learning model that uses URL features, Google Page Rank, and page features, to detect phishing websites. Ardi {\em et al}.~\cite{Ardi:2016} proposed an approach that stores cryptographic hash of whitelisted web pages and compares them with newly visited web pages to detect phishing attacks.

However, the main benefit of using models trained only on URL based features is that they can determine the label of a new URL before the page is loaded by the web browser, thus preventing other potential dangers like drive-by download and cryptojacking attacks. Further, most page based detection techniques work only after the entire web page is rendered in the browser. Hence, there is a possibility of users divulging sensitive data on the web page before it is detected as phishing.

\subsection{Deep Learning}
Traditionally, machine learning models were trained on bag-of-words (BoW) like features extracted from the URL string~\cite{Garera:2007, Ma:2009, Le:2011,tupsamudre:2019}. Le \emph{ et al}.~\cite{le2018urlnet}, proposed a deep learning based solution called URLNet which eliminates the need of manual feature engineering and showed that it significantly outperforms the traditional BoW based approaches. However, on top of the problems faced by traditional machine learning algorithms (as described in subsection~\ref{sec:rel_ml}) such as obtaining representative URL samples to train the model, URLNet suffers from two additional shortcomings, that may hinder its deployment at client-side: a) lack of explainability; it is difficult to provide justification as to why the URL is classified as phishing or benign and b) excessive consumption of memory; it requires millions of parameters for decision-making. Moreover, recently it was demonstrated that deep learning based URL classifiers are susceptible to adversarial attacks~\cite{cmc:adversarial}. 


\subsection{Search Engine}
Another simple yet potent approach to determine whether a given website is phishing or not is to search its full URL on popular search engines such as Google, Bing or Yahoo~\cite{Huh:2012,search:2019}. Most of the times, URLs of legitimate websites retrieve large number of search results and are ranked first, whereas URLs of phishing websites either return zero results or are not ranked at all. However, querying search engines for every URL incurs additional overhead and may be impractical due to the rate-limiting applied by search engine providers. Paid options may be provided with higher rate-limits by the search engine, however, searching every URL would still compromise the user's privacy.

Now that we have discussed the related work on the basis of the underlying core approach, we describe some additional related work, starting with a client-side deployment archetype and then on the topic of user education.


\subsection{Browser Extensions}
Popular browser extensions such as Windows Defender by Microsoft~\cite{WindowsDefender} and Suspicious Site Reporter by Google~\cite{SuspiciousReporter} employ a small whitelist of trusted domains to determine if the user is on a safe website or not. If the domain is not present in the whitelist, then the URL and additional details ({\em e.g.}, web page DOM) are sent to company's servers for further analysis. Other extensions such as Online Security by Avast~\cite{avastOnline} and Anti-Phishing by Netcraft~\cite{Netcraft} do not maintain any local databases and contact their data servers for every URL visited by the user. Transferring the browsing data to the external servers threatens the privacy of the user. Recently, Mozilla removed Avast Online Security extension from its store when it discovered that the extension transmitted data that allows reconstruction of the entire web browsing history of the user~\cite{mozillaAvast}.

\subsection{Educating Users}
Researchers have also explored various training methods to teach users about the importance of various security indicators and to recognize phishing attacks. For instance, numerous educational games such as Anti-Phishing Phil~\cite{Sheng:2007}, NoPhish~\cite{Gamze:2014} and Phishy~\cite{CJ:2018} have been developed to educate users about phishing URLs. The main purpose of these games is to inform users about different URL obfuscation techniques. Although, educating users to recognize phishing attacks could potentially complement the existing phishing detection methods, the extent to which these training methods are effective against real attacks have not been yet demonstrated.\\

\textbf{PhishMatch Differentiators:}
In our system, we utilize a global whitelist and personalized whitelists for the user to efficiently detect benign URLs. We then develop an optimized Aho-Corasick automaton to store popular brand names and domain names that enables the detection of various cybersquatting techniques in the input URL more efficiently. We also create an $n$-gram based fast approximate matching algorithm to detect misspelled domains. Recently, Tupsamudre {\em et al}.~\cite{tupsamudre:2019} showed that models trained using segmented bag-of-words (SBoW) features consume relatively less space and are more effective in detecting phishing URLs containing phishy words. Hence, we train an SBoW based machine learning (ML) model and use it only when the URL does not contain any popular brand name or misspelled domain name. We also use the search engine queries to determine whether a URL is phishy or not. However, keeping in mind that search engine is an expensive resource, we use it judiciously. In our system, a search engine is queried only if the ML model classifies the URL as phishing with a low confidence score (< 0.9). When a phishy URL is detected, appropriate warnings are displayed to the user that explain why the URL is classified as phishy. Our analysis (Section~\ref{sec:experiments}) shows that PhishMatch requires only around 20 MB memory and classifies a URL with high accuracy. PhishMatch classifies at least 90\% of the benign URLs in just 3.6 milliseconds, and at least 90\% of the phishing URLs in 8.3 milliseconds.
Hence, PhishMatch is fast, lightweight, and suitable for deployment at client-side.

\section{PhishMatch}
\label{sec:phishmatch}
\begin{figure*}[t]
\centering
\includegraphics[scale=0.8]{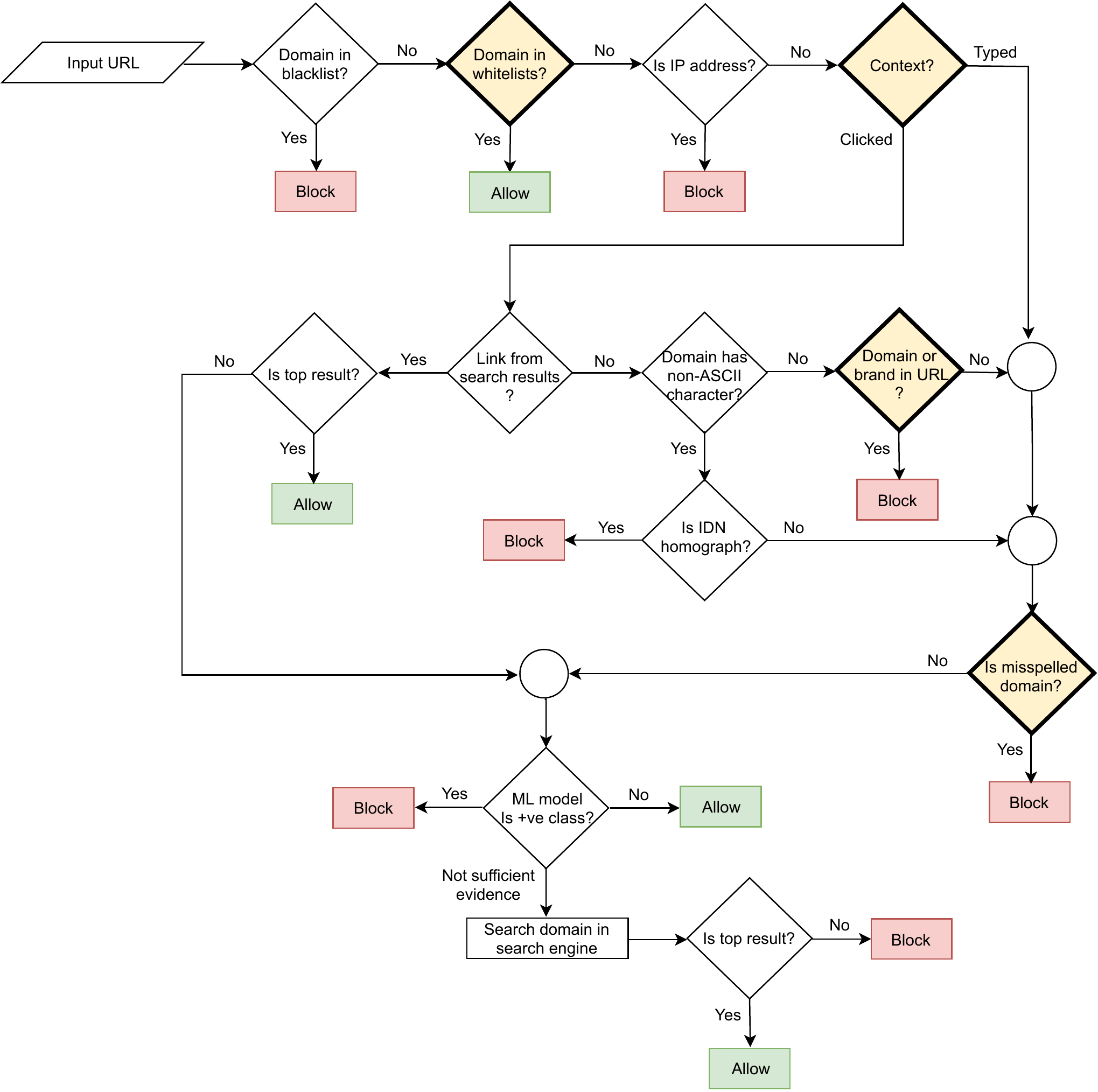}
\caption{The figure depicts the flowchart of PhishMatch. The components that employ novel algorithms are shaded and highlighted with a bold outline.}
\label{fig:phishmatch}
\end{figure*}
The flowchart of our proposed system PhishMatch is shown in Figure~\ref{fig:phishmatch}. Some of the components test the input URL for phishiness ({\em e.g.,} blacklist and pattern matching algorithms), whereas other components test the URL for benignness ({\em e.g.}, whitelists and search engine). The components that employ novel algorithms are shaded and highlighted with a bold outline in the figure. PhishMatch is invoked each time a web request is generated to visit a URL. If the URL is deemed safe to visit then the request is allowed, otherwise, the request is blocked and an appropriate warning message is shown to the user.


An input URL is evaluated by PhishMatch as follows. First, the URL is checked against a blacklist that consists of domains determined to be unsafe by PhishMatch in the past. If the input domain is found in the blacklist, the request is blocked and an appropriate warning message is issued to the user. Otherwise, the domain is checked in whitelists ({\em session}, {\em local}, {\em community} and {\em global}). If the result is positive, the user is allowed to access the requested web resource. If the domain is not found in any of the whitelists, we ascertain the context in which the URL is being visited, {\em i.e.}, whether it is clicked or typed. 

When the user uses a popular search engine ({\em e.g.}, Google Search) to find specific information and clicks a link from the top search results, then the resource pointed by the URL is fetched. If the URL is not clicked from the top search results, the domain is checked for the presence of international character(s). If the result is positive, the domain is checked for IDN homograph attack, otherwise the URL is checked for the presence of popular domain name or brand name. If the result of either tests is positive, the web request is blocked and an appropriate warning message is displayed to the user. If the result is negative or the URL is typed, then the domain is checked for misspelling(s). If the domain is found to be a misspelled version of some popular domain, the request is blocked and an appropriate warning message is issued to the user. Otherwise, the phishiness of URL is determined using a machine learning model trained on segmented bag-of-word (SBoW) features extracted from hostnames.

If there is not enough evidence (low confidence score) to classify the input hostname as benign or phish, then PhishMatch searches the domain using a popular search engine ({\em e.g.}, Google Search). If the domain appears in the top search engine results, the web request is allowed, otherwise it is blocked. Whenever the domain or hostname is blocked, it is added to the blacklist (if not already present), otherwise it is added to the local whitelist via session whitelist (if not already present). Further, whenever the URL is blocked, the user can still visit it by overriding the warning system of PhishMatch. This action removes the domain from the blacklist and adds it to the session whitelist. If the user visits such a domain frequently, it will eventually make its way to the local whitelist of the user. In the following sections, we describe each of these components in more detail. We also provide information about relevant Chrome APIs~\cite{chromeAPI} required to implement PhishMatch extension for Chrome browser. We note that extensions written for Chrome browser can also be ported to Mozilla Firefox, Microsoft Edge and Opera with just few changes~\cite{compatibility}. 

\subsection{Blacklist}
The first component of PhishMatch is blacklist, which consists of domains categorized as phishing by the system previously. This blacklist is not intended as a replacement to the existing blacklists employed by web browsers ({\em e.g.}, Google's Safe Browsing~\cite{safeBrowsing}), rather its objective is to extend the detection capability of the existing blacklists. The PhishMatch blacklist is implemented using a hashmap (key-value pairs), with the suspicious domain as key and details about the component 
that determined the corresponding domain to be a phish (in the past) as value. The component details are stored in order to construct a relevant warning message in case the user visits the blacklisted domain again. This blacklist is updated with a new domain whenever a new phishing domain is detected by the system.


Chrome browser extensions employ background scripts, that monitor different events and react with specified instructions. The background script of the PhishMatch extension intercepts the request for each web page using the \texttt{OnBeforeRequest} event defined in the \texttt{chrome.webRequest} API. Subsequently, it extracts the input domain from the requested URL using the \texttt{URI.js} library~\cite{URIlib} and looks it up in the blacklist. If the domain is found in the blacklist, then the web request is blocked and an appropriate warning message is displayed to the user. Otherwise, the extension checks if the domain is present in one of the whitelists.

\subsection{Whitelist}
Whitelist is the second component of PhishMatch and its aim is to minimize the number of false positives (benign URLs classified as phishing). We employ four different whitelists, {\em local}, {\em community}, {\em global} and {\em session}. All whitelists store only domain part of the URL. We now describe different whitelists and some novel heuristics that we have designed for their creation and maintenance. The related pseudocode can be found in the appendix as Algorithms~\ref{LWAlgo},~\ref{CWAlgo}, and~\ref{SWAlgo}.

\subsubsection{Local Whitelist}
The local whitelist is a personalized list of safe domains created from the browsing history of the user. Popular browser extensions such as Google's Suspicious Site Reporter~\cite{SuspiciousReporter} consider every domain in the user's browsing history safe to visit, whereas other extensions ({\em e.g.}, Auntie Tuna~\cite{Ardi:2016}) require the feedback of the user to determine whether a visited domain should be in the whitelist or not. However, whitelisting every domain in the browsing history or relying on the user can be dangerous as prior research~\cite{Dhamija:2006, alsharnouby:2015} shows that users cannot reliably distinguish a legitimate website from a spoofed website. In our approach, we analyze several attributes of each domain present in the browsing history of the user and develop a criterion to determine if the previously visited domain should be present in the local whitelist or not. 

Browser extensions can access the browsing history of Chrome users using the \texttt{chrome.history} API. This API provides access to two tables, History table and Visit table. History table contains information about each URL visited by the user such as the page title, last visit time, the number of visits to the URL and the number of times the URL is typed. Visit table contains information about each visit to the URL such as the time when the visit occurred, visit id of the referrer and transition type of the visit ({\em e.g.}, link or typed). While History table contains a unique record for every URL, Visit table contains an entry for every URL visit made by the user. In order to determine whether a domain should be in the local whitelist or not, we compute the following four attributes for each domain from the History and Visit tables.

\begin{itemize}[labelindent=0em,noitemsep,nolistsep,leftmargin=*]
\item Age : The time difference (in days) between the recent visit time and oldest visit time to the domain is referred as the age of the domain.
\item Visits : The number of unique days on which the user visited the domain is referred to as visits. Multiple visits to the domain on the same day is counted as 1.
\item Recency : The time difference (in days) between the current time and recent visit time is referred as the recency of the domain.
\item Typed : It is a binary feature which indicates if the domain is `ever' visited by typing the URL. If at least one visit to the domain involves user typing the URL, its value is true, otherwise it is false.
\end{itemize}

\begin{figure}[h] 
	\centering
	\includegraphics[scale=0.6]{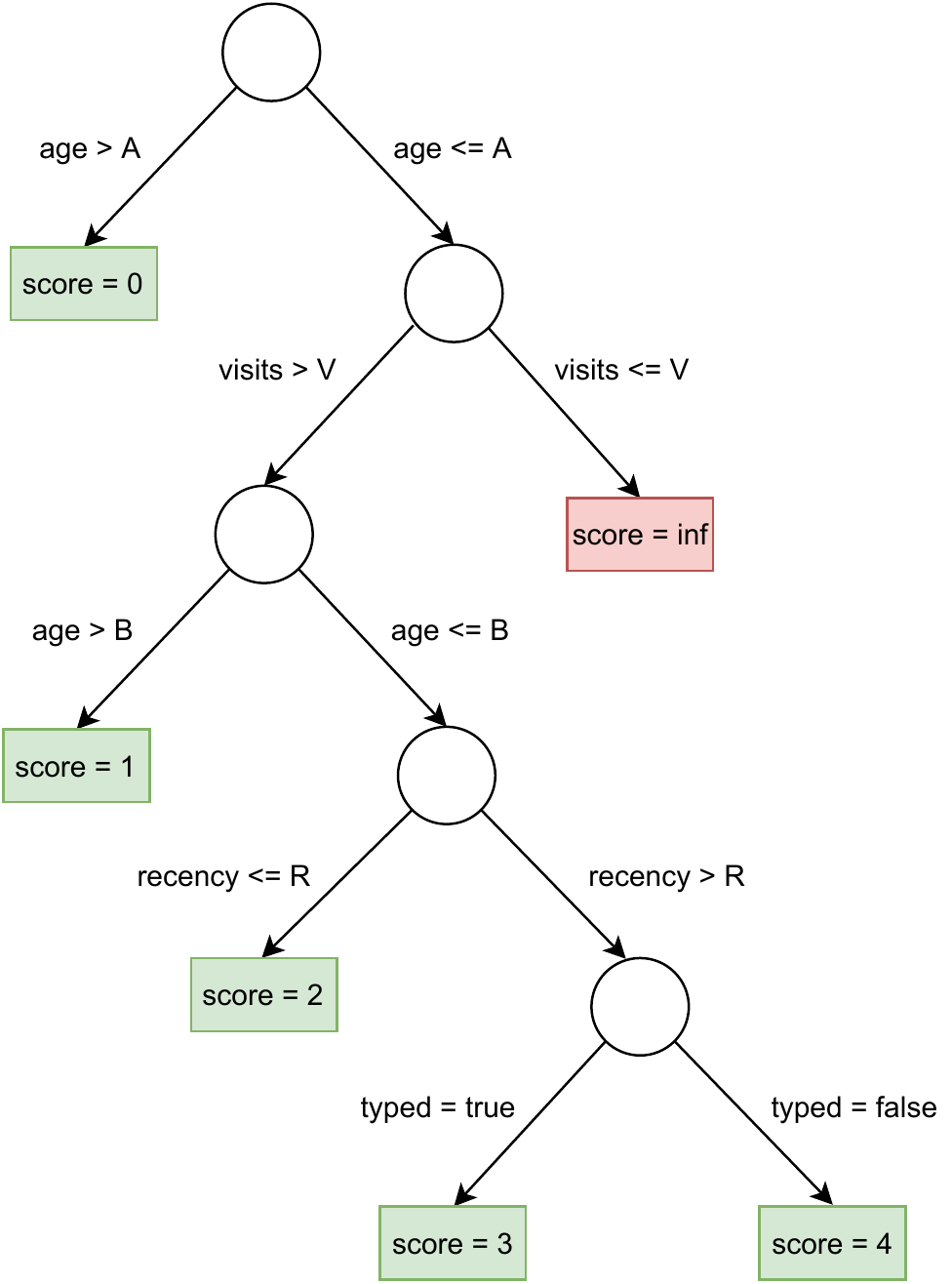}
	\caption{Decision tree for scoring user-visited domains based on four features, age, visits, recency and typed.}
	\label{fig:lw}
\end{figure}

Thus, each domain $d_i$ is associated with four features, $<$$a_i$, $v_i$, $r_i$, $t_i$$>$, where $a_i$ is the age of the domain, $v_i$ is the number of unique visits to the domain, $r_i$ is the recency and $t_i$ indicates if the domain is ever typed or not. We define a scoring function (depicted in Figure \ref{fig:lw} as decision tree) which takes as input the four attributes of the domain $d_i$ and outputs a score $s_i \in \{0,1,2,3,4,inf\}$. The score is computed as follows. Most of the phishing domains have a shorter lifetime, typically few hours~\cite{APWGReport2014}. Therefore, if the age of the domain is large enough, say, greater than $A_L = 45$ days, we consider it safe and assign it the score of 0. If the age is smaller than $A_L$ days, we check it for the number of unique visits. If the domain is visited, say, on more than $V_L = 10$ different days (which implies that the age is greater that $V_L$ days), we consider it safe and add it to the local whitelist. The intuition is that if the domain is visited frequently, and the user has not faced any issues, then the website can be considered safe to visit.
However, we assign frequently visited domains different scores based on age, recency and typed features.

If the age of the domain is, say, greater than $B_L = 30$ days, we assign it the score of 1. Otherwise, we check if the domain is visited recently, say, within past $R_L = 7$ days. The age of the domain is still greater than $V_L$ days, hence if the domain has been visited frequently in the recent days, then the website can still be deemed safe. If the domain has not been visited recently, we check if there is at least one instance where a URL containing the domain is typed by the user. The typed URL indicates that the user has visited the domain voluntarily and hence, we assign it the score of 3. If the domain is never typed, we assign it the score of 4. If the domain does not meet the age and the number of visits criteria, we assign it the highest score of $\inf$. Thus, higher the score, less trustworthy is the domain. $A_L$, $B_L$, $V_L$, $R_L$, where $A_L > B_L > V_L$, are the parameters of the scoring function for creating the local whitelist. Note that the age of a domain last visited $r$ days ago but also on a total of $v$ number of days, will be at least $v + r$ days. Thus, if $B_L$ is not sufficiently large, specifically if the condition $B_L >= R_L + V_L$ is not satisfied, all the domains with age $<= B_L$ but number of visits $> V_L$ will fall in the recently visited category. For the experiments reported in Section~\ref{sec:experiments}, we have set $A_L = 45$, $B_L = 30$, $V_L = 10$ and $R_L = 7$.

The local whitelist $W_L$ is created immediately after the extension is installed. Only the domains with score $s_i \in \{0,1,2,3,4\}$ are placed in the local whitelist. At the end of every browsing session, the whitelist is saved locally in the variable $W^{prev}_L$ (using the \texttt{chrome.storage} API). At the start of every new browsing session, the local whitelist $W_L$ is updated using the saved local whitelist $W^{prev}_L$ and saved session whitelist $W^{prev}_S$ from the previous browsing session. More information about the session whitelist can be found in section~\ref{sec:SW}. If the domain is deleted from the Chrome's browsing history $H$, we refrain from including it in the updated local whitelist $W_L$. However, if the domain is present in the browsing history, we calculate the updated domain score $s^{new}_i$ using the decision tree depicted in Figure~\ref{fig:lw}. There are three cases of interest.


\begin{itemize}[labelindent=0em,noitemsep,nolistsep,leftmargin=*]
\item \textbf{Case 1:} The domain is present in both the previous local whitelist $W^{prev}_L$ and session whitelist $W^{prev}_S$, which implies that the domain was visited in the last browsing session and hence, there is a possibility of improving its score. The domain is stored in the updated local whitelist $W_L$ and its score is updated if it has improved.
\item \textbf{Case 2:} The domain is present only in the previous local whitelist $W^{prev}_L$ in which case the domain and its score is retained in $W_L$. Thus, once a domain enters the local whitelist, it stays there until it is removed from the browsing history. As Chrome browser stores browsing history of past 90 days, if a domain is not accessed within last 90 days, it is removed from the browsing history and, therefore, from $W_L$.
\item \textbf{Case 3:} The domain is present only in the previous session whitelist $W^{prev}_S$, which implies that the domain was visited in the last browsing session, and was either deemed safe to visit by some component of PhishMatch or the user trusts the domain and chose to ignore the warning given by the system. Such domains are given a score of 8 and added to the updated local whitelist $W_L$.
\end{itemize}

Since we do not have to consider all domains in the browsing history for updating the local whitelist, the process of updation is more efficient than creation. The local whitelist $W_L$ is implemented using a hashmap, with domain $d_i$ as key and score $s_i \in \{0,1,2,3,4,8\}$ as its value. Note that higher score indicates lower trust.

\subsubsection{Community Whitelist}
The local whitelist $W^j_L$ of each user $U^j$ can be combined to create a more effective whitelist for the whole community. For instance, if the users are employees of an organization, their local whitelists can be transmitted to an internal server for creating a more elaborate and reliable whitelist for the entire organization. For creating the community whitelist, we relax the criteria used to create the local whitelist ($A_C = A_L/2$, $B_C = B_L/2$ and $V_C = V_L/2$) and create a new relaxed local whitelist $W^j_{RL}$ locally from the browsing history $H^j$ of each user $U^j$ using the same scoring function (Figure~\ref{fig:lw}). We also update the relaxed whitelist $W^j_{RL}$ whenever the local whitelist $W^j_L$ is updated. Thus, we have $W^j_L \subset W^j_{RL} \subset H^j$. The relaxed whitelist $W^j_{RL}$ consisting of domain and score as key-value pairs is transmitted to the server. We refrain from sending the complete URL or any of its attributes (e.g., exact age or frequency) due to privacy concerns. 

The community whitelist $W_C$ is implemented as a hashmap, with domain $d_i$ as key and community score $W_C[d_i]$ as value. Initially, the community whitelist is empty. We iterate over relaxed whitelists $(W^1_{RL}, W^2_{RL}, \ldots W^N_{RL})$ received from $N$ users and compute the community score $W_C[d_i]$ for each domain $d_i$. If the domain $d_i$ is encountered for the first time {\em i.e.}, it is not present in the community whitelist, then we penalize the corresponding community score $W_C[d_i]$ with 16 for each of the remaining $N-1$ users. If the domain is already present in the community whitelist, we nullify the penalty of 16 from its community score $W_C[d_i]$ for the current user $U^j$. Subsequently, we update the community score $W_C[d_i]$ using the score in the relaxed local whitelist $W^j_{RL}$. 

The maximum score of a domain in a relaxed whitelist is 8. Further, we increase the community score of a domain by 16 if it is not present in the relaxed whitelist of a given user. Therefore, smaller the score, the more popular is the website domain. Top $k$ domain entries with the smallest community scores are selected and returned to form the community whitelist. 
Once the community whitelist is created, one may update it periodically, for instance once in a week.

\subsubsection{Global Whitelist}
Global whitelist consists of the most popular domains visited by users across the globe. To construct global whitelist, we use the recently developed tranco-list~\cite{pochat2019tranco}, which is also available as an online service. Tranco-list combines the domain lists from four different providers: Alexa, Cisco Umbrella, Majestic and Quantcast, using an improved ranking function which is more robust and resistant to adversarial manipulation. It consists of top one million domains obtained by aggregating all four rankings averaged over the past 30 days. We use the first 50,000 domains from tranco-list as our global whitelist. Comparatively, anti-phishing Chrome extensions by Google (Suspicious Site Reporter~\cite{SuspiciousReporter}) and Microsoft (Windows Defender~\cite{WindowsDefender}) use domain whitelists of size 5,000 and 30,000 respectively. The global whitelist $W_G$ is stored using an Aho-Corasick pattern matching machine described in section~\ref{sec:aho}.

\subsubsection{Session Whitelist}~\label{sec:SW}
Session whitelist consists of domains visited in the current browsing session that are determined to be safe for visit. It is implemented as a hashmap, with domain $d_i$ as key and $true$ as value. At the start of the browsing session, the session whitelist $W_S$ is empty. If the domain is classified as safe by any component of PhishMatch or user visits the URL by overriding the warning of PhishMatch, it is added to the session whitelist. Hence, session whitelist serves as a fast cache for future lookups of the recently visited domains. At the end of browsing session, the session whitelist is stored locally using the \texttt{chrome.storage} API and retrieved in the next browsing session for updating the local whitelist $W_L$.

To summarize, if the requested domain is not found in the blacklist, it is first checked in the session whitelist $W_S$ followed by local whitelist $W_L$, community whitelist $W_C$, and the global whitelist $W_G$. If the domain is found in any of the whitelists, the request is allowed. Otherwise, the URL is subjected to additional tests as described in the next sections.

\subsection{IP address}
It has been established in the literature~\cite{Garera:2007,whittaker:2009}, that if the URL's hostname is an IP address, they are not trustworthy in general. Hence, we classify such URLs as phishing and give a warning message to the user. IP address is represented in either hex or the usual dot-decimal notation. To reliably detect whether the hostname is an IP address or not, we use \texttt{is(`ip')} function from the \texttt{URI.js} library~\cite{URIlib}. We note that not all IP addresses are unsafe. This is especially true in the development and testing environment where developers work on machines setup on the internal network. If an IP address is visited frequently over a period of time, it will eventually be added to the local whitelist and all future web requests to it will be allowed. Alternatively, organizations can proactively whitelist a list of IP addresses to enable access to server machines on their internal network.

\subsection{Context}
If the input domain is not present in any of the whitelists, we determine the context in which the URL is being visited. Each URL visit in the Chrome browsing history is associated with one of the transition types such as link (click), typed and reload. However, when we intercept the web request in Chrome browser using the \texttt{onBeforeRequest} event handler in the \texttt{chrome.webRequest} API, the corresponding web request object does not contain details about the transition type. Therefore, in order to infer the transition type of the URL, we use certain heuristics. Specifically, we rely on the \texttt{initiator} and \texttt{tabId} fields of the web request object. The \texttt{initiator} indicates the origin where the request was initiated and \texttt{tabId} indicates the ID of the tab in which the request takes place. 

We observed that if the user clicks a URL and opens the new web page in the same tab, then the $initiator$ field of the corresponding web request object is set to the URL of the previous web page (where the request was made). In this case, it is certain that the transition type is link. However, if the user opens the URL in a new tab (with right-click) or clicks the URL from an external application ({\em e.g.}, email client), then the $initiator$ field is undefined. The user can also visit a web page by typing its URL in the address bar of the browser. It can happen in two ways. First, the user creates a new tab (\texttt{chrome://newtab/}) and then types the URL or second, the user types the URL in one of the existing tabs. In both cases, the $initiator$ field in the request object is undefined. Therefore, the $initiator$ field alone is not sufficient to determine the transition type of the requested URL.

To determine whether the transition type is link or typed when the $initiator$ field is undefined, we monitor tab creation and updation events using the \texttt{chrome.tabs} API. Specifically, we maintain a hashmap $tabUrl$ which consists of $tabId$ as key and latest URL associated with that tab as value. All tabs have unique $tabId$ within a browser session. Further, the Id of the tab remains the same even if the URL associated with the tab is changed. We observed that if the link is opened in a new tab, then the \texttt{onBeforeRequest} event is handled before the tab creation event. Therefore, in this case, the key $tabId$ in the hashmap $tabUrl$ does not exist yet and the transition type is determined as link. On the other hand, the presence of key $tabId$ implies that the web request has taken place in the existing tab and the transition type of the request is typed. We note that if the page is refreshed or back button is pressed, then the corresponding domain will be found in the session whitelist, and the request will be allowed directly.

The pseudocode for determining the transition type is described as Algorithm~\ref{ContextAlgo} in the appendix. 

\subsection{Popular Search Engine}
If the transition type of the URL request is $link$, we determine if the URL is clicked from a popular search engine results page. It has been observed that, users rely on search engines to find information about specific websites on the internet. Popular search engines such as Google Search detect phishing web pages during crawling phase (using proprietary mechanisms) and avoid indexing them \cite{whittaker:2009}. Hence, the search engine results can be trusted. 

Browser extensions can access the content of web pages through content scripts. As Google Search is the most popular search engine, we created a content script for parsing Google Search results pages. Similar scripts can also be created for other search engines such as Bing or Yahoo search. The content script is injected on every Google Search page immediately after the DOM is constructed but before resources such as images are loaded. The content script extracts and stores two key information in a data structure, the number of results returned by the search engine and URLs on the current results page. This data is communicated via the \texttt{chrome.runtime} API to the main background script. 

We trust URLs from only those searches that retrieve a large number of search results, say more than 10,000 and only top results, say top 20. Therefore, if the number of results returned by the search engine is greater than a pre-defined threshold (10,000), our script stores the top 20 in a hashmap against senders $tabId$ as key. If the requested URL is found in the hashmap, that implies the URL is clicked from the search result page and hence, can be considered as safe to visit.


\subsection{IDN Attack}
If the URL is not clicked from the search engine results page, we check the domain for the presence of international characters which can be visually similar to other characters and deceive users~\cite{UnicodeSecurity}. Modern web browsers generally convert domains with Unicode characters (\texttt{paypal.com}) to the Punycode encoding (\texttt{xn-{}-pypal-4ve.com}), a representation of domains with the limited ASCII character subset $\{a$-$z \ 0$-$9$ . -$\}$. Once, a Punycode domain is detected , we call the \texttt{isIDNAttack} procedure (described as Algorithm~\ref{IDNAlgo} in the appendix).

We begin by converting the domain back into Unicode using \texttt{toUnicode()} function from the \texttt{punycode.js} library~\cite{Punycode}. For security purposes, Unicode consortium maintains what is known as confusables file that provides a mapping of source characters to their lookalike Latin (ASCII) letters among others~\cite{UnicodeConfusables}. For instance, Cyrillic small letter `a' (U+0430) maps to visually similar Latin small letter `a' (U+0061). However, the confusables map alone is not sufficient to detect IDN homograph attacks. 

To evade detection, attackers often employ accented characters to create visually similar domains and such mappings are not present in the confusables file. For instance, in fake domain \texttt{paypaĺ.com}, letter `l' is substituted with accented character `ĺ' (U+013a). Therefore, we first apply NFKD normalization~\cite{UnicodeNormalization} which decomposes an accented character (`ĺ') in the Unicode domain into a letter (`l') and acute accent ($'$), and then use the confusables map to substitute a non-ASCII character to ASCII character(s). Finally, we retain only the set of ASCII characters that are allowed in the domain and check if the resulting ASCII domain string is present in any whitelist. If yes, then the domain is identified as an instance of IDN homograph attack and an appropriate warning is displayed to the user, otherwise we check the resulting ASCII string for misspelling(s) of a popular domain.

\subsection{Aho-Corasick Machine}~\label{sec:aho}
If the input domain is not an instance of IDN attack, we check the URL for the presence of popular brand name and domain name from our global whitelist. One simple approach is to build a hash table out of the popular brand names in advance, and tokenize each input URL string using special characters ({\em e.g.}, `.', `-') and lookup the resulting tokens in the hash table. However, brand names may not always appear as separate tokens in the phishing URLs. For instance, in the hostname \texttt{auth-signpaypalalertcustomerapp.apacayus} \texttt{.com}, the brand \texttt{paypal} is embedded in the subdomain \texttt{signpaypalalertcustomerapp}. Hence, this approach could result in false negatives (missed detection). Therefore, one has to search for the occurrence of each brand name at all possible positions in the URL string which is inefficient as it requires $O(l \cdot \sum_{j=1}^{m} |y_j|)$ time, where $l$ is the number of characters in the text (URL), $m$ is the number of keywords (brand names and domains) in the dictionary $D$ and $|y_j|$ is the length of the $j^{th}$ keyword. Moreover, this approach does not scale well with large dictionaries. We note that the number of domains in our global whitelist is $50,000$ and, therefore, the size of the dictionary (brand names and domains) is at least $50,000$. 

To address these problems, we use Aho-Corasick pattern matching algorithm~\cite{Aho:1975}, which examines every character in the text O(1) times and detects patterns (if any) simultaneously in just $O(l \cdot \sum_{k=1}^{r} |p_k|)$ time, where $r$ is the number of patterns found in the input text of length $l$ and $|p_k|$ is the length of the $k^{th}$ matched pattern. It works by constructing a pattern matching machine from a finite set of keywords $Y = \{y_1, y_2, \ldots y_m\}$ to be matched, where each keyword $y$ is composed of symbols from an input alphabet $\Sigma$.

The machine is constructed~\cite{Aho:1975} using a trie to store the set of keywords $Y$. It is represented using three data-structures, a $goto$ function that maps a state $s$ and an input symbol $a \in \Sigma$ to either another state $s'$ or the message fail, a $fail$ function to be consulted whenever $goto$ function outputs the fail message, and an $output$ function that maps a state $s$ to a set of outputs $\{o_1, o_2, \ldots o_t\}$ representing the set of matched keywords.

\begin{figure}[!t]
\begin{subfigure}[b]{\linewidth}
  \includegraphics[scale=0.35]{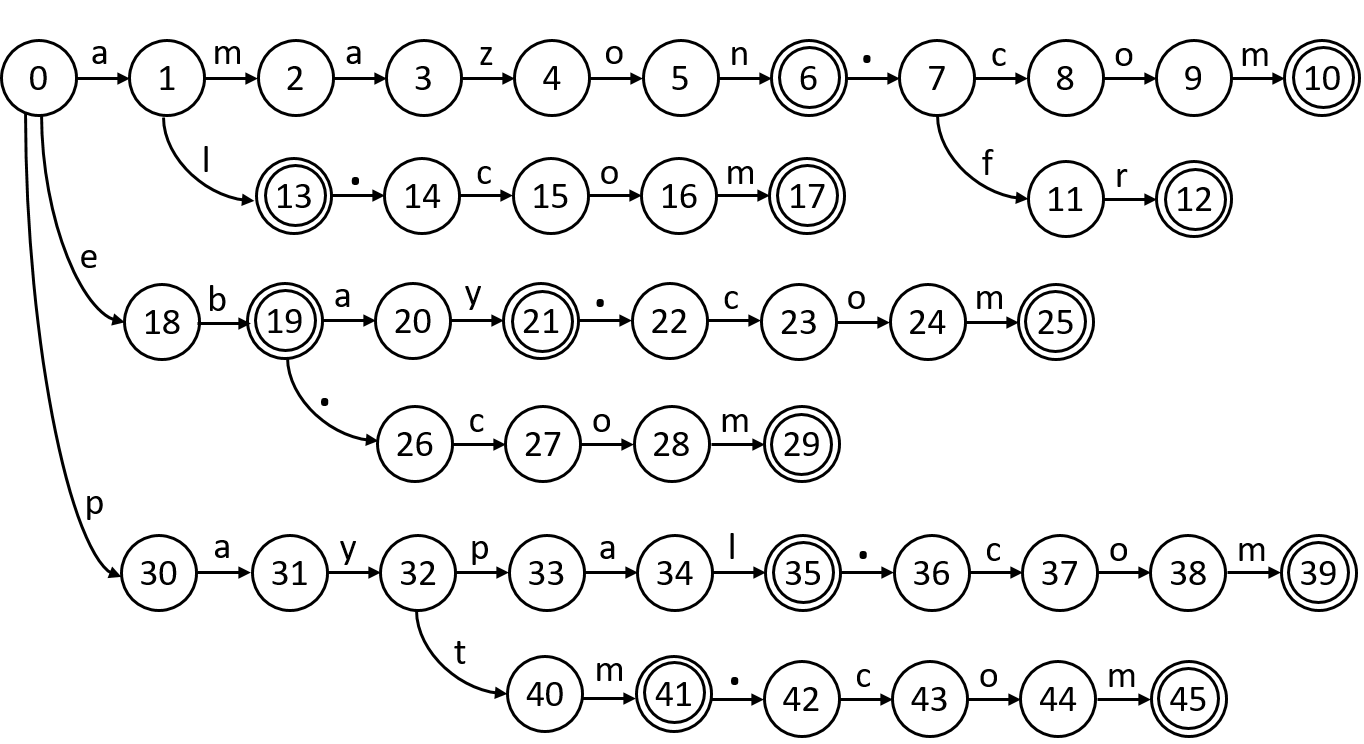}
  \caption{Goto function}~\label{fig:goto_unopt}
\end{subfigure}

\begin{subfigure}[b]{\linewidth}
  \centering
  \includegraphics[scale=0.4]{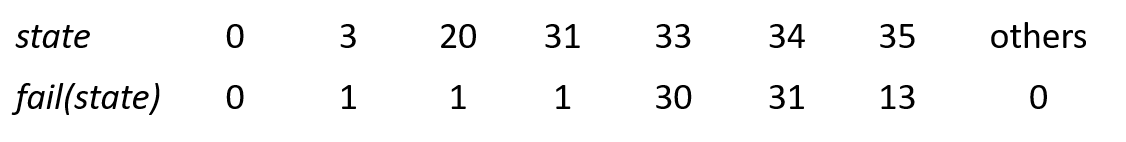}
  \caption{Fail function}~\label{fig:fail_unopt}
\end{subfigure}

\begin{subfigure}[b]{\linewidth}
  \centering
  \includegraphics[scale=0.4]{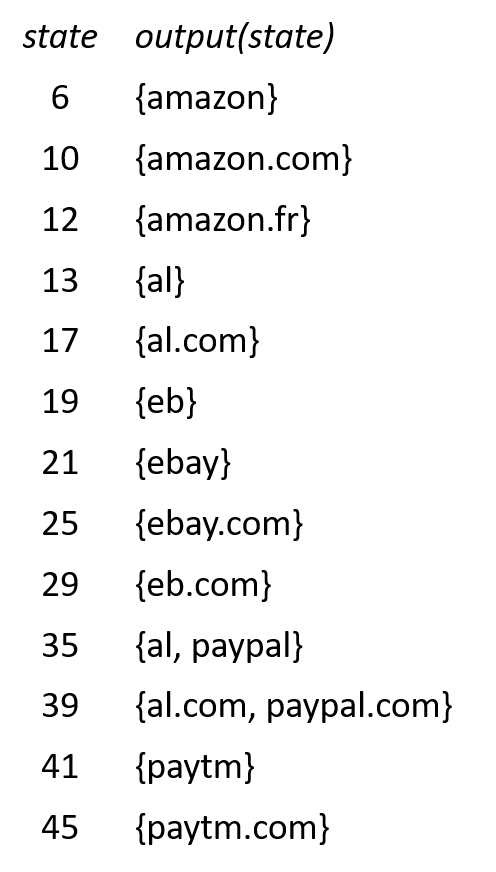}
  \caption{Output function}~\label{fig:output_unopt}
\end{subfigure}

\begin{subfigure}[b]{\linewidth}
  \centering
  \includegraphics[scale=0.4]{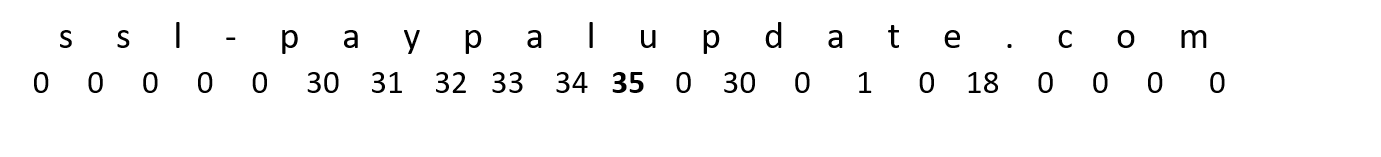}
  \caption{Pattern matching using Aho-Corasick machine}~\label{fig:match}
\end{subfigure}
\caption{Unoptimized pattern matching machine $M$. \\ \centering $\Sigma = \{a$-$z$, $0$-$9$, -, .$\}$}
\label{fig:aho_unopt}
\end{figure}

Figure~\ref{fig:aho_unopt} shows an Aho-Corasick pattern matching machine $M$ constructed out of set $Y = \{$\texttt{amazon.com}, \texttt{amazon.fr}, \texttt{al.com}, \texttt{eb.com}, \texttt{ebay.com}, \texttt{paypal.com}, \texttt{paytm.com}$\}$ consisting of seven domains. The states where either brand name ({\em e.g.}, \texttt{paypal}) or domain name ({\em e.g.}, \texttt{paypal.com}) terminates are designated as output states (double circled in Figure~\ref{fig:goto_unopt}) that give the output function in Figure~\ref{fig:output_unopt}. Figure~\ref{fig:match} depicts the behavior of the pattern matching machine $M$ on the input string \texttt{ssl-paypalupdate.com}. The machine starts at state 0 and follows goto and fail functions, making state transitions and matching patterns at any output state it encounters. This matching process continues until all characters of the input string are read. 

Aho-Corasick algorithm in its original form is fast but consumes a lot of memory. During implementation, $goto$ and $fail$ functions can be combined and stored in a two-dimensional array $A$~\cite{Aho:1975}, where each row corresponds to a state and each column corresponds to an input symbol. Thus, given a state $s$ and an input symbol $a \in \Sigma$, one can determine the next state in constant time by accessing the entry $A[s][a]$. 
In such an implementation, the memory requirement (in bits) of each state (excluding the memory needed to store output) is:
\begin{equation}~\label{eq:unopt_state}
|\Sigma| \cdot sizeof(int)
\end{equation}
Therefore, the memory requirement of the entire pattern matching machine is at least:
\begin{equation}
S \cdot |\Sigma| \cdot sizeof(int)
\end{equation}
where $S$ is the number of states in the machine, $\Sigma$ is the set of input symbols and $sizeof(int)$ is the number of bits required to represent a state number. However, states in the pattern matching machine are densely connected at the top and sparsely at the bottom. Based on this observation, researchers proposed several optimization techniques to reduce the size of the pattern matching machine~\cite{Tuck:2004, Zha:2008}. These techniques can be broadly grouped into two classes, a) reducing the size of each state and b) reducing the number of states.
\\
\\
\textbf{Bitmap State} : Tuck {\em et al.}~\cite{Tuck:2004} showed that the size of each state can be reduced significantly using a bitmap representation. Each bit $b$ in a bitmap indicates whether the transition on the corresponding input symbol $a \in \Sigma$ is valid {\em i.e.} $goto(s, a) \neq fail$ or not. The corresponding valid transitions from state $s$ are stored in an array called $nextarr$. As before, the failure pointer is required in case there is no transition from state $s$ on some input symbol $a \in \Sigma$. With these modifications, the size of a state (in bits) in the Aho-Corasick pattern matching machine is equal to the sum of memory required for its $bitmap$, $nextarr$ and $fail$:
\begin{align}~\label{eq:opt_state}
|\Sigma| + |nextarr| \cdot sizeof(int) + sizeof(int)
\end{align}
where, $|nextarr| \le |\Sigma|$ as the number of transitions from any given state is at most $|\Sigma|$. Typically, $|nextarr|$ is much smaller than $|\Sigma|$. 

\begin{figure}[!t]
\begin{subfigure}[b]{\linewidth}
\centering
  \includegraphics[scale=0.4]{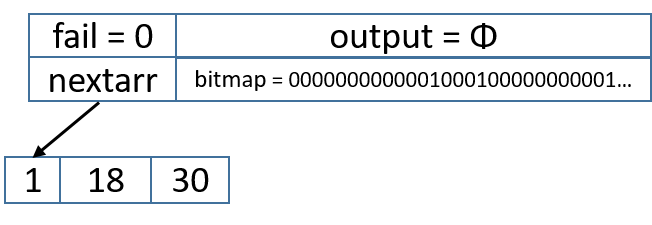}
  \caption{Bitmap state representation of start state 0~\cite{Tuck:2004}}~\label{fig:bitmap}
\end{subfigure}

\begin{subfigure}[b]{\linewidth}
\centering
  \includegraphics[scale=0.4]{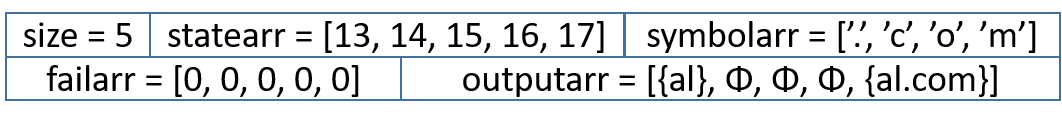}
  \caption{Path compressed state representation of states 13-17~\cite{Tuck:2004}}~\label{fig:compressed}
\end{subfigure}

\begin{subfigure}[b]{\linewidth}
  \centering
  \includegraphics[scale=0.4]{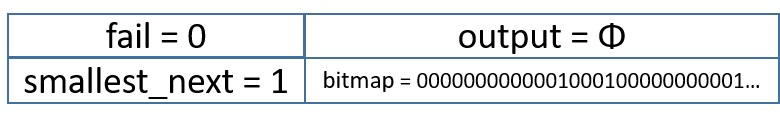}
  \caption{Bitmap state representation of start state 0 in lex trie}~\label{fig:lex}
\end{subfigure}

\begin{subfigure}[b]{\linewidth}
  \centering
  \includegraphics[scale=0.4]{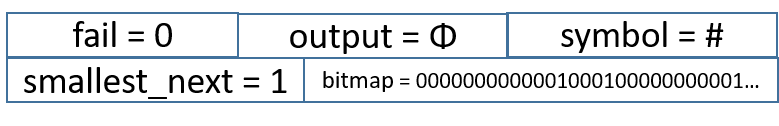}
  \caption{Bitmap-lex state representation of start state 0}~\label{fig:bitmap_lex}
\end{subfigure}

\begin{subfigure}[b]{\linewidth}
  \centering
  \includegraphics[scale=0.4]{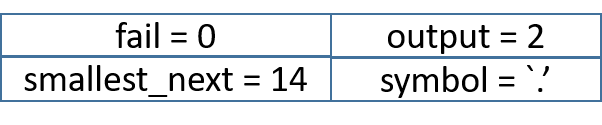}
  \caption{Lex state representation of state 13}~\label{fig:lex2}
\end{subfigure}
\caption{Different representations of states of machine $M$ depicted in Figure~\ref{fig:aho_unopt}.}
\end{figure}

The structure of a bitmap state is illustrated in Figure~\ref{fig:bitmap} which corresponds to the start state 0 of machine M in Figure~\ref{fig:aho_unopt}. Bit numbers are mapped to symbols in $\Sigma = \{$a-z, 0-9, -, .$\}$ in the order of their ASCII codes. Therefore, bit 0 corresponds to `-', bit 1 corresponds to `.', bits numbered 2-11 corresponds to 10 digits, and bits from 12 to 37 corresponds to 26 letters. Bits corresponding to the symbols which have valid transitions from state 0 are set to 1 and the rest of the bits are kept 0. 
The offset to access the next state from $nextarr$ is calculated by counting the number of set bits in the bitmap prior to the bit corresponding to the current symbol. 
The failure pointer and the output pointers work as previously. 

If the size of an integer is 64 bits and the number of symbols in $\Sigma$ is 38 (as depicted in Figure~\ref{fig:aho_unopt}), then the memory required to store the start state in the unoptimized Aho-Corasick machine (Equation~\ref{eq:unopt_state}) is $38 \cdot 64 = 2,432$ bits. Since, the number of valid transitions from the start state is 3 (size of $nextarr$), the storage required for its bitmap representation (Equation~\ref{eq:opt_state}) is just $38 + 3 \cdot 64 + 64 = 294$ bits. 

As the size of bitmap is $|\Sigma|$ bits, the determination of offset requires $O(|\Sigma|)$ time in the worst case. To make the process of determining offset more efficient, Zha {\em et al.}~\cite{Zha:2008} suggested to maintain multiple level of summaries for each bitmap state. One possibility is to partition the bitmap into $k$ nearly equal-sized blocks and to store the running count of set bits till that block. Using $k$ summaries, we can determine the offset with at most $\frac{|\Sigma|}{k} + 1$ additions.
\\
\\
\textbf{Path Compressed State} : While bitmap reduces the memory requirement of states in the pattern matching machine, it still results in wasted space for states at the bottom. Therefore, Tuck {\em et al.}~\cite{Tuck:2004} introduced path compression that involves merging a chain of states at the bottom of the pattern matching machine into a single large composite state, thereby eliminating the need of bitmap. An example of a composite state is shown in Figure~\ref{fig:compressed}. The path compressed state consists of the input symbols and the corresponding sequence of states. It also stores the failure pointer and output set for each state. However, we need to keep track of whether the state is bitmap or path compressed. The memory requirement of failure pointer also increases because we must now include an offset within the state to account for the possibility of failure transitions into the middle of a path compressed state. Further, in case of path compressed state, we need to store its size and input symbols on which the state transitions occur~\cite{Tuck:2004}. 

\subsubsection{Improvements}
We build upon the optimizations performed by Tuck {\em et al}.~\cite{Tuck:2004} and further reduce the memory consumption of the Aho-Corasick pattern matching machine. In particular, we propose three improvements; the first two are generic and applicable to any set of keywords, while the third improvement exploits the structure of domains. The improvements are as follows:
\\
\\
\textbf{Lex Trie} : We observed that if the child (next) states of each state in the trie are numbered consecutively in the order in which the corresponding input symbols appear in the bitmap, then instead of storing all child (next) states in the array $nextarr$ we need to store only the child (next) state with the smallest state number. As a result, the memory required for the bitmap state is greatly reduced. More formally, if there are $k$ valid transitions from a state $s$ on input symbols $a_1, a_2, \ldots a_k$ as indicated by set bits $b_{a_1}, b_{a_2}, \ldots, b_{a_k}$ in the bitmap, where $a_i < a_{i+1}$ for $1 \le i < k$, and the corresponding next states are numbered as $s_{a_{i+1}} = s_{a_{i}} + 1$, then we need to store only the smallest next state $s_{a_1}$ corresponding to the input symbol $a_1$. The next transition state from state $s$ on any input symbol $a_i \in \Sigma$ can be determined by computing offset using bitmap of state $s$ and adding it to the smallest transition state $s_{a_1}$. With these modifications, the size of any state (in bits) is equal to the sum of memory required for its $bitmap$, $smallest\_next$ and $fail$.
\begin{align}~\label{eq:lex_state}
|\Sigma| + sizeof(int) + sizeof(int)
\end{align}
We give an example. Suppose that we number the next states from state 0 on input symbols `a', `e' and `p' consecutively as 1, 2 and 3 respectively. The start state 0 can now be represented as shown in Figure~\ref{fig:lex}. Note that the array $nextarr$ (Figure~\ref{fig:bitmap}) has been replaced by the element $smallest\_next$ which stores the smallest next state from state 0, {\em i.e.}, state 1. Therefore, the storage required for the lex representation of the start state is just $38 + 64 + 64 = 166$ bits. If the machine is in state 0 and it reads letter `p', then the next state is determined as follows:
\begin{enumerate}[labelindent=0em,noitemsep,nolistsep,leftmargin=*]
\item First, bit 27 corresponding to letter `p' in the bitmap of state 0 is consulted (Figure~\ref{fig:lex}). As it is set to 1, we know that there exists a valid transition from the start state on letter `p'. 
\item Second, we count all set bits prior to bit 27 in the bitmap. The number such bits is 2 (bits 12 and 16).
\item Third, we add the count to the smallest next state. Adding 2 to the smallest next state 1, we get the next state 3, {\em i.e.}, $goto(0$, `$p$'$) = 3$.
\end{enumerate}
In order to create a pattern matching machine where the child states of each node are numbered consecutively, we first sort all keywords in lexicographic order and then construct the trie level by level starting from state 0 until all keywords are inserted. 
We call the resulting trie as $lex \ trie$. 
\\\\
\textbf{Bitmap-lex State} : There is a potential to save more space and simplify the path compression implementation proposed by Tuck {\em et al}.~\cite{Tuck:2004}. The number of valid transitions from a state $s$ is referred to as its degree. Instead of compressing a chain of states into a composite state which results in an additional overhead, we eliminate bitmap for states that either have no transition (degree 0) or have only one valid transition (degree one). To achieve this, we associate a variable $symbol$ with each state that can take one of the following three values: an input symbol (in $Sigma$) or special characters `$\#$' or `$\$$'. 
\begin{itemize}[labelindent=0em,noitemsep,nolistsep,leftmargin=*]
\item If state $s$ has more than one transitions, we set the value of $symbol$ as `$\#$' which indicates that state $s$ requires bitmap. Such state is referred as \textit{bitmap-lex} state. Bitmap required by bitmap-lex state is stored in a hashmap $bitmap\_table$ which is indexed by the state number.
\item If state $s$ has only one valid transition on some input symbol $a \in \Sigma$, we set $symbol$ as $a$. The corresponding next state is stored in the variable $smallest\_next$. As state $s$ does not require bitmap, we refer to it simply as \textit{lex} state. 
\item If state $s$ is a leaf node and has no valid transitions, we set $symbol$ as `$\$$' and $smallest\_next$ as $null$. 
\end{itemize}
Each state object consists of three pieces of information: $symbol$ which indicates whether the state is bitmap-lex or not, $smallest\_next$ that contains the smallest next state in the lex trie and $fail$ which points to a failover state when there is no valid transition on a given input symbol. We do not need to store an offset for the failure pointer as required by the path compression implementation~\cite{Tuck:2004}, since none of the states are composite. Further, we do not store the actual pattern and store only its length. The matched pattern can be easily extracted using the position of the text where the pattern is found and the length of the matched pattern.

The examples of bitmap-lex state for start state 0 and lex state for state 13 from pattern matching machine $M$ (Figure \ref{fig:aho_unopt}) are given in Figure~\ref{fig:bitmap_lex} and Figure~\ref{fig:lex2} respectively. Since state 0 has more than one transitions as indicated by symbol `$\#$', it requires bitmap. Since state 13 has only one valid transition on symbol `.' , we do not store its bitmap. Further, brand $al$ terminates at state 13, hence, we store its length 2 as the state's output. 

Both bitmap-lex state and path compressed state eliminate the need of bitmap for sequential states. However, bitmap-lex state requires additional memory for storing the variable $symbol$, whereas path compressed state requires additional memory for storing failure pointers, a variable indicating whether the state is bitmap or path compressed, size of path compressed state and input symbols on which state transitions occurs. Further, the number of states in the path compressed implementation is less than the bitmap-lex implementation since a chain of sequential states in the former is coalesced into a composite state. Therefore, determining the condition where the bitmap-lex state representation is more memory efficient than the path compressed representation is not straightforward and depends on the particular set of keywords to be matched. Yet, we can prove that under certain circumstances, the bitmap-lex state representation consumes less memory than the bitmap state representation.

Now, we determine the condition as to when this proposed implementation is more memory efficient than the bitmap implementation. Although maintaining the variable $symbol$ for each state requires additional memory, we also save memory by not requiring bitmap for all states. If bitmap is stored for each state, then the memory requirement is $S \cdot |\Sigma|$ bits, where $S$ is the total number of states in the pattern matching machine and $\Sigma$ is the input alphabet. Let $f$ be the fraction of states with $degree > 1$. For each of these states, the proposed implementation requires memory of $|\Sigma|$ bits for storing bitmap in a hashmap and additional memory $\alpha$ bits to store the special character `$\#$'. The remaining $1-f$ states have degree at most one and need only $\alpha$ bits for storing the transition character. Therefore, the total memory requirement of our implementation is
\begin{equation}
S \cdot f \cdot (c \cdot |\Sigma| + \alpha) + S \cdot (1-f) \cdot \alpha
\end{equation}
We want the memory requirement of our implementation to be less than the bitmap implementation. Thus,
\begin{equation}
S \cdot f \cdot (c \cdot |\Sigma| + \alpha) + S \cdot (1-f) \cdot \alpha < S \cdot |\Sigma|
\end{equation}
where $c \ge 1$ is the constant associated with the hashmap implementation. Simplifying the equation, we get
\begin{align}
\nonumber f \cdot c \cdot |\Sigma| + f \cdot \alpha + \alpha - f \cdot \alpha < |\Sigma| \\
\nonumber f \cdot c \cdot |\Sigma| + \alpha < |\Sigma|\\
f < \frac{|\Sigma| - \alpha}{c \cdot |\Sigma|} = \gamma
\end{align}
We built an Aho-Corasick pattern matching machine~\cite{Aho:1975} using top 50,000 brand names and domains from tranco-list \cite{pochat2019tranco}. We found that around 95\% of the states either have no transition at all (degree 0) or has only one valid transition (degree one). Thus, we can get rid of bitmap for all such states. For the built trie, we have $f \sim 0.05$, $|\Sigma| = 38$, $\alpha = 8$ bits (assuming only ASCII letters), $c = 2$ (at most twice memory requirement for hashmap), and therefore $\gamma = 30/76 \sim 0.395 > f$. Hence, in this case our proposed implementation is more efficient than the bitmap implementation. Note that the memory required for storing fail pointers and output is same in both implementations.

The bitmap implementation requires $|\Sigma|/2$ time to access the bitmap. With our proposed implementation, $f$ fraction of states require O(1) time to access the array and $|\Sigma|/2$ time to access the bitmap whereas the remaining $1-f$ states require only O(1) time to access to the array. Therefore, the expected time required is
\begin{align*}
& f \cdot (|\Sigma|/2 + 1) + (1-f) \\
= & f \cdot |\Sigma|/2 + 1 
\end{align*}
We have $f \sim 0. 05$ and $|\Sigma| = 38$, the expected time is $0.05 \cdot 38/2 + 1 = 1.95 < |\Sigma|/2$. Hence, in this case, our proposed implementation is also faster than the bitmap implementation.

\begin{figure}[!t]
\begin{subfigure}[b]{\linewidth}
  \includegraphics[scale=0.35]{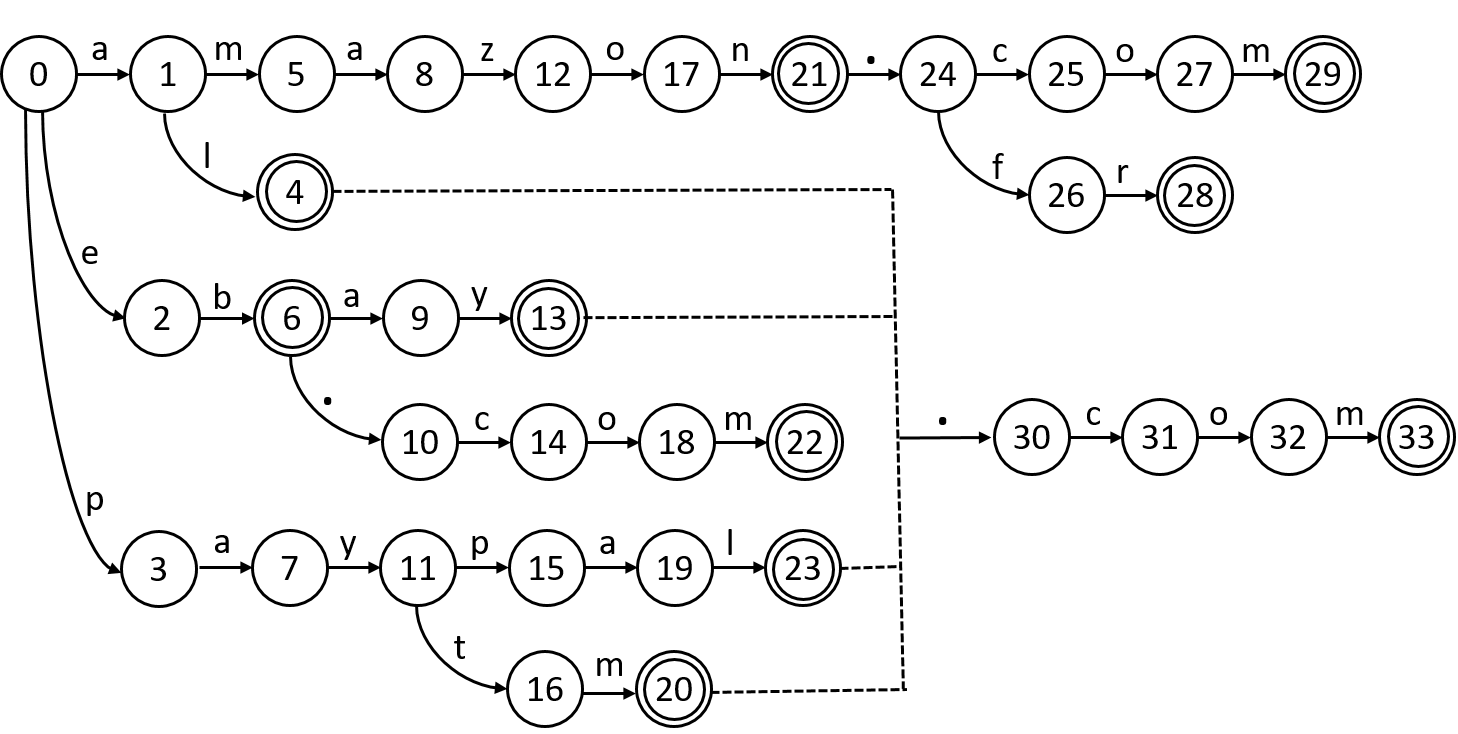}
  \caption{Goto function}~\label{fig:trie_opt}
\end{subfigure}

\begin{subfigure}[b]{\linewidth}
  \centering
  \includegraphics[scale=0.4]{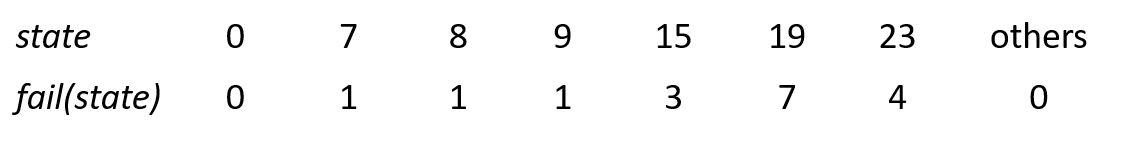}
  \caption{Fail function}~\label{fig:fail_opt}
\end{subfigure}

\begin{subfigure}[b]{\linewidth}
  \centering
  \includegraphics[scale=0.4]{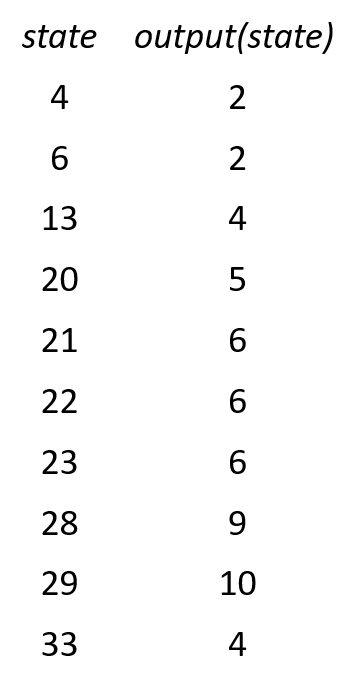}
  \caption{Output function}~\label{fig:output_opt}
\end{subfigure}

\begin{subfigure}[b]{\linewidth}
  \centering
  \includegraphics[scale=0.4]{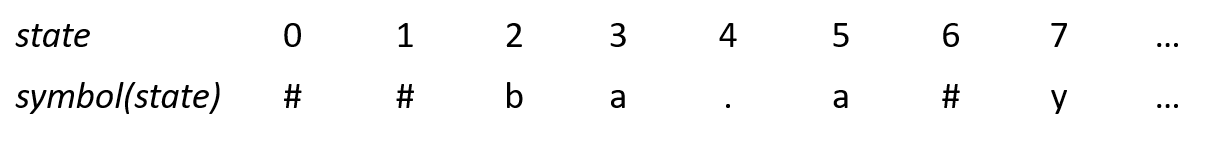}
  \caption{Symbol function}~\label{fig:symbol_opt}
\end{subfigure}

\begin{subfigure}[b]{\linewidth}
  \centering
  \includegraphics[scale=0.4]{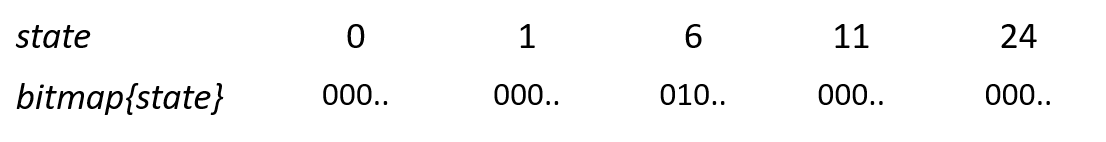}
  \caption{Bitmap Hashmap}~\label{fig:bitmap_opt}
\end{subfigure}

\begin{subfigure}[b]{\linewidth}
  \centering
  \includegraphics[scale=0.4]{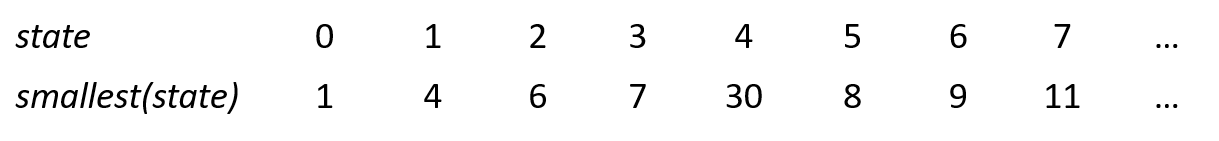}
  \caption{Smallest next function}~\label{fig:small_opt}
\end{subfigure}

\begin{subfigure}[b]{\linewidth}
  \centering
  \includegraphics[scale=0.4]{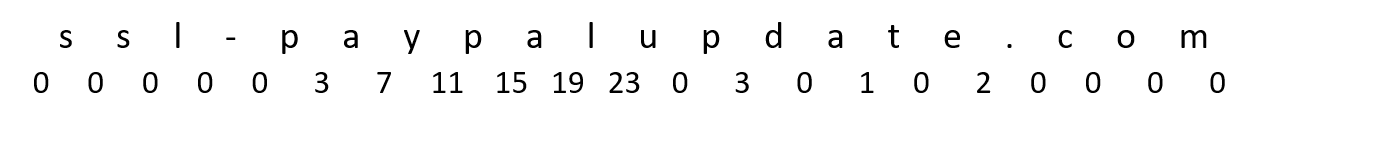}
  \caption{Pattern matching using optimized Aho-Corasick machine}~\label{fig:match_opt}
\end{subfigure}

\begin{subfigure}[b]{\linewidth}
  \centering
  \includegraphics[scale=0.4]{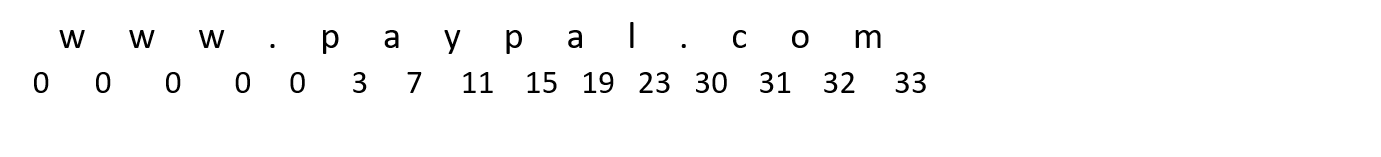}
  \caption{Pattern matching using optimized Aho-Corasick machine}~\label{fig:match2_opt}
\end{subfigure}

\caption{Optimized pattern matching machine $M_{opt}$.}~\label{fig:aho_opt}
\end{figure}

\noindent
\textbf{TLDs Compression} : We observed that the number of domains in our global whitelist is 50,000, however, the number of unique TLDs is only 635. The most common TLD \texttt{com} appears in more than 28,000 domains. Therefore, the memory consumption of the pattern matching machine can be reduced significantly, if one could avoid storing the same TLD multiple times. To minimize the memory usage due to repeated storing of TLDs, we propose the following modifications. 
\begin{itemize}[labelindent=0em,noitemsep,nolistsep,leftmargin=*]
\item \textbf{Lex Trie} : We split each domain $d$ in the global whitelist $D$ into brand $b$ (SLD) and TLD $t$, and create a lex trie $B_{trie}$ consisting of only brands and a separate lex trie for each TLD $t$ (stored in a dictionary $T_{trie}$ indexed by $t$). Then, we concatenate the lex trie $B_{trie}$ with lex tries in $T_{trie}$ by adding appropriate edges between the end nodes of $B_{trie}$ and the start nodes of $T_{trie}$. More specifically, for a domain $d = b.t$, we add an edge from a node in $B_{trie}$ where the brand $b$ ends to the start node of the trie $T_{trie}[t]$ containing $t$, and label the edge with symbol `.'. Therefore, starting from the start node of $B_{trie}$, there is a unique path in the concatenated trie that spells out the domain $d = b.t$. The start state of the concatenated trie is the start state of $B_{trie}$.
However, with these modifications, we need to ensure that the resulting trie is also a lex trie. Note that, a trie is referred as lex trie if all child nodes of a state are numbered consecutively in the order in which the corresponding input symbols appear in the bitmap.

The concatenated trie violates the property of lex trie in two cases: a) if a brand is registered in multiple TLDs and b) if a brand is a prefix of another longer brand. To ensure that all child nodes of each state in the concatenated lex trie are numbered consecutively, these brands are stored along with the TLDs in $B_{trie}$. Out of 50,000 domains, the number of brands with a single TLD is 43,924 ($\sim87.84\%$). Out of these 43,924 brands, the number of brands that appear as a prefix of some longer brand is 4,073. Therefore, we can optimize the TLDs of the remaining 39,851 ($\sim79.70\%$) brands. Hence, the potential for memory saving is significant. 

For ease of reference, we refer to a brand in our global whitelist that is registered in multiple TLDs as $mbrand$. We refer to a brand registered in a single TLD but that is a prefix of some longer brand as $pbrand$. We refer to a brand registered in single TLD that is not a prefix of any other brand as $nbrand$. All these definitions are with respect to the given global whitelist. Consider the global whitelist consisting of six domains shown in Figure~\ref{fig:aho_unopt}. In this case, an example of $mbrand$ is \texttt{amazon}, $pbrand$ is \texttt{eb} and $nbrand$ is \texttt{paypal}.

With these improvements the trie in Figure~\ref{fig:goto_unopt} is converted into an optimized lex trie depicted in Figure~\ref{fig:trie_opt}. The number of states in the optimized lex trie is reduced from 46 to 34. The edges from the end nodes of $B_{trie}$ to the start nodes of $T_{trie}$ are shown using dotted lines. The TLD \texttt{com} (states 30-33) is shared by four brands \texttt{al}, \texttt{ebay}, \texttt{paypal} and \texttt{paytm}. As brand \texttt{amazon} is registered in multiple TLDs ($mbrand$), \texttt{com} and \texttt{fr}, these two domains are stored separately in $B_{trie}$ to ensure that the child nodes of state 24 (where the string \texttt{amazon.} ends) are numbered consecutively. As brand \texttt{eb} is a prefix of longer brand \texttt{ebay} ({\em i.e.}, \texttt{eb} is a $pbrand$), its TLD \texttt{com} is also stored separately, to ensure that the child nodes of state 6 (where the brand \texttt{eb} ends) are numbered consecutively. It is possible to optimize the storage of these TLDs further, however the benefits are not significant (more details are given in the appendix in Figure~\ref{fig:aho_opt2}). 

\item \textbf{Output} : Because $mbrands$ and $pbrands$ are inserted into $B_{trie}$ along with their TLDs, the states where such brands end store the corresponding brand information and the states where their TLDs end store the corresponding domain information. For instance, as shown in Figure~\ref{fig:output_opt}, state 21 stores the length of the brand \texttt{amazon} and state 29 stores the length of the domain \texttt{amazon.com}. For $nbrands$, the brand part is inserted into $B_{trie}$ which is concatenated with the TLD trie in $T_{trie}$. The state in $T_{trie}[t]$ where TLD $t$ ends stores the length of the string $.t$ as output. For instance, the TLD \texttt{com} is stored in a separate trie containing states 30-33. As the TLD \texttt{com} ends in state 33, it stores the length of \texttt{.com} (four) as the output. Because, multiple brands are concatenated to a single TLD trie, during pattern matching process it is important to keep the track of which brand is matched before the TLD is found. If during pattern matching process, a pattern is found that starts with symbol `.', then the pattern must be a TLD which is shared by multiple brands. The previous match then must be a brand, as it is not possible to reach the symbol `.' in the trie before outputting a brand.

\item \textbf{Fail} : After creating the concatenated lex trie, we compute the failure pointer for each of its states. We reset the failure pointer of the start state in $B_{trie}$ to itself. We also reset the failure pointers of states in $B_{trie}$ which ends in `.' (where TLD begins) to the start state of $B_{trie}$. Similarly, we reset the failure pointers of start state of each TLD trie in $T_{trie}$ to the start state of $B_{trie}$. The failure pointers for the remaining states in the concatenated trie are computed as described in the original Aho-Corasick Algorithm~\cite{Aho:1975}. While computing these pointers, the output of state $s$ and $fail(s)$ are merged. However, we are interested only in the longest match, hence every state is associated with at most one output pattern. For instance, in the original implementation (Figure~\ref{fig:output_unopt}), state 35 is associated with two outputs \texttt{al} and \texttt{paypal}. In our implementation, we only store the length of the longest brand \texttt{paypal} with state 23 (Figure~\ref{fig:output_opt}). As each state is associated with at most one output, we store state and output as key-value pair in a hashmap $output\_map$.

\end{itemize}

\subsubsection{Optimized Machine} To create our optimized Aho-Corasick pattern matching machine from a set of domains $D$, we first extract from $D$ a set $B$ of brands, and a set $T$ of TLDs by splitting a domain on the first dot. We then extract a set $D_M$ of domain names registered in multiple TLDs, and a set $D_P$ of domain names whose brands are a prefix of other brands by creating a temporary trie out of $D$ and traversing it in a depth-first search manner. Once, we have the required sets, the proposed optimizations can be applied appropriately. The optimized Aho-Corasick pattern matching machine is constructed offline (out of global whitelist) and only the related data-structures are distributed along with the extension. 

The procedure $OptimizedAhoCorasick$ to create the optimized Aho-Corasick pattern matching machine from a set of domains $D$ is given in the appendix as Algorithm~\ref{AhoAlgo}.

After using this algorithm, the pattern matching machine of Figure~\ref{fig:aho_unopt} is transformed into an optimized machine as shown in Figure~\ref{fig:aho_opt}. Originally, the machine consisted of 46 states, after optimization the number of states is reduced to 34. As the number of states with degree greater than one is 5, the bitmap is required only for 5 states. The rest of the states are either leaf nodes or have only one valid transition. Figures~\ref{fig:match_opt} and \ref{fig:match2_opt} depict the state transitions of the optimized machine on the input strings \texttt{ssl-paypalupdate.com} and \texttt{www.paypal.com} respectively. Thus, the optimized machine still examines every character in the input text O(1) times, and detects patterns (if any) simultaneously in $O(l\cdot\Sigma_{k=1}^r|p_k|)$ time, where $r$ is the number of patterns found in the input text of length $l$ and $|p_k|$ is the length of the $k^{th}$ matched pattern.
Algorithm~\ref{AhoMatch} in the appendix describes how the optimized pattern matching machine can be used to find patterns in the input string $x$. 

\subsubsection{Memory Consumption} We implemented five variations of Aho-Corasick algorithm, namely Original~\cite{Aho:1975}, Bitmap and Bitmap+Path compression by Tuck~\cite{Tuck:2004}, and Bitmap-lex and Bitmap-lex + TLD compression proposed in this paper, for different whitelist sizes and measured the memory consumption in MB. All implementations were done in JavaScript and the related data-structures were loaded in Chrome browser's memory. The memory consumption of each implementation is shown in Table~\ref{tab:aho}. 

Bitmap-lex implementation requires less memory than bitmap implementation, as it eliminates the need of bitmap for degree 0-1 states. We note that around 95\% of the states in original Aho-Corasick have degree 0 or 1. Although, bitmap and path compression requires ($\sim20\%$ ) less memory than bitmap alone, it requires additional overhead to store composite states and hence, it consumes more memory than our bitmap-lex implementation. Combining bitmap-lex along with TLDs compression results in the least memory consumption compared to all other implementations. It consumes only 8.705 MB memory for storing a whitelist of 50,000 domains compared to 16.158 MB required to implement Tuck's proposal, resulting in approximately $46.12\%$ reduction in memory. These memory calculations do not include the memory required for storing output information as it consumes similar amount of memory in all variations. Therefore, the proposed optimized Aho-Corasick implementation is more space efficient than the state-of-the-art~\cite{Tuck:2004} without compromising the time complexity of the algorithm.

\begin{table}[h]
\centering
\scriptsize
\caption{Memory requirements (in MB) of different Aho-Corasick implementations for different whitelist sizes.}\label{tab:aho}
\begin{tabular}{|l|l|l|l|l|l|}
\hline
\textbf{Size} & \textbf{Original}  & \textbf{Bitmap}  & \textbf{Bitmap + Path} & \textbf{Bitmap-} & \textbf{Bitmap-lex}\\
 & \cite{Aho:1975} & \cite{Tuck:2004} & \textbf{compression} \cite{Tuck:2004} & \textbf{lex} & \textbf{+ TLD }\\
\hline
5,000 & 13.584 & 2.040 & 1.639 & 1.210 & 0.985\\
10,000 & 26.874 & 4.032 & 3.248 & 2.394 & 1.801\\
20,000 & 53.640 & 8.046 & 6.479 & 4.776 & 3.831\\
30,000 & 80.396 & 12.060 & 9.715 & 8.648 & 5.548\\
40,000 & 106.904 & 16.038 & 12.927 & 9.527 & 7.626\\
50,000 & 133.674 & 20.052 & 16.158 & 11.713 & 8.705\\
\hline
\end{tabular}
\end{table}

\subsubsection{Detection of Phishing URLs}
The optimized Aho-Corasick pattern matching machine $A_{opt}$ constructed using the top 50,000 domains from tranco-list~\cite{pochat2019tranco} can be used to check if the URL $u$ is benign or phishing. If the input domain is present in the global whitelist, it would be returned as one of the match results by $A_{opt}$, and hence the URL is deemed safe to visit. Otherwise, if the brand name (SLD) is found as an exact match, then the domain is classified as an instance of TLDsquatting. If a brand name is found as a substring in the domain, it is an instance of combo-squatting, and if the subdomain contains either a popular brand or domain names, it is an instance of subdomain spoofing. Otherwise, we check if the directory contains popular domain or brand names, in which case the URL is flagged suspicious. The global whitelist contains certain brands ({\em e.g.}, \texttt{buzz}, \texttt{sports}, \texttt{health}) that are very common and likely to appear as substring of other brands. We consider a brand in our global whitelist as common if it appears in more than 20 different brands in the top 100,000 domains of the tranco-list. If such brands are detected by our pattern matching machine, we do not classify the URL as phishing. This helps in reducing the number of false positives.
The pseudocode for this procedure is described as Algorithm~\ref{MatchAlgo} in the appendix.

\subsection{Approximate Matching Algorithm}
If the URL does not contain any popular brand name or it is typed, then we extract the domain part from the URL and check whether it is a misspelling of some popular domain in the global whitelist. Prior research shows that most of the misspelled domains occur within Damerau-Levenshtein distance one from popular domains \cite{banerjee:2008}. The edit operations considered in Damerau-Levenshtein distance are insertion, deletion, substitution and swapping of adjacent characters. 

\begin{figure}[!t]
	\centering
	  \includegraphics[scale=0.5]{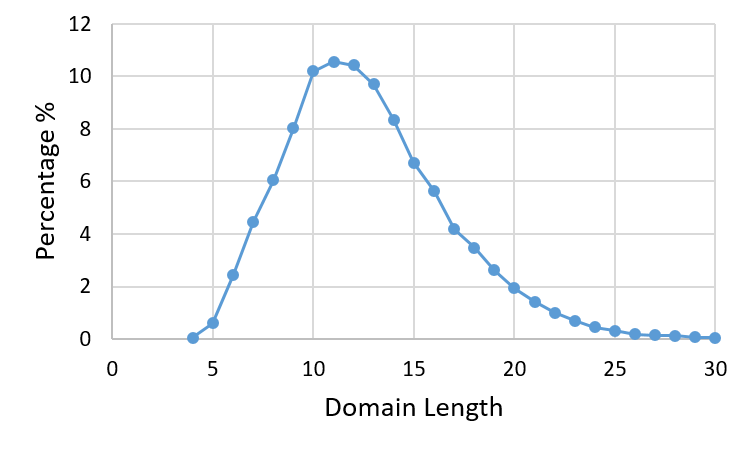}
	  \caption{Length distribution in global whitelist.}~\label{fig:len_dist}
\end{figure}

Typosquatters are targeting longer domains as most of the short domains are already in use \cite{agten2015seven}. Further, the probability of making spelling mistakes increases with the domain length. Therefore, we check if the input domain $d$ of length $l$ is within distance one of the shorter popular domains (length $ \le 10$) or within distance two of the longer popular domains (length $> 10$). A simple technique to find the candidate popular domains is as follows:
\begin{enumerate}[labelindent=0em,noitemsep,nolistsep,leftmargin=*]
\item First, retain only those domains in the global whitelist whose length $l'$ lies within the tolerance error $e$ of the input domain $d$ of length $l$, {\em i.e.}, $l - e \le l' \le l+e$.
\item Second, filter the shortlisted domains by unigrams, as the number of distinct characters between $d$ and a popular domain cannot be more than $e$.
\item Finally, compute the Damerau-Levenshtein distance between $d$ and each domain in the filtered set, and return domains within distance $e$ of $d$.
\end{enumerate}
We refer to this technique as $basic\ filter$. We note that most of the domains ($\sim$90\%) in the global whitelist have length between 6 to 20 (Figure~\ref{fig:len_dist}). If the input domain $d$ has length 12, then it needs to be compared with $\sim$50\% of the global whitelist for similarity (around 25,000 popular domains are of length 10, 11, 12, 13 and 14) which is inefficient.

Various $n$-gram based techniques have been proposed for detecting and correcting spelling errors~\cite{robertson1998applications, ahmed2009revised, islam2012text}. Motivated by these works, we design a novel $n$-gram based technique that shortlists a set of popular domains (in the global whitelist) within distance $e \le 2$ of the input domain efficiently. Suppose that the Damerau-Levenshtein distance between a misspelled domain $d$ and the popular target domain $p$ is $e \le 2$. We observe that if we divide the misspelled domain $d$ into $e+1$ nearly equal parts, then for all edit operations except swap, by pigeonhole principle, at least one part of the string (and the corresponding $n$-grams) will be free from error. This can be used for much better filtering of the candidate domains and, therefore, more efficient deduction of the domain $p$ that was misspelled as $d$ due to error in typing by the user or deliberately by an attacker to phish the user. 

\begin{figure}[!t]
\begin{subfigure}[b]{\linewidth}
\centering
  \includegraphics[scale=0.4]{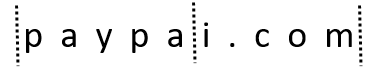}
  \caption{Substitution of l with i}~\label{fig:typore}
\end{subfigure}

\begin{subfigure}[b]{\linewidth}
\centering
  \includegraphics[scale=0.4]{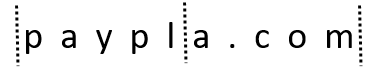}
  \caption{Swapping a and l}~\label{fig:typosw}
\end{subfigure}

\begin{subfigure}[b]{\linewidth}
\centering
  \includegraphics[scale=0.4]{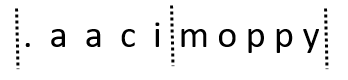}
  \caption{Sorted misspelling paypai.com}~\label{fig:typores}
\end{subfigure}

\begin{subfigure}[b]{\linewidth}
\centering
  \includegraphics[scale=0.4]{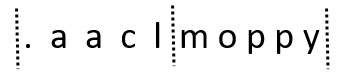}
  \caption{Sorted domain paypal.com and its misspelling paypla.com}~\label{fig:typosws}
\end{subfigure}
\caption{Misspellings of domain paypal.com}~\label{fig:typo}
\end{figure}

Now, consider the misspelled domain \texttt{paypla.com} where adjacent letters `a' and `l' from \texttt{paypal.com} are swapped.
Dividing \texttt{paypla.com} into two equal parts yields \texttt{paypl} and \texttt{a.com} (Figure~\ref{fig:typosw}). In this case, both parts contain an error. Hence, neither the trigrams \texttt{\{pay, ayp, ypl\}} extracted from the first part, not the trigrams \texttt{\{a.c, .co, com\}} extracted from the second part are entirely contained in the trigrams \texttt{\{pay, ayp, ypa, pal, al., l.c, .co, com\}} extracted from the target domain \texttt{paypal.com}. 
Therefore, to handle the swap operation, we slightly alter the $n$-gram extraction step. We observe that sorting the original string and any of its permutations produces the same string. Hence, we first sort the input domain and popular domains, and then extract $n$-grams from the sorted strings. 

Thus, if the Damerau-Levenshtein distance between a misspelled domain $d$ and the target domain $p$ is $e \le 2$, then if we sort $d$ before dividing it into $e+1$ nearly equal parts, then for all edit operations, by pigeonhole principle, at least one part of the sorted string (and the corresponding $n$-grams) will be free from error which can be used to deduce the sorted target domain $p$ the user intended to type (visit).
Figure~\ref{fig:typo} depicts some misspellings of the domain \texttt{paypal.com} along with their sorted versions. 

Suppose that the number of domains in a global whitelist is $m$, where each domain is a string over the input alphabet $\Sigma$. We create $|\Sigma|^n$ buckets, one corresponding to each $n$-gram, and add a domain $d$ to a bucket $g$, if $d$ contains $n$-gram $g$. A domain of length $l$ is thus present in at most $l-n+1$ buckets. If the average length of a domain is $l_{avg}$, then the total memory requirement is O($|\Sigma|^n + m \cdot (l_{avg}-n+1))$ and the average size of a $n$-gram bucket is ~$\frac{m \cdot ( l_{avg}-n+1)}{|\Sigma|^n}$. Thus, choosing a higher value of $n$ yields a large number of $n$-gram buckets (thus requiring more memory), but results in faster computation due to fewer candidate domains in each $n$-gram bucket. For instance, using bigrams ($n=2$) reduces the memory requirement at the expense of computation time whereas using quadgrams ($n=4$) reduces the computation time at the expense of more memory. The use of trigrams ($n=3$) achieves the balance between memory and computation; hence we employ trigrams in our experiments. For $\Sigma = \{a-z, 0-9, -, .\}$ and $m = 50,000$, the number of trigram buckets is $|\Sigma|^3 = 38^3 = 54,872$ and the average size of a trigram bucket is approximately $\frac{50000 \cdot (l_{avg}-2)}{54872} \sim 0.91 \cdot (l_{avg}-2)$. Therefore, the trigram based filtering could greatly speed up the typosquatting detection process.

Our $n$-gram based approximate matching algorithm consists of two steps, an offline preprocessing step and a matching step (Algorithm~\ref{TypoAlgo} in the appendix). In the preprocessing step, we process each domain in the global whitelist $W_G$ and create three dictionaries, a length dictionary $L$, a trigram domain dictionary $T$ and a trigram frequency dictionary $F$. We note that $W_G$ is stored using an optmized Aho-Corasick automata, however the implementation of the approximate matching algorithm is much easier if we store it using an array. Therefore, we maintain the global whitelist in the array $D$. The length dictionary $L$ consists of length $l$ as key and indices of domains with length $l$ as value. The trigram dictionary $T$ consists of trigram $t$ as key and indices of domains containing trigram $t$ as its value. Note that we first sort the domain before extracting its trigrams. Further, we do not store trigrams of domains with length less than 6 or length greater than 22. The trigram frequency dictionary $F$ consists of trigram $t$ as key and the count of the number of times trigram $t$ appears in $D$ as value. 

The matching step is performed when the input domain $d$ is received for the typosquatting check. Depending on the length of the input domain $d$, we employ two different filtering strategies to shortlist domains from the global whitelist $D$ which are potentially similar to $d$. As noted before, most of the domains in the global whitelist are concentrated between length 6 and 20 (Figure~\ref{fig:len_dist}). For such domains, we first apply trigram filtering and retrieve a list of candidate domains $D_T$. Then we filter the shortlisted domains in $D_T$ by length and unigrams, and compute Damerau-Levenshtein distance between the remaining candidate domains and the typed domain $d$. If the distance is less than or equal to $e$, we return all such domains. If the length of the input domain $d$ is less than 6 or greater than 20, we skip the trigram based filtering step. If the length of the input domain $d$ is less than or equal to 10 we find popular domains within edit distance $e=1$, otherwise we find popular domains within edit distance $e = 2$. 

In the trigram based filtering, we sort the input domain $d$ and divide it into $e+1$ near equal parts. For each part $d_q$, we first we extract the set of trigrams from $d_q$ and sort these trigrams by frequency using trigram frequency dictionary $F$. Then we determine domains in $D$ containing all trigrams of $d_q$ by taking the intersection of domain indices $T[t]$ corresponding to each trigram $t$ in $d_q$ in the increasing order of their frequency. If the frequency of some trigram $t'$ in $d_q$ is 0, it implies that $t'$ does not appear in any of the domains in $D$. In this case, we do not consider the remaining trigrams in $d_q$ as the result of their intersections is always an empty set. Similarly, we do not consider the remaining trigrams in $d_q$ once the intersection till a trigram $t'$ is deemed empty. Lastly, we take union of candidate domains found using $d_q$ with candidate domains found using previous $q-1$ parts. The part with no error will retrieve the index of the correct domain. Finally, the resulting set $D_T$ of candidate domains is returned.

The proposed approach is efficient, but requires additional memory for storing a global whitelist $D$, trigram dictionary $T$, frequency dictionary $F$ and length dictionary $L$. The space required for storing a list $D$ of $m$ domains is $O(\sum_{j=1}^{m} |d_j|)$, where $|d_j|$ is the length of $j^{th}$ domain in the list. If $\Sigma$ is the set of input symbols, the number of possible trigrams is $|\Sigma|^3$. As the dictionary $F$ stores the frequency of each trigram, the size of $F$ is proportional to $O(|\Sigma|^3)$. The number of trigrams in a domain $d_j \in D$ is $|d_j|-2$. Therefore, the corresponding index $j$ appears against $|d_j|-2$ trigrams in the dictionary $T$. Therefore, the size of $T$ is $O(|\Sigma|^3 + \sum_{j=1}^{m} |d_j|)$. As the maximum number of characters in any domain is 253, the number of keys in $L$ never exceeds 253. The dictionary $L$ stores the indices of domains in $D$ having length $l$ against key $l$, the memory requirement of $L$ is $O(m)$, where $m \gg 253$. Therefore, the total memory requirement for detecting typosquatting domains is $O(|\Sigma|^3 + \sum_{j=1}^{m} |d_j|)$. All required data-structures are created offline using the global whitelist of size $m = 50,000$ and input alphabet of size $|\Sigma| = 38$, and shipped as a part of the extension. The memory consumption of the resulting dictionaries is around 5 MB. Hence, the typosquatting detection algorithm is fast as well as lightweight.

\subsection{Machine Learning Model}~\label{sec:ml}
If the URL does not contain any popular brand name or the input domain is not a misspelling of any popular domain name, we resort to the machine learning approach. The problem of detecting phishing URLs is formulated as a binary classification task in which phishing URLs are treated as positive instances and benign URLs are treated as negative instances. Prior research shows that the presence of certain words such as $account$, $secure$ and $confirm$ in URLs are highly indicative of phishing attacks \cite{Ma:2009}\cite{Le:2011}\cite{tupsamudre:2019}. To find such phishy words, we extract the segmented bag-of-words (SBoW) features~\cite{tupsamudre:2019} from the labelled URL dataset and employ a logistic regression model to learn their weights. The words with positive weights are phishy, whereas the words with negative weights are benign. We have implemented the classifier in Python and ported the resulting model to JavaScript.

\subsubsection{Learning}
We use logistic regression model as it is effective, computationally efficient and consumes less memory as compared to more advanced techniques such as neural networks, which makes it suitable for deployment at client-side. The logistic regression model predicts the probability of a positive class, given a feature vector $x$ as follows:
\begin{equation}\label{eq:cost}
p(y=1|x;\theta, b) = \sigma(\theta^{T} x + b) = \frac{1}{1+\exp^{-(\theta^{T} x + b)}}
\end{equation}
where, $\theta \in \mathbb{R}^{n}$ and bias $b$ are the parameters (weights) of the logistic regression model, and $\sigma(\cdot)$ is the sigmoid function defined as $\sigma(z) = 1/(1+\exp^{-z})$. The output of the sigmoid function $\sigma(\cdot)$ is interpreted as the probability that the example belongs to positive class. The parameters $\theta$ and bias $b$ are estimated by minimizing the loss function ${\cal L}(\theta, b)$ using a gradient descent algorithm:
\begin{equation}\label{eq:loss}
{\cal L}(\theta, b) = \sum_{i=1}^{m} \log p(y_i|x_i) + \lambda \sum_{j=1}^{n} |\theta_j|
\end{equation}
where $y_i$ is the label of sample $x_i$ in the training dataset, $m$ is the number of training examples and $n$ is the size of a feature vector. The first term in equation~\ref{eq:loss} measures the conditional log-likelihood that the training examples are correctly classified and the second term penalizes weights with large magnitudes. This is known as $l_1$ norm regularization and it has many beneficial properties over SVM and Naive Bayes classifiers while working with large feature dimensions. First, it prevents overfitting. Second, it retains only the most relevant features and thus aids in feature selection by setting weights of unimportant features as zero. Due to these properties, logistic regression is one of the most frequently used models for phishing detection~\cite{Garera:2007}\cite{Ma:2009}\cite{tupsamudre:2019}. 

\begin{table*}[t]
	\centering
	\scriptsize
	\captionsetup{font=scriptsize}
	\caption{Examples of Segment bag-of-words (SBoW) features and numerical features.}\label{tab:features}
	\begin{tabular}{|l|l|l|}
	\hline
	\textbf{Hostname} & \textbf{SBoW} & \textbf{Numerical}\\
	\hline
	\texttt{signin-ppl-users-542544558966554-com.umbler.net} & $sub =\{$\texttt{signin, ppl, users, 542544558966554, com}$\}$, & $nchar = 47$, $ndot = 2$, \\
	& $dom =\{$\texttt{umbler}$\}$, $TLD =\{$\texttt{net}$\}$ & $nhyphen = 4$\\
	\hline
	\texttt{www.nwolb.co.uk.secureonlinelogin.s-secureuk.com} & $sub = \{$\texttt{www, nwo, lb, co, uk, secure, online, login}$\}$, & $nchar = 48$, $ndot = 6$, \\
	& $dom = \{$\texttt{s, secure, uk}$\}$, $TLD = \{$\texttt{com}$\}$ & $nhyphen = 1$ \\
	\hline
	\texttt{www.onlinenewspapers.com} & $sub = \{$\texttt{www}$\}$, $dom = \{$\texttt{online, news, papers}$\}$, & $nchar = 24$, $ndot = 2$, \\
	& $TLD=\{$\texttt{com}$\}$ & $nhyphen = 0$\\
	\hline
	\end{tabular}
\end{table*}

\subsubsection{Datasets}
PhishTank~\cite{PhishTank} and DMOZ~\cite{DMOZ} are the most popular data sources for obtaining phishing and benign URLs respectively~\cite{Ma:2009}\cite{Le:2011}\cite{tupsamudre:2019}. PhishTank is a collaborative platform where suspicious URLs are submitted by human verifiers and are marked as phish if they are voted by at least two other members of the community. The data in PhishTank is available free of cost through the PhishTank's website and API. We scraped 150,000 unique URLs from PhishTank website.
DMOZ is a large open human-edited directory of the web containing over five million URLs from different categories such as arts, business, news and others. We crawled 150,000 unique URLs from DMOZ website.
Hence, our balanced dataset consists of 300,000 unique URLs of which 150,000 URLs are phishing and 150,000 URLs are benign.

We note that DMOZ stopped functioning and has not been updated since 2017. However, we believe that DMOZ is still a valuable resource for learning brand agnostic features ({\em e.g.}, phishy words) and differentiating phishing URLs from benign URLs. Further, we observed that PhishTank data consists of full URLs of phishing websites, while data in DMOZ comprises of mostly hostnames. To avoid any bias during training, we extract features only from hostnames.

\subsubsection{Features}
To improve the detection of phishing URLs, we rely on the positional segmented bag-of-words (SBoW) features~\cite{tupsamudre:2019} which are extracted (from hostnames) as follows. First, we decompose hostname into subdomain, domain and TLD components. Second, we tokenize each of the components using special characters `.' and `-'. Third, we apply a word segmentation algorithm on each token to recover the individual words
using python's WordSegment module~\cite{pythonWord}.
We also extract three numerical features from each hostname, namely the number of characters ($nchar$), number of dots ($ndot$) and number of hyphens ($nhyphen$). It has been observed that phishing URLs have unusually long subdomains and many hyphens in the hostnames.

The main benefit of using positional bag-of-words features is that they aid in detecting obfuscation techniques where TLDs ({\em e.g.}, \texttt{com}) or phishy words ({\em e.g.}, \texttt{secure}) appear in the unexpected parts of the URLs. For instance, the word \texttt{com} is more likely to appear in the TLD part of the hostname rather than the subdomain part in which case the URL is a potential phish. Further, we use segmented bag-of-word (SBoW) features to retrieve phishy words (\texttt{update} and \texttt{secure}) embedded in combosquatting domains (\texttt{ssl-paypalupdate.com}) and phishing URLs containing unrelated domains (\texttt{secureuk.com}). 

Table~\ref{tab:features} shows positional SBoW features and numerical features for three different hostnames. The first two instances belong to phishing category and the third one belongs to benign. After applying the word segmentation algorithm, the token \texttt{secureonlinelogin} in the second instance is further divided into three words \texttt{\{secure, online, login\}}.
As shown in Table~\ref{tab:ml}, the total number of features (SBoW + numerical) in the training set were 80,016 of which only 7,917 features ($\sim$10\%) were deemed relevant (non-zero weights). 4,862 features were positive (phishy) and the rest 3,055 were negative (benign). The table also shows popular phishy words in subdomain, domain and TLD parts of the training URLs. In case of subdomains, the TLD \texttt{com} and words \texttt{pay}, \texttt{update}, \texttt{login} and \texttt{app} were highly suggestive of phishing attacks. Many phishing websites were deployed on the free web-hosting provider \texttt{000webhostapp} and hence it was the most important domain related feature in determining phishing URLs. Further, the TLD examples show that some country code TLDs are notorious for hosting the phishing content. Finally, all numerical features were given positive weights as phishing hostnames were relatively long and contained more special characters (dots and hyphens). 

\begin{table}[t]
	\centering
	\scriptsize
	\captionsetup{font=scriptsize}
	\caption{The number of features in each component, the number of relevant features, and features with positive (+ve) coefficients along with examples. }\label{tab:ml}
	\begin{tabular}{|l|l|l|l|l|}
	\hline
	\textbf{Feature} & \textbf{\#Features} & \textbf{\#Relevant} & \textbf{+ve} & \textbf{Examples}\\
	\hline
	Subdomain & 19,283 & 548 & 255 & \texttt{pay, com, update,}\\
	& & & & \texttt{login, app}\\
	\hline
	Domain & 60,076 & 7,124 & 4,437 & \texttt{000webhostapp,}  \\
	& & & & \texttt{login, migration}\\
	& & & & \texttt{support, wallet}\\
	\hline
	TLD & 654 & 242 & 167 & \texttt{bd, ng, ke, ve,}\\
	& & & & \texttt{ga}\\
	\hline
	Numerical & 3 & 3 & 3 & \texttt{nchar, ndot,}\\
	& & & & \texttt{nhyphen}\\
	\hline
	\textbf{Total} & 80,016 & 7,917 & 4,862 & \\
	\hline
	\end{tabular}
\end{table}

\subsubsection{Performance}
We randomly split our dataset consisting of 300,000 URLs into 240,000 training URLs (80\%) and 60,000 test URLs (20\%). We extract SBoW and numerical features from the training set, and train a logistic regression model to learn the association between these features and two categories of URLs (phish and benign). Subsequently, we predict the category of test URLs using the trained model. The misclassification rate (MCR) and false negative rate (FNR) of the classifier is 5.42\% and 3.23\% respectively. MCR measures the fraction of URLs classified incorrectly, whereas false negative rate (FNR) measures the fraction of phishing instances that are incorrectly detected as benign. Specifically,
\begin{equation}
MCR = \frac{FN + FP}{TP + FN + TN + FP} \cdot 100
\end{equation}
\begin{equation}
FNR = \frac{FN}{TP + FN} \cdot 100
\end{equation}
where $TP$ (true positives) is the number of phishing URLs correctly detected as phishing, $TN$ (true negatives) is the number of benign URLs correctly detected as benign, $FN$ (false negatives) is the number of phishing URLs incorrectly detected as benign and $FP$ (false positives) is the number of benign URLs incorrectly detected as phishing. 

\subsubsection{Deployment}
We port the trained logistic regression model comprising of relevant features and their weights to JavaScript, and integrate it with the PhishMatch extension. We note that the extraction of tokens (words delimited by `.' and `-') and numerical features from the hostname is easy, however segmentation of the extracted tokens requires a word segmentation algorithm and a large word corpus 
which would consume a large amount of memory in the browser. Alternatively, we can build an Aho-Corasick pattern matching machine out of the segmented word features retained in the model and store the corresponding weights (parameters) as the output. The resulting machine requires less memory as we need to store only 548 subdomain features, 7124 domain features and 242 TLD features. We can then detect the phishy (benign) words by running the subdomain, domain and TLD components of the input URL through these machines. This approach is approximate but efficient. 
PhishMatch classifies the URL as phishing or benign when the model predicts its class with a high confidence score (for example, we have used probability $p \ge 0.9$ as threshold in our experiments).


\subsection{Search Engine}
If the machine learning model cannot classify the input URL with high confidence, we extract the input domain and search it using Google Search, the most popular search engine on the internet. This component exploits the fact that most of the times, benign websites that are safe for users to visit are ranked first in the search results, on the other hand, phishing websites have lower ranking or are not ranked at all~\cite{Huh:2012}. Upon searching the domain, if Google Search returns a large number of results and if the searched domain is present in the top results, then we categorize the domain as benign, otherwise we label it as phishing. 

We construct a search query comprising of a domain name and send a synchronous XMLHttpRequest to the Google server. We parse the search results page and determine the number of results returned and the rank of the domain in the search results. For a domain to be labelled as benign, we require the number of results to be at least 10,000 and the domain should be found within top 20 results. The number of free queries to search engines are often limited and therefore needs to be used judiciously. Google Search API allows only 100 queries per day for free whereas Bing Search API allows 5000 free queries per month. Hence, we query search engine only when the machine learning model fails to classify the input URL with high confidence. If more queries are required, then one can use multiple search engines (Google, Bing, Yahoo) in a round robin fashion.

\subsection{Warning Messages}
The effective detection of phishing URLs is crucial but it is equally important to provide explanation to the user as to why the URL is classified as benign or phishing. If the warning messages are not comprehensible, there is a possibility of user overriding the system’s warning, making the detection system futile. PhishMatch enables the generation of explainable warning messages to caution users about phishing URLs. Depending on the PhishMatch component where the URL is detected as benign or phishing, an appropriate message can be constructed. Some sample messages given by PhishMatch are shown in Table~\ref{tab:messages}.

\begin{table*}[h]
	\centering
	\scriptsize
	\caption{Messages and recommendations generated by PhishMatch upon encountering different benign and phishing URLs related to \texttt{paypal}. We can also include links to various educational resources such as short text/video tutorials and awareness games~\cite{Sheng:2007}~\cite{Gamze:2014}~\cite{CJ:2018} to teach users about different components of a URL and the (underlined) detected obfuscation technique.}\label{tab:messages}
	\renewcommand{\arraystretch}{1.2}
	\begin{tabular}{|l|l|l|}
	\hline
	\textbf{URL} & \textbf{Component} & \textbf{Message(s) and Recommendation (if any)}\\
	\hline
	\texttt{https://www.paypal.com/in/home} & Whitelist & The URL is safe to visit because:\\
					& & 1. The domain \texttt{paypal.com} is present in all three whitelists, {\em i.e.}, local, community and global\\				
	\hline
	\texttt{https://www.paypal.com/in/signin} & User initiated & The URL is safe to visit because:\\
					& Google Search & 1. The URL is clicked from \texttt{Google Search} results retrieved by the user query\\
							  && 2. The number of results retrieved is \texttt{1,58,00,00,000} and the rank of the URL is \texttt{1}\\					
	\hline
	\texttt{http://paypal.com.elvalorsocial.com/} & Optimized  & The URL is not safe to visit because:\\
																	   &Aho-Corasick& 1. The domain \texttt{elvalorsocial.com} is not present in local, community or global whitelists \\
																	   && 2. The brand \texttt{paypal} is present in its subdomain\\
																	   && 3. Probable \underline{subdomain-spoofing} instance of legitimate domain \texttt{paypal.com}\\
																	   && \textbf{Recommendation} : Visit \texttt{paypal.com}\\
	\hline															
	\texttt{http://ssl-paypalupdate.com/login} & Optimized & The URL is not safe to visit because:\\
									 &Aho-Corasick& 1. The domain \texttt{ssl-paypalupdate.com} is not present in local, community or global whitelists\\
									 && 2. The brand \texttt{paypal} is present in the domain along with phishy words \texttt{ssl} and \texttt{update}\\
									 && 3. Probable \underline{combosquatting} instance of legitimate domain \texttt{paypal.com}\\    
									 && \textbf{Recommendation} : Visit \texttt{paypal.com}\\
							
	\hline
	\texttt{http://paypal.net/home}  & Optimized & The URL is not safe to visit because:\\
						   &Aho-Corasick& 1. The domain \texttt{paypal.net} is not present in local, community or global whitelists\\
						   && 2. The brand \texttt{paypal} is registered in different TLD \texttt{net}\\
						   && 3. Probable \underline{wrongTLDSquatting} instance of legitimate domain \texttt{paypal.com}\\  
						   && \textbf{Recommendation} : Visit \texttt{paypal.com}\\  
	\hline
	\texttt{http://papyal.com/login} & Misspelled  & The URL is not safe to visit because:\\
						   & domain & 1. The domain \texttt{papyal.com} is not present in local, community or global whitelists\\
						   && 2. The domain \texttt{pa}\textbf{py}\texttt{al.com} is a \underline{misspelling of domain} \texttt{paypal.com}\\    
						   && \textbf{Recommendation} : Visit \texttt{paypal.com}\\
								   
	\hline
	\texttt{http://www.pay}$_{.}$\texttt{pal.com/webapp/login} & IDNAttack   & The URL is not safe to visit because:\\
						   & and Optimized & 1. The domain \texttt{pay}$_{.}$\texttt{pal.com} is not present in local, community or global whitelists\\
						   & Aho-Corasick & 2. The domain \texttt{pay}$_{.}$\texttt{pal.com} looks similar to domain \texttt{paypal.com}\\   
						   && 3. Probable \underline{IDN Homograph attack} instance of legitimate domain \texttt{paypal.com}\\ 
						   && \textbf{Recommendation} : Visit \texttt{paypal.com}\\
	\hline
	\texttt{http://secureonlinelogin.s-secureuk.com/} &ML model & The URL is not safe to visit because:\\
						   && 1. The domain \texttt{s-secureuk.com} is not present in local, community or global whitelists\\
						   && 2. The \underline{hostname} contains words \texttt{secure}, \texttt{online} and \texttt{login}, suggestive of phishing attack\\
	\hline
	\texttt{http://securecuserver.co.uk/login.php} & System initiated & The URL is not safe to visit because:\\
					& Google Search & 1. The domain \texttt{securecuserver.co.uk} is not present in local, community or global whitelists\\
					& & 2. The number of results obtained after searching the domain using \texttt{Google Search} is only \texttt{377}\\	
	\hline				    
	\end{tabular}
\end{table*}

If the input domain is determined as benign, the URL visit is allowed and the reason (presence in whitelists, top search result, etc.) along with additional information (such as features of the whitelisted domain, number of retrieved results and the rank of the URL in the returned search results, etc.) can be accessed by clicking on the extension icon.


If the input domain is not present in any whitelist and the URL contains a popular brand name, PhishMatch notifies the user about the appropriate squatting type depending on where the brand is detected, and recommends the user to visit the website related to the detected brand. If the input domain is detected as a misspelling of some popular domain, it creates a justification for labelling the URL as phishing (by highlighting the error in the misspelled domain) and suggests the user to visit the popular domain instead. If the hostname is classified as phishing by SBoW based machine learning (ML) model, it indicates to the user the specific words in the hostname that are suggestive of phishing attack. We can also include links to various educational resources such as short text/video tutorials and awareness games~\cite{Sheng:2007}~\cite{Gamze:2014}~\cite{CJ:2018} to teach users about different components of a URL and the detected obfuscation technique.


\section{Experiments}
\label{sec:experiments}
To evaluate the efficacy of the proposed system PhishMatch, we collected browsing history from 33 Chrome browser users within our organization. With user's consent, we extracted History and Visit tables from their Chrome history database (stored as SQLite file on the local machine). All users gave consent to use their browsing data for research purpose(s). The URLs visited by users are evaluated by multiple anti-phishing softwares at the proxy server within our organization ({\em e.g.}, Symantec’s Blue Coat, Forcepoint’s Web Filter), hence we consider them as benign. To evaluate the effectiveness of PhishMatch, we crawled around 18,000 unique new confirmed phishes from the PhishTank repository~\cite{PhishTank}. We also trained different URLNet models~\cite{le2018urlnet} and compared their performance against PhishMatch. 
This dataset is different from the dataset used for training and evaluating the SBoW based machine learning model described in section~\ref{sec:ml} (See Table~\ref{tab:dataset} for detailed information on the different datasets used).

\begin{table*}[h]
	\centering
	\scriptsize
	\caption{Different datasets used to train and test PhishMatch and URLNet models. Positive class represents phishing samples.}\label{tab:dataset}
	\begin{tabular}{|l|l|r|c|l|}
	\hline
	\textbf{Dataset} & \textbf{Source} & \textbf{\#Samples} & \textbf{Class Label} & \textbf{Usage}\\
	\hline
	\textbf{PhishTank (Train)} & Crawled from PhishTank website & 120,000 & +ve & Train and validate SBoW based ML model and URLNet models\\
	\textbf{PhishTank (Test)} & Crawled from PhishTank website & 30,000 & +ve & Test SBoW based ML model\\
	\textbf{DMOZ (Train)} & Crawled from DMOZ website & 120,000 & -ve & Train and validate SBoW based ML model and URLNet models\\
	\textbf{DMOZ (Test)} & Crawled from DMOZ website & 30,000 & -ve & Test SBoW based ML model\\
	\textbf{BH-Train-60} & First 60 days history data of 25 users &  60,137 & -ve & Train URLNet models and create personalized whitelists for PhishMatch\\
	\textbf{PhishMatch (New)} & Crawled from PhishTank website & 18,000 & +ve & Test PhishMatch and URLNet models\\
	\textbf{BH-Test-30} & Remaining 30 days history data of 25 users & 102,141 & -ve & Test PhishMatch and URLNet models\\
	\hline
	\end{tabular}
\end{table*}

\subsection{Procedure}
\noindent 
\textbf{Data.} Chrome browser maintains the history of last 90 days. Out of 33 users, 25 users had the complete browsing history of past 90 days, so we evaluated the efficacy of PhishMatch on the URL data of these 25 users. The browsing data of each user was split into two parts. We used the URL data of the first 60 days for training and the URL data of the remaining 30 days for testing. All 18,000 new confirmed phishes collected from the PhishTank~\cite{PhishTank} were used for testing.
\\\\
\textbf{PhishMatch.} We developed a prototype PhishMatch extension for Chrome browser version 90.0.4430.212 on a 64-bit Windows 10 machine 
(8 GB memory and AMD Ryzen 5 3500U CPU). 
The extension operates as per the flowchart described in Figure~\ref{fig:phishmatch}. To measure the memory requirement of the PhishMatch extension, we loaded all data structures and functions related to whitelists, Aho-Corasick pattern matching machine, $n$-gram based approximate matching and the parameters of the machine learning model in the browser memory. We developed two models of PhishMatch:
\begin{itemize}
	\item PhishMatch (Full) - It utilizes all the components described in Section~\ref{sec:phishmatch}.
	\item PhishMatch (W/o-BH) - It utilizes only the global whitelist and does not initialize the local and community whitelists with user history data. All other components are included completely. (W/O-BH is short for \emph{Without Browsing History}.)
\end{itemize}

The memory consumed by the local and community whitelists in PhishMatch (Full) is in the order of a few KBs, hence, the total memory consumption of both versions of PhishMatch is around 20 MB. 

We simulated the use of PhishMatch extension on the test data of each user on both models of PhishMatch. We initialized the blacklist to empty for both the models. For PhishMatch (W/o-BH), we initialized the local and community whitelists to empty. For PhishMatch (Full), we used the browsing data of first 60 days (train data) to create the local $W_L$ whitelist and relaxed local whitelist $W_{RL}$ for each user. The parameters for creating the local whitelist were $A_L= 45$ days, $B_L = 30$ days, $V_L = 10$ days and $R_L = 7$ days, and the parameters used for creating the relaxed local whitelist were $A_C= A_L/2 \sim 22$ days, $B_C = B_L/2 = 15$ days, $V_C = V_L/2 = 5$ days and $R_C = 7$ days. The relaxed local whitelist of each user was combined and the top $k=100$ domains were used as community whitelist. All other components were implemented the same for both the models.

We used top 50,000 domains from the latest tranco-list ~\cite{pochat2019tranco} as our global whitelist. The Aho-Corasick pattern matching machine and $n$-gram based dictionaries were created using these 50,000 domains. Time taken for feature extraction (456 seconds) and training (338 seconds) of the Machine Learning component, along with the creation of the optimized Aho-Corasick machine (21 seconds) and $n$-gram based dictionaries (11 seconds) is added to get the total training time (826 seconds, or 13 minutes 46 seconds) of PhishMatch (W/o-BH) as mentioned in Table~\ref{tab:cmpresults}. 
To this, we added the average time taken to create local whitelist for each user (22 seconds) and the time taken to create community whitelist (18 seconds) to arrive at the training time (862 seconds, or 14 minutes 26 seconds) for PhishMatch (Full).

The test set of each user contains the browsing data of last 30 days. Each day is considered as a new session and the local whitelist of each user is updated using the domains accessed in the previous session.
To speed up the detection process, if some domain is determined as benign, it is added to the session whitelist of the user. However, if the domain is visited again in the current session, we attribute its detection to the PhishMatch component, which first identified it. Similarly, if some domain is identified as phishing, it is added to the user's blacklist for faster detection. If the blacklisted domain is visited again in the subsequent sessions, we attribute its detection to the PhishMatch component, which first identified it.

Next, we simulated PhishMatch models on 18,000 new confirmed phishes. Again, we initialized the blacklist to empty. Further, we did not use any local or community whitelist for W/o-BH model. However, we created the global whitelist as described above and used it in our simulation. For PhishMatch (Full), the community whitelist was created using the relaxed local whitelist of all 25 users. Each URL is evaluated as per the flowchart given in Figure~\ref{fig:phishmatch}. 
\\\\
\textbf{URLNet.} URLNet~\cite{le2018urlnet} is an end-to-end deep learning framework that learns a nonlinear embedding for Malicious URL Detection directly from the URL. Specifically, it applies Convolutional Neural Networks to both characters and words of the URL string to learn the URL embedding in a jointly optimized framework. This approach allows the model to capture several types of semantic information and is shown to outperform the existing BoW based models. There are three different URLNet models:
\begin{itemize}
\item URLNet(Character-Level) - only the Character-level CNN,
\item URLNet(Word-Level) - only the Word-level CNN,
\item URLNet(Full) - which is the end-to-end framework combining both character-level and word-level CNNs.
\end{itemize}
The source code to train and test URLNet models is available on github~\cite{URLNetCode}. We trained all three URLNet models on two different datasets. The first dataset is the same one as that used to train our SBoW based machine learning model and comprises of 240,000 unique PhishTank and DMOZ URLs (section~\ref{sec:ml}). The second dataset also uses PhishTank and DMOZ URLs, but in addition, it employs URLs from the training set of each user. The number of unique URLs in the combined training set of all users (BH-Train-60) is 60,137. Therefore, the size of the second dataset is around 300,137. The hyper-parameters for training each model were chosen as recommended in the URLNet paper~\cite{le2018urlnet}. The training accuracy of all models varied between 95.30\% and 98.50\%.

The efficacy of the resulting models were compared with PhishMatch using the test set (last 30 days browsing data of each user) and 18,000 new confirmed phishing URLs. 
URLNet models evaluate each URL independently. Hence, evaluating URLNet models on the test data of each user and then combining the performance numbers is equivalent to performance number obtained by evaluating models on the combined set (BH-Test-30) containing test data of all users. The total number of URLs in the combined set (BH-Test-30) is 102,141.

All URLNet models were trained and tested on a 64-bit Windows 10 machine (8 GB memory and AMD Ryzen 5 3500U CPU). We measured the resource consumption of each URLNet model by loading it in the memory. Each model required more than 200 MB memory, at least 10 times more than that of PhishMatch. We also measured the time taken to train each URLNet model (see Table~\ref{tab:cmpresults}). The fastest model took more than 1.5 hours to train, which is about 6 times the time taken to train PhishMatch.

\subsection{Results}
\noindent
\textbf{Browsing History.} We simulated the use of both models of PhishMatch on the test set of each user individually. 
Tables~\ref{tab:simbenignglobal} and~\ref{tab:simbenign} show the performance of both PhishMatch models on the combined test set containing 102,141 test URLs (BH-Test-30) along with the average time taken to classify URLs in each category (including the time spent in all the previous components of PhishMatch). In \textbf{PhishMatch (W/o-BH)}, around 79.44\% of the test URLs were found in the global whitelist. Additional 3.21\% URLs where inferred to be visited through user-initiated Google searches. Around 1.03\% of the URLs were detected as IP addresses. The remaining 16.31\% test URLs were checked for different cybersquatting types. A total of 7.36\% of the URLs were identified as the instances of TLDsquatting, combosquatting, subdomain spoofing, directory spoofing and misspelled domains. The remaining 8.95\% test URLs were evaluated using the SBoW based machine learning (ML) model which correctly classified 6.05\% of them as benign, misclassified 1.84\% of them as phishy and gave low confidence score on 1.06\% of the URLs which were then searched using Google search. All queries returned more than 10,000 search results and the domains were found within top 20 search results. The total number of true negatives (TNs) therefore turned out to be 89.77\%.

We also measured the time taken by the URLs classified by each component. Average time taken by PhishMatch (W/o-BH) to classify each URL is 16.5 milliseconds. However, note that a majority of the URLs (79.44\%) were found in the global whitelist within 4.8 ms. Only the URLs that reached the Google Search component of the PhishMatch ($< 1.06\%$) took 1078.8 ms to be classified while the URLs classified by all other components took less than the average time. This is because search engine queries are inherently time consuming. However, note that once determined as benign, the URL will then be added in the session whitelist of the user, and all future lookups for that URL will be extremely fast. Since the time spent by the Google Search component is much longer than the time spent by any other component of PhishMatch to classify a URL, average time may not give a complete picture of the time consumption merits of PhishMatch. Hence, we also report median (50 percentile), third quartile (75 percentile), and ninth decile (90 percentile) times (4.8 ms, 4.8 ms, and 5 ms respectively) in Table~\ref{tab:simbenignglobal}.

\textbf{PhishMatch (Full)} found 97.50\% of the test URLs in different whitelists. 
Additional 1.51\% URLs were inferred to be visited through user-initiated Google searches. Thus, around 99.01\% test URLs were found either in whitelists or were visited using Google Search engine. The remaining 0.99\% test URLs were checked for the presence of popular brand names or domain names.
A very small percentage of the test URLs (total 0.16\%) were identified as the instances of TLDsquatting, combosquatting, subdomain spoofing, directory spoofing and misspelled domains. The remaining 0.84\% test URLs were evaluated using the SBoW based machine learning (ML) model out of which about 0.49\% URLs were determined as benign. About 0.16\% of the URLs were classified as phishing with low confidence ($< 0.9$) and were then searched using Google search. All queries returned more than 10,000 search results and the domains were found within top 20 search results. 
The number of true negatives (TNs), therefore, increased to 99.66\%. 
Detailed information regarding the efficacy of whitelists on test URLs of each user are given in Table~\ref{tab:simwhitelist} in the appendix.


On average, PhishMatch (Full) took 5.4 milliseconds to classify each URL. Majority of the URLs ($97.5\%$) were found in one of the whitelists within 3.6 ms, and only those URLs that reached the Google search component ($< 0.16\%$) took 1089.6 ms on average. The median, third quartile, and ninth decile times of PhishMatch (Full) are 3.6 ms each.

We also evaluated the set of 102,141 URLs (BH-Test-30) using URLNet models. Table~\ref{tab:cmpresults} shows the performance of three different URLNet models trained on two different datasets. The URLNet models trained only on the first dataset comprising of PhishTank and DMOZ URLs correctly classified less than 10\% of the test URLs as benign (TN). This is because the distribution of test URLs (visited by the real users) and the distribution of DMOZ URLs (on which the URLNet models were trained) were different. When the URLNet models were trained using the user browsing data (BH-Train-60) in addition to PhishTank and DMOZ data, their performance on BH-Test-30 data improved considerably. Among all three URLNet models, URLNet (Full) performed the best (97.76\%). However, it did not perform better than PhishMatch (Full). The table also shows the memory consumption of each URLNet model and both PhishMatch models. The memory requirement of URLNet models is at least 10 times more than that required by PhishMatch models. URLNet (Full) takes about 10.4 milliseconds on average to classify a URL.


\begin{table}[h]
	\centering
	\scriptsize
	\caption{The performance of PhishMatch (W/o-BH) components on 102,141 test URLs (BH-Test-30) visited by 25 users. The URLs found in whitelists, those visited from top Google Search results and those classified as benign by the ML model (marked by *) are counted towards true negatives (TNs).}\label{tab:simbenignglobal}
	\begin{tabular}{|p{1.7cm}|l|l|l|l|l|}
	\hline
	\textbf{Component} & \textbf{URLs} & \textbf{\%} & \textbf{TN} & \textbf{FP} & \textbf{Avg Time}\\
	\hline
	\textbf{Whitelists*} & 81,145 & \textbf{79.44\%} & $\surd$ & & \textbf{4.8 ms}\\
	\textbf{User initiated Google Search*} & 3,282 & 3.21\% & $\surd$ & & 3.8 ms\\
	\textbf{IP Address} & 1054 & 1.03\% & & $\surd$ & 1.2 ms\\
	\textbf{TLDsquatting} & 59 & 0.06\% & & $\surd$ & 5 ms\\
	\textbf{IDN \linebreak Homograph} & 0 & 0\% & & $\surd$ & 11 ms\\
	\textbf{Combosquatting} & 4800 & 4.70\% & & $\surd$ & 5 ms\\
	\textbf{Subdomain Spoofing} & 194 & 0.19\% & & $\surd$ & 5 ms\\
	\textbf{Directory Spoofing} & 1806 & 1.77\% & & $\surd$ & 5 ms\\
	\textbf{Misspelled \linebreak Domains} & 656 & 0.64\% & & $\surd$ & 16.25 ms\\
	\textbf{ML model*} & 6184 & 6.05\% & $\surd$ & & 8.3 ms\\
	\textbf{Google Search*} & 1086 & 1.06\% & $\surd$ & & 1078.8 ms\\ 
	\textbf{Misclassified by ML model} & 1875 & 1.84\% & & $\surd$ & 8.3 ms\\
	\hline
	\textbf{Total} & 102,141  & 100\% & \textbf{89.77\%} & \textbf{10.23\%} & \textbf{16.5 ms}\\
	\hline
	\multicolumn{5}{|l|}{\textbf{Median Time}} & \textbf{4.8 ms}\\
	\hline
	\multicolumn{5}{|l|}{\textbf{Third Quartile Time}} & \textbf{4.8 ms}\\
	\hline
	\multicolumn{5}{|l|}{\textbf{Ninth Decile Time}} & \textbf{5 ms}\\
	\hline
	\end{tabular}
\end{table}

\begin{table}[h]
\centering
\scriptsize
\caption{The performance of PhishMatch (Full) components on 102,141 test URLs (BH-Test-30) visited by 25 users. The URLs found in whitelists, those visited from top Google Search results and those classified as benign by the ML model (marked by *) are counted towards true negatives (TNs).}\label{tab:simbenign}
\begin{tabular}{|p{1.7cm}|l|l|l|l|l|}
\hline
\textbf{Component} & \textbf{URLs} & \textbf{\%} & \textbf{TN} & \textbf{FP} & \textbf{Avg Time}\\
\hline
\textbf{Whitelists*} & 99,592 & \textbf{97.50\%} & $\surd$ & & \textbf{3.6 ms}\\
\textbf{User initiated Google Search*} & 1,542 & 1.51\% & $\surd$ & & 3.8 ms\\
\textbf{IP Address} & 0 & 0\% & & $\surd$ & 1.2 ms\\
\textbf{TLDsquatting} & 9 & 0.009\% & & $\surd$ & 5 ms\\
\textbf{IDN \linebreak Homograph} & 0 & 0\% & & $\surd$ & 11 ms\\
\textbf{Combosquatting} & 50 & 0.049\% & & $\surd$ & 5 ms\\
\textbf{Subdomain Spoofing} & 6 & 0.006\% & & $\surd$ & 5 ms\\
\textbf{Directory Spoofing} & 90 & 0.089\% & & $\surd$ & 5 ms\\
\textbf{Misspelled \linebreak Domains} & 13 & 0.013\% & & $\surd$ & 17.2 ms\\
\textbf{ML model*} & 503 & 0.49\% & $\surd$ & & 8.1 ms\\
\textbf{Google Search*} & 164 & 0.16\% & $\surd$ & & 1089.6 ms\\ 
\textbf{Misclassified by ML model} & 172 & 0.17\% & & $\surd$ & 8.1 ms\\
\hline
\textbf{Total} & 102,141  & 100\% & \textbf{99.66\%} & \textbf{0.34\%} & \textbf{5.4 ms}\\
\hline
\multicolumn{5}{|l|}{\textbf{Median Time}} & \textbf{3.6 ms}\\
\hline
\multicolumn{5}{|l|}{\textbf{Third Quartile Time}} & \textbf{3.6 ms}\\
\hline
\multicolumn{5}{|l|}{\textbf{Ninth Decile Time}} & \textbf{3.6 ms}\\
\hline
\end{tabular}
\end{table}

\begin{table}[h]
\centering
\scriptsize
\caption{The performance of PhishMatch (both) on 18,000 new confirmed phishes from PhishTank~\cite{PhishTank}. The URLs found in the whitelists and those misclassified by the ML model (marked by *) are counted towards false negatives (FNs).}\label{tab:simphish}
\begin{tabular}{|p{1.7cm}|l|l|l|l|l|}
\hline
\textbf{Component} & \textbf{URLs} & \textbf{\%} & \textbf{FN} & \textbf{TP} & \textbf{Avg Time}\\
\hline
\textbf{Whitelists*} & 60 & 0.33\% & $\surd$ & & 4.8 ms\\
\textbf{IP Address} & 284 & 1.58\% & & $\surd$ & 1.2 ms\\
\textbf{TLDsquatting} & 148 & \textbf{0.82\%}  & & $\surd$ & \textbf{5} ms\\
\textbf{IDN \linebreak Homograph} & 22 & 0.12\%  & & $\surd$ & 11 ms\\
\textbf{Combosquatting} & 3,717 & \textbf{20.65\%}  & & $\surd$ & \textbf{5 ms}\\
\textbf{Subdomain Spoofing} & 1,947 & \textbf{10.82\%}  & & $\surd$ & \textbf{5ms}\\
\textbf{Directory Spoofing} & 2,924 & \textbf{16.24\%}  & & $\surd$ & \textbf{5ms}\\
\textbf{Misspelled \linebreak Domains} & 801 & 4.45\%  & & $\surd$ & 15.4 ms\\
\textbf{ML model} & 7,179 & \textbf{39.88\%} & & $\surd$ & \textbf{8.3 ms}\\
\textbf{Google Search} & 738 & 4.10\% & & $\surd$ & 1078.8 ms\\
\textbf{Misclassified by ML model*} & 180 & 1.00\% & $\surd$ & & 8.3 ms\\ 
\hline
\textbf{Total} & 18,000 & 100\% & \textbf{1.33\%} & \textbf{98.66\%} & \textbf{50.8 ms}\\
\hline
\multicolumn{5}{|l|}{\textbf{Median Time}} & \textbf{5 ms}\\
\hline
\multicolumn{5}{|l|}{\textbf{Third Quartile Time}} & \textbf{8.3 ms}\\
\hline
\multicolumn{5}{|l|}{\textbf{Ninth Decile Time}} & \textbf{8.3 ms}\\
\hline
\end{tabular}
\end{table}

\begin{table*}[h]
\centering
\scriptsize
\caption{Performance of URLNet and PhishMatch models on the browsing history test data (BH-Test-30) of 25 users comprising of 102,141 URLs, and PhishTank dataset~\cite{PhishTank} comprising of new 18,000 phishing URLs. TN indicates true negatives, TP indicates true positives and MCR is misclassification rate. The table also shows the approximate memory consumption, the training time, and average test time of URLNet and PhishMatch models along with the median, third quartile, and ninth decile times of PhishMatch models.}\label{tab:cmpresults}
\renewcommand{\arraystretch}{1.2}
\begin{tabular}{|l|l|l|l|l|l|l l|}
\hline
\textbf{Model (PhishTank + DMOZ)} & BH-Test-30 (TN) & Phishing (TP) & MCR & Memory (MB) & Training Time & \multicolumn{2}{c|}{Avg Test Time}\\
	& & & & & (HH:MM:SS)  & Benign & Phishing \\
\hline
URLNet (Character-Level) & 6.77\% & 97.05\% & 79.70\% & 225 MB & 01:33:32 & 5.2 ms & 5.2 ms \\
URLNet (Word-Level) & 8.5\% & 96.85\% & 78.26\% & 240 MB & 01:53:47 & 6.4 ms & 6.4 ms \\
URLNet (Full) & 9.67\% & 98.15\% & 77.07\% & 240 MB & 03:08:14 & 10.3 ms & 10.3 ms\\
\hline
\textbf{Model (PhishTank + DMOZ + BH-Train-60)} & BH-Test-30 (TN) & Phishing (TP) & MCR & Memory (MB) & Training Time & \multicolumn{2}{c|}{Avg Test Time}\\
	& & & & & (HH:MM:SS) & Benign & Phishing \\
\hline
URLNet (Character-Level) & 97.67 & 95.50\% & 2.66\% & 212 MB & 01:46:31 & 5.1 ms & 5.1 ms\\
URLNet (Word-Level) & 97.53 & 94.90\% & 2.86\% & 204 MB & 02:50:05 & 6.5 ms & 6.5 ms\\
URLNet (Full) & 97.76 & 96.70\% & 2.40\% & 220 MB & 04:23:22 & 10.4 ms & 10.4 ms\\
\hline
PhishMatch (W/o-BH) & 89.77\% & 98.66\% & 8.89\% & 20 MB & \textbf{00:13:46} & 16.5 ms & 50.8 ms \\
Median Time & & & & & & 4.8 ms & \textbf{5 ms}\\
Third Quartile Time & & & & & & 4.8 ms & 8.3 ms\\
Ninth Decile Time & & & & & & 5 ms & 8.3 ms\\
\hline
PhishMatch (Full) & \textbf{99.66\%}& \textbf{98.66\%} & \textbf{0.49\%} & \textbf{20 MB} & 00:14:26 & 5.4 ms & 50.8 ms \\
Median Time & & & & & & 3.6 ms & \textbf{5 ms}\\
Third Quartile Time & & & & & & 3.6 ms & 8.3 ms\\
Ninth Decile Time & & & & & & \textbf{3.6 ms} & 8.3 ms\\
\hline
\end{tabular}
\end{table*}

\noindent
\textbf{PhishTank.}
On the PhishTank dataset comprising of 18,000 confirmed phishing URLs, both PhishMatch models showed same performance. More specifically, no URL was found in any of the local or community whitelists of PhishMatch (Full).
Table~\ref{tab:simphish} shows the performance of different PhishMatch components on the PhishTank dataset comprising of 18,000 confirmed phishing URLs. Overall, our system correctly identified 98.66\% of the URLs as phish. 
A small fraction (0.33\%) of the URLs were found in the global whitelist (false positives) and 1\% of the URLs were misclassified by the SBoW based machine learning (ML) model as benign. 

Of the URLs classified as phishing, about 1.58\% URLs were identified as IP addresses, 0.82\% URLs were identified as instances of TLDsquatting and 0.12\% URLs were identified as instances of IDN homograph attacks. Combosquatting instances were relatively more prevalent (20.65\%). The top five brands targeted through combosquatting were \texttt{paypal}, \texttt{oneplus}, \texttt{itau}, \texttt{allegro} and \texttt{bethesda}. In 10.82\% of the cases, popular domain names were used in the subdomain part of the phishing URLs. The most common domains targeted through subdomain spoofing were \texttt{amazon.co.jp}, \texttt{facebook.com}, \texttt{paypal.com}, \texttt{apple.com} and \texttt{icloud.com}. The number of phishing URLs that contained a popular domain name in their directory was 16.24\%. The most common domain names that appeared in the directory part were \texttt{bankofamerica.com}, \texttt{alibaba.com}, \texttt{paypal.com}, \texttt{wellsfargo.com} and \texttt{dropbox.com}. 

The optimized Aho-Corasick pattern matching machine aids in detecting TLDsquatting, combosquatting, subdomain spoofing and directory spoofing attacks, which adds to 48.53\% of the phishing URLs. Misspelled domains comprised 4.45\% of the phishing URLs. The most common misspelled domains were \texttt{google.com} and \texttt{tumblr.com}. Thus, 52.98\% of the phishing URLs were identified correctly before reaching the machine learning component (Figure~\ref{fig:phishmatch}). The SBoW based machine learning model correctly classified 39.88\% of the phishing URLs with high confidence ($\ge 0.9$) while 4.10\% of the URLs were confirmed using Google Search engine due to a low confidence score. 

We note that not all phishing URLs employ cybersquatting techniques. On the PhishTank dataset of 18,000 confirmed phishing URLs of various kinds, 9,559 ($53.11\%$) of the URLs were classified as employing one of the cybersquatting techniques, misspellings of popular domain names, or IDN Homograph attacks. Of the remaining 8,441 URLs, which can be assumed to not be employing these type of obfuscations, we correctly classify 8,201 (97.16\%) as phishing. Hence, PhishMatch is able to efficiently determine phishing URLs of all kinds.

We evaluated the set of 18,000 phishing URLs using three different URLNet models trained on two different sets of data (Table~\ref{tab:cmpresults}). The URLNet models trained on the first dataset comprising of PhishTank and DMOZ URLs correctly classified a significant of phishing URLs as phishing (TP). However, they suffered significantly in identifying the benign URLs. The URLNet models trained on the second dataset that consisted of user browsing data (BH-Train-60) in addition to PhishTank and DMOZ URLs, improved in identifying the benign URLs, but their performance in detecting phishing URLs dropped. The best performing URLNet (Full) model identified 98.15\% of the phishing URLs when trained on the first dataset, however it identified 96.70\% of the phishing URLs when trained on the second dataset. PhishMatch, on the other hand, correctly identified 98.66\% of the URLs as phish.

The average time taken to classify a phishy URL by PhishMatch is 50.8 ms. However, the URLs classified by the optimized Aho-Corasick pattern matching machine (48.53\%) take 5 ms on average and by the SBoW based machine learning model (39.88\%) take 8.3 ms. Thus, the majority of the URLs (88.41\%) took less than 8.3 ms to be classified correctly, and only those URLs that reached the Google Search component (4.10\%) took 1078.8 ms each. Note that once determined as phishy, the URL will then be added to the blacklist, and all future lookups will be faster. 
The median, third quartile and ninth decile times of PhishMatch are 5 ms, 8.3 ms, and 8.3 ms respectively.
In comparision, URLNet (Full) takes about 10.4 ms on average to classify every URL.



\section{Conclusion}
In this paper, we proposed a layered anti-phishing defense system called PhishMatch, which employs multiple components such as blacklist, whitelists, string matching algorithms, machine learning and search engine(s), to determine whether a URL is phishing or not. 
PhishMatch employs four different whitelists, namely local, community, global and session. We described procedures for creating and updating personalized local whitelist using the browsing history of the user. We also gave a method to combine the relaxed local whitelist of each user to create a more reliable whitelist for the whole community. To speed up the domain lookup process, we maintained a session whitelist comprising of recently visited safe domains. Our simulation results on the browsing history of 25 users reveal that around 97.50\% of the domains visited by users can be found in one of the whitelists. Therefore, whitelists play a significant role in the reduction of the number of false positives (benign URLs classified as phishing).

We also suggested heuristics to determine if the URL is clicked or typed. URLs clicked from the top search engine results are considered to be safe. After classifying such URLs as benign, the number of true negatives increased to 99.01\%. If the URL is not present in any whitelist and if it is not clicked from the top search results, then we determine whether it contains a popular brand name or domain name from the global whitelist. To enable faster detection of such patterns in the URL, we stored the global whitelist using Aho-Corasick pattern matching machine. Finding patterns using the state machine is fast but it consumes a lot of memory. Therefore, we suggested several improvements to reduce the memory consumption of the state machine. The resulting optimized pattern matching machine consumes $\sim$46\% less memory than the state-of-the-art~\cite{Tuck:2004}. Our experiment shows that around 48.53\% of the phishing URLs were detected by the optimized pattern matching machine. Further, 1.58\% of the phishing URLs contained IP address and 0.12\% were detected as instances of IDN homograph attack. To speed up the detection process, we also maintained a blacklist of previously detected phishing domains.

If the URL is typed or does not contain any popular brand name, then we check if it is a misspelling of some popular domain using a novel $n$-gram based approximate pattern matching algorithm. In our simulation, the algorithm identified around 4.45\% of the phishing URLs as a misspelling of some popular domain name. If the URL does not contain any popular brand name or a misspelled domain name, then we use a machine learning model trained using SBoW features extracted from the hostnames of benign and phishing URLs. Our experiment shows that about 43.98\% of the phishing URLs and 0.65\% of the benign URLs were correctly detected using the machine learning model and Google Search engine. If the confidence score of the model for a given URL is less than 0.9, we extract the domain part of the URL and search it using Google Search. If the number of search results is greater than 10,000 and the domain appears in the top 20 results, then we consider the domain to be safe. Only 0.16\% of the benign domains and 4.10\% of the phishing domains were searched using Google Search engine. Thus, in majority of the cases the label of the URL is determined on the client-side.

Consequently, PhishMatch correctly identified 98.66\% of the phishing URLs correctly (true positives) and 99.66\% of the benign URLs (true negatives). The performance of PhishMatch was found to be better than the state-of-the-art deep learning based URLNet models~\cite{le2018urlnet}. 
The PhishMatch extension identifies at least 90\% of the benign URLs within 3.6 ms, and at least 90\% of the phishing URLs within 8.3 ms. 
Depending on the component where the URL is detected as phishy, PhishMatch displays a comprehensible warning message that explains why the URL was classified as phishy. PhishMatch also displays a recommendation if one of the cybersquatting techniques is detected. This explainability aspect of PhishMatch makes it much more likely that the user will not bypass the warning.
Moreover, the PhishMatch extension consumes only around 20 MB memory whereas URLNet models consumes more than 200 MB memory. 
Thus, the proposed system is not only accurate but also fast and lightweight. 

\subsection{Future Work}
Phishing attacks are continuously evolving. Therefore, it is important to update the SBoW based machine learning model as and when new phishing URLs are available. In the current work, we employed batch learning algorithm to learn the parameters of the model (logistic regression with gradient descent). However, for the task of detecting phishing URLs, online learning methods seem to be more appropriate as they can adapt to changes in phishing attacks and their features over time~\cite{Ma:2009}. We wish to explore this thread further in future.

Cross site scripting (XSS) is a type of security vulnerability in web applications by which criminals may launch a variant of phishing attack. It focuses on injection of malicious scripts to vulnerable and benign websites. Hence, only URL based approach fails to detect such attacks. However, our anti-phishing solution is not designed to replace the current solutions~\cite{xss:2012,xss:zhou,xss:2021}
to prevent XSS attacks, but can be used to efficiently complement them. It will be interesting to include another layer in PhishMatch that detects these attacks after the URL is deemed safe to load in the browser for further protecting the user.

We also aim to conduct a large-scale study to evaluate the usability of PhishMatch extension and the effect of warning messages on the behaviour of users. Further, we plan to make the source code of extension available for further research.

\bibliographystyle{plain}
\bibliography{PhishMatch}

\begin{thebibliography}{10}

\bibitem{Garera:2007}
{A Framework for Detection and Measurement of Phishing Attacks}.
\newblock In {\em Proceedings of the 2007 ACM Workshop on Recurring Malcode},
  WORM '07, pages 1--8, New York, NY, USA, 2007. ACM.

\bibitem{Punycode}
{A robust Punycode converter that fully complies to RFC 3492 and RFC 5891.}
\newblock \url{https://github.com/bestiejs/punycode.js/}, 2020.

\bibitem{avastOnline}
{Avast Online Security - Google Chrome}.
\newblock
  \url{https://chrome.google.com/webstore/detail/avast-online-security/gomekmidlodglbbmalcneegieacbdmki?hl=en},
  2020.

\bibitem{compatibility}
{Browser Extensions}.
\newblock
  \url{https://developer.mozilla.org/en-US/docs/Mozilla/Add-ons/WebExtensions},
  2020.

\bibitem{FBI}
{BUSINESS E-MAIL COMPROMISE THE 12 BILLION DOLLAR SCAM}.
\newblock \url{https://www.ic3.gov/media/2018/180712.aspx}, 2020.

\bibitem{chromeAPI}
{Chrome APIs}.
\newblock \url{https://developer.chrome.com/extensions/api_index}, 2020.

\bibitem{UnicodeConfusables}
{confusables.txt}.
\newblock \url{https://www.unicode.org/Public/security/13.0.0/confusables.txt},
  2020.

\bibitem{mozillaAvast}
{Mozilla removes Avast and AVG extensions from add-on portal over snooping
  claims}.
\newblock
  \url{https://www.zdnet.com/article/mozilla-removes-avast-and-avg-extensions-from-add-on-portal-over-snooping-claims/},
  2020.

\bibitem{Netcraft}
{Netcraft Extension}.
\newblock \url{https://toolbar.netcraft.com/}, 2020.

\bibitem{pythonWord}
{Python Word Segmentation}.
\newblock \url{http://www.grantjenks.com/docs/wordsegment/}, 2020.

\bibitem{SuspiciousReporter}
{Suspicious Site Reporter - Google Chrome}.
\newblock
  \url{https://chrome.google.com/webstore/detail/suspicious-site-reporter/jknemblkbdhdcpllfgbfekkdciegfboi},
  2020.

\bibitem{UnicodeNormalization}
{UNICODE NORMALIZATION FORMS}.
\newblock \url{https://unicode.org/reports/tr15/}, 2020.

\bibitem{UnicodeSecurity}
{UNICODE SECURITY CONSIDERATIONS}.
\newblock \url{http://unicode.org/reports/tr36/}, 2020.

\bibitem{URIlib}
{URI.js}.
\newblock \url{https://medialize.github.io/URI.js/}, 2020.

\bibitem{URLNetCode}
{URLNet}.
\newblock \url{https://github.com/Antimalweb/URLNet}, 2020.

\bibitem{WindowsDefender}
{Windows Defender Browser Protection - Google Chrome}.
\newblock
  \url{https://chrome.google.com/webstore/detail/windows-defender-browser/bkbeeeffjjeopflfhgeknacdieedcoml},
  2020.

\bibitem{dom:2019}
M.A. Adebowale, K.T. Lwin, E.~Sánchez, and M.A. Hossain.
\newblock Intelligent web-phishing detection and protection scheme using
  integrated features of images, frames and text.
\newblock {\em Expert Systems with Applications}, 115:300--313, 2019.

\bibitem{agten2015seven}
Pieter Agten, Wouter Joosen, Frank Piessens, and Nick Nikiforakis.
\newblock {Seven Months' Worth of Mistakes: A Longitudinal Study of
  Typosquatting Abuse}.
\newblock In {\em Proceedings of the 22nd Network and Distributed System
  Security Symposium (NDSS 2015)}. Internet Society, 2015.

\bibitem{ahmed2009revised}
Farag Ahmed, Ernesto William~De Luca, and Andreas N{\"u}rnberger.
\newblock Revised n-gram based automatic spelling correction tool to improve
  retrieval effectiveness.
\newblock {\em Polibits}, (40):39--48, 2009.

\bibitem{Aho:1975}
Alfred~V Aho and Margaret~J Corasick.
\newblock {Efficient String Matching: An Aid to Bibliographic Search}.
\newblock {\em Communications of the ACM}, 18(6):333--340, 1975.

\bibitem{alsharnouby:2015}
Mohamed Alsharnouby, Furkan Alaca, and Sonia Chiasson.
\newblock {Why Phishing still Works: User Strategies for Combating Phishing
  Attacks}.
\newblock {\em International Journal of Human-Computer Studies}, 82:69--82,
  2015.

\bibitem{APWGReport2014}
{Anti-Phishing Work Group}.
\newblock {Global Phishing Survey: Trends and Domain Name Use in 2H2014}.
\newblock
  \url{https://docs.apwg.org/reports/APWG\_Global\_Phishing\_Report\_2H\_2014.pdf},
  2020.

\bibitem{APWGReport}
{Anti-Phishing Work Group}.
\newblock {Phishing Activity Trends Report: 4th Quarter 2021}.
\newblock \url{https://docs.apwg.org/reports/apwg_trends_report_q4_2020.pdf},
  2021.

\bibitem{Ardi:2016}
Calvin Ardi and John Heidemann.
\newblock {AuntieTuna: Personalized Content-Based Phishing Detection}.
\newblock In {\em Proceedings of the NDSS Workshop on Usable Security}, San
  Diego, California, USA, February 2016. The Internet Society.

\bibitem{banerjee:2008}
A.~{Banerjee}, D.~{Barman}, M.~{Faloutsos}, and L.~N. {Bhuyan}.
\newblock Cyber-fraud is one typo away.
\newblock In {\em IEEE INFOCOM 2008 - The 27th Conference on Computer
  Communications}, pages 1939--1947, April 2008.

\bibitem{Gamze:2014}
Gamze Canova, Melanie Volkamer, Clemens Bergmann, and Roland Borza.
\newblock Nophish: An anti-phishing education app.
\newblock In Sjouke Mauw and Christian~Damsgaard Jensen, editors, {\em Security
  and Trust Management}, pages 188--192, Cham, 2014. Springer International
  Publishing.

\bibitem{chen:2010}
Teh-Chung Chen, Scott Dick, and James Miller.
\newblock Detecting visually similar web pages: Application to phishing
  detection.
\newblock {\em ACM Trans. Internet Technol.}, 10(2), June 2010.

\bibitem{CJ:2018}
Gokul CJ, Sankalp Pandit, Sukanya Vaddepalli, Harshal Tupsamudre, Vijayanand
  Banahatti, and Sachin Lodha.
\newblock {PHISHY - A Serious Game to Train Enterprise Users on Phishing
  Awareness}.
\newblock In {\em Proceedings of the 2018 Annual Symposium on Computer-Human
  Interaction in Play Companion Extended Abstracts}, CHI PLAY '18 Extended
  Abstracts, pages 169--181, New York, NY, USA, 2018. ACM.

\bibitem{Dhamija:2006}
Rachna Dhamija, J.~D. Tygar, and Marti Hearst.
\newblock Why phishing works.
\newblock In {\em Proceedings of the SIGCHI Conference on Human Factors in
  Computing Systems}, CHI ’06, page 581–590, New York, NY, USA, 2006.
  Association for Computing Machinery.

\bibitem{DMOZ}
DMOZ.
\newblock \url{http://dmoz-odp.org/}, 2020.

\bibitem{Fu:2006}
Anthony~Y. Fu, Xiaotie Deng, Liu Wenyin, and Greg Little.
\newblock {The Methodology and an Application to Fight Against Unicode
  Attacks}.
\newblock In {\em Proceedings of the Second Symposium on Usable Privacy and
  Security}, SOUPS '06, pages 91--101, New York, NY, USA, 2006. ACM.

\bibitem{safeBrowsing}
{Google Safe Browsing}.
\newblock \url{https://safebrowsing.google.com/}, 2020.

\bibitem{Holgers:2006}
Tobias Holgers, David~E. Watson, and Steven~D. Gribble.
\newblock Cutting through the confusion: A measurement study of homograph
  attacks.
\newblock In {\em Proceedings of the Annual Conference on USENIX '06 Annual
  Technical Conference}, ATEC '06, pages 24--24, Berkeley, CA, USA, 2006.
  USENIX Association.

\bibitem{Huh:2012}
Jun~Ho Huh and Hyoungshick Kim.
\newblock Phishing detection with popular search engines: Simple and effective.
\newblock In Joaquin Garcia-Alfaro and Pascal Lafourcade, editors, {\em
  Foundations and Practice of Security}, pages 194--207, Berlin, Heidelberg,
  2012. Springer Berlin Heidelberg.

\bibitem{islam2012text}
Aminul Islam, Evangelos Milios, and Vlado Ke{\v{s}}elj.
\newblock Text similarity using google tri-grams.
\newblock In {\em Canadian Conference on Artificial Intelligence}, pages
  312--317. Springer, 2012.

\bibitem{Khonji:2013}
M.~Khonji, Y.~Iraqi, and A.~Jones.
\newblock {Phishing Detection: A Literature Survey}.
\newblock {\em IEEE Communications Surveys Tutorials}, 15(4):2091--2121, Fourth
  2013.

\bibitem{Kintis:2017}
Panagiotis Kintis, Najmeh Miramirkhani, Charles Lever, Yizheng Chen, Rosa
  Romero-G\'{o}mez, Nikolaos Pitropakis, Nick Nikiforakis, and Manos
  Antonakakis.
\newblock {Hiding in Plain Sight: A Longitudinal Study of Combosquatting
  Abuse}.
\newblock In {\em Proceedings of the 2017 ACM SIGSAC Conference on Computer and
  Communications Security}, CCS '17, pages 569--586, New York, NY, USA, 2017.
  ACM.

\bibitem{Le:2011}
A.~Le, A.~Markopoulou, and M.~Faloutsos.
\newblock {PhishDef: URL Names Say it All}.
\newblock In {\em 2011 Proceedings IEEE INFOCOM}, pages 191--195, April 2011.

\bibitem{le2018urlnet}
Hung Le, Quang Pham, Doyen Sahoo, and Steven~CH Hoi.
\newblock {URLNet: Learning a URL Representation with Deep Learning for
  Malicious URL Detection}.
\newblock {\em arXiv preprint arXiv:1802.03162}, 2018.

\bibitem{ml:li2019}
Yukun Li, Zhenguo Yang, Xu~Chen, Huaping Yuan, and Wenyin Liu.
\newblock A stacking model using url and html features for phishing webpage
  detection.
\newblock {\em Future Generation Computer Systems}, 94:27--39, 2019.

\bibitem{Ma:2009}
Justin Ma, Lawrence~K. Saul, Stefan Savage, and Geoffrey~M. Voelker.
\newblock {Beyond Blacklists: Learning to Detect Malicious Web Sites from
  Suspicious URLs}.
\newblock In {\em Proceedings of the 15th ACM SIGKDD International Conference
  on Knowledge Discovery and Data Mining}, KDD '09, pages 1245--1254, New York,
  NY, USA, 2009. ACM.

\bibitem{McGrath:2008}
D.~Kevin McGrath and Minaxi Gupta.
\newblock {Behind Phishing: An Examination of Phisher Modi Operandi}.
\newblock In {\em Proceedings of the 1st Usenix Workshop on Large-Scale
  Exploits and Emergent Threats}, LEET'08, pages 4:1--4:8, Berkeley, CA, USA,
  2008. USENIX Association.

\bibitem{Moore:2008}
Tyler Moore and Richard Clayton.
\newblock {Financial Cryptography and Data Security}.
\newblock chapter Evaluating the Wisdom of Crowds in Assessing Phishing
  Websites, pages 16--30. Springer-Verlag, Berlin, Heidelberg, 2008.

\bibitem{xss:2012}
Angelo~Eduardo Nunan, Eduardo Souto, Eulanda~M. dos Santos, and Eduardo
  Feitosa.
\newblock Automatic classification of cross-site scripting in web pages using
  document-based and url-based features.
\newblock In {\em 2012 IEEE Symposium on Computers and Communications (ISCC)},
  pages 000702--000707, 2012.

\bibitem{PhishTank}
PhishTank.
\newblock \url{https://www.antiphishing.org/resources/apwg-reports/}, 2020.

\bibitem{pochat2019tranco}
Victor~Le Pochat, Tom Van~Goethem, Samaneh Tajalizadehkhoob, Maciej
  Korczy{\'n}ski, and Wouter Joosen.
\newblock {Tranco: A Research-Oriented Top Sites Ranking Hardened Against
  Manipulation}.
\newblock In {\em Proceedings of the 22nd Network and Distributed System
  Security Symposium (NDSS 2019)}. Internet Society, 2019.

\bibitem{prakash:2010}
P.~{Prakash}, M.~{Kumar}, R.~R. {Kompella}, and M.~{Gupta}.
\newblock {PhishNet: Predictive Blacklisting to Detect Phishing Attacks}.
\newblock In {\em 2010 Proceedings IEEE INFOCOM}, pages 1--5, March 2010.

\bibitem{Quinkert:2019}
F.~{Quinkert}, T.~{Lauinger}, W.~{Robertson}, E.~{Kirda}, and T.~{Holz}.
\newblock It's not what it looks like: Measuring attacks and defensive
  registrations of homograph domains.
\newblock In {\em 2019 IEEE Conference on Communications and Network Security
  (CNS)}, pages 259--267, June 2019.

\bibitem{search:2019}
Routhu~Srinivasa Rao and Alwyn~Roshan Pais.
\newblock Jail-phish: An improved search engine based phishing detection
  system.
\newblock {\em Computers \& Security}, 83:246--267, 2019.

\bibitem{cmc:adversarial}
Bader Rasheed, Adil Khan, S.~M.~Ahsan Kazmi, Rasheed Hussain, Md.~Jalil Piran,
  and Doug~Young Suh.
\newblock Adversarial attacks on featureless deep learning malicious urls
  detection.
\newblock {\em Computers, Materials \& Continua}, 68(1):921--939, 2021.

\bibitem{robertson1998applications}
Alexander~M Robertson and Peter Willett.
\newblock Applications of n-grams in textual information systems.
\newblock {\em Journal of Documentation}, 54(1):48--67, 1998.

\bibitem{ml:Sahingoz2019}
Ozgur~Koray Sahingoz, Ebubekir Buber, Onder Demir, and Banu Diri.
\newblock Machine learning based phishing detection from urls.
\newblock {\em Expert Systems with Applications}, 117:345--357, 2019.

\bibitem{Sheng:2007}
Steve Sheng, Bryant Magnien, Ponnurangam Kumaraguru, Alessandro Acquisti,
  Lorrie~Faith Cranor, Jason Hong, and Elizabeth Nunge.
\newblock {Anti-Phishing Phil: The Design and Evaluation of a Game That Teaches
  People Not to Fall for Phish}.
\newblock In {\em Proceedings of the 3rd Symposium on Usable Privacy and
  Security}, SOUPS '07, pages 88--99, New York, NY, USA, 2007. ACM.

\bibitem{sheng:2009}
Steve Sheng, Brad Wardman, Gary Warner, Lorrie Cranor, Jason Hong, and
  Chengshan Zhang.
\newblock {An Empirical Analysis of Phishing Blacklists}.
\newblock In {\em Sixth conference on email and anti-spam (CEAS)}. California,
  USA, 2009.

\bibitem{xss:2021}
Iram Tariq, Muddassar~Azam Sindhu, Rabeeh~Ayaz Abbasi, Akmal~Saeed Khattak,
  Onaiza Maqbool, and Ghazanfar~Farooq Siddiqui.
\newblock Resolving cross-site scripting attacks through genetic algorithm and
  reinforcement learning.
\newblock {\em Expert Systems with Applications}, 168:114386, 2021.

\bibitem{Tuck:2004}
N.~{Tuck}, T.~{Sherwood}, B.~{Calder}, and G.~{Varghese}.
\newblock {Deterministic Memory-efficient String Matching Algorithms for
  Intrusion Detection}.
\newblock In {\em IEEE INFOCOM 2004}, volume~4, pages 2628--2639 vol.4, March
  2004.

\bibitem{tupsamudre:2019}
Harshal Tupsamudre, Ajeet~Kumar Singh, and Sachin Lodha.
\newblock Everything is in the name -- a url based approach for phishing
  detection.
\newblock In Shlomi Dolev, Danny Hendler, Sachin Lodha, and Moti Yung, editors,
  {\em Cyber Security Cryptography and Machine Learning}, pages 231--248, Cham,
  2019. Springer International Publishing.

\bibitem{Vanhoenshoven}
Frank Vanhoenshoven, Gonzalo Nápoles, Rafael Falcon, Koen Vanhoof, and Mario
  Köppen.
\newblock Detecting malicious urls using machine learning techniques.
\newblock In {\em 2016 IEEE Symposium Series on Computational Intelligence
  (SSCI)}, pages 1--8, 2016.

\bibitem{Verizon}
Verizon.
\newblock {2021 Data Breach Investigations Report}.
\newblock \url{https://verizon.com/dbir/}, May 2021.

\bibitem{Verma:2017}
Rakesh Verma and Avisha Das.
\newblock What's in a url: Fast feature extraction and malicious url detection.
\newblock In {\em Proceedings of the 3rd ACM on International Workshop on
  Security And Privacy Analytics}, IWSPA '17, pages 55--63, New York, NY, USA,
  2017. ACM.

\bibitem{Wang:2006}
Yi-Min Wang, Doug Beck, Jeffrey Wang, Chad Verbowski, and Brad Daniels.
\newblock {Strider Typo-patrol: Discovery and Analysis of Systematic
  Typo-squatting}.
\newblock In {\em Proceedings of the 2Nd Conference on Steps to Reducing
  Unwanted Traffic on the Internet - Volume 2}, SRUTI'06, pages 5--5, Berkeley,
  CA, USA, 2006. USENIX Association.

\bibitem{whittaker:2009}
Colin Whittaker, Brian Ryner, and Marria Nazif.
\newblock Large-scale automatic classification of phishing pages.
\newblock In {\em NDSS '10}, 2010.

\bibitem{Xiang:2011}
Guang Xiang, Jason Hong, Carolyn~P. Rose, and Lorrie Cranor.
\newblock {CANTINA+: A Feature-Rich Machine Learning Framework for Detecting
  Phishing Web Sites}.
\newblock {\em ACM Trans. Inf. Syst. Secur.}, 14(2):21:1--21:28, September
  2011.

\bibitem{Zha:2008}
{Xinyan Zha} and S.~{Sahni}.
\newblock {Highly Compressed Aho-Corasick Automata for Efficient Intrusion
  Detection}.
\newblock In {\em 2008 IEEE Symposium on Computers and Communications}, pages
  298--303, July 2008.

\bibitem{Zhang:2007}
Yue Zhang, Jason~I. Hong, and Lorrie~F. Cranor.
\newblock Cantina: A content-based approach to detecting phishing web sites.
\newblock In {\em Proceedings of the 16th International Conference on World
  Wide Web}, WWW '07, pages 639--648, New York, NY, USA, 2007. ACM.

\bibitem{xss:zhou}
Yun Zhou and Peichao Wang.
\newblock An ensemble learning approach for xss attack detection with domain
  knowledge and threat intelligence.
\newblock {\em Computers \& Security}, 82:261--269, 2019.

\end{thebibliography}
\newpage
\appendix{}
\begin{algorithm}[h]
	\scriptsize
	\caption{Local Whitelist Algorithm}~\label{LWAlgo}
	\begin{algorithmic}[1]
	\Procedure{$CreateLocalWhitelist$}{} \\
	\textbf{Input:} browsing history $H$ of the user \\
	\textbf{Output:} local whitelist $W_L$ for the user
	\State $W_L = \{\}$ 
	\For {$d_i \in H$}
		\State <$a_i$, $v_i$, $r_i$, $t_i$> = $GetDomainFeatures(d_i, H)$
		 \State /* compute score using decision tree given in Figure~\ref{fig:lw} */
		\State $s_i$ $=$ $getDomainScore(a_i, v_i, r_i, t_i)$
		\If {$s_i  \neq \inf$}
			\State $W_L[d_i] = s_i$
		\EndIf
	\EndFor
		\State return $W_L$
	\EndProcedure
	\\
	\Procedure{$UpdateLocalWhitelist$}{} \\
	\textbf{Input:} browsing history $H$ of the user, and local whitelist $W^{prev}_L$ and session whitelist $W^{prev}_S$ from the previous browsing session\\
	\textbf{Output:} updated local whitelist $W_L$ for the user
	\State $W_L = \{\}$
	\For {$d_i \in W^{prev}_L \cup W^{prev}_S$}
		\If {$d_i \notin H$}
			\State continue /* do nothing, deletion of old domain */		
		\Else
			\State <$a_i$, $v_i$, $r_i$, $t_i$> = $GetDomainFeatures(d_i, H)$
			\State /* compute score using decision tree given in Figure~\ref{fig:lw} */
			\State $s^{new}_i$ $=$ $GetDomainScore(a_i, v_i, r_i, t_i)$
			\If {$d_i \in W^{prev}_L$ and $d_i \in W^{prev}_S$}
				\State $W_L[d_i] = min(s^{new}_i, W^{prev}_L[d_i])$  /* improve score */
			\break
			\ElsIf {$d_i \in W^{prev}_L$}
				\State $W_L[d_i] = W^{prev}_L[d_i]$ /* retain the existing domain */
			\break
			\ElsIf {$d_i \in W^{prev}_S$}
				\State $W_L[d_i] =   min(s^{new}_i, 8)$  /* addition of new domain */
			\EndIf
		\EndIf	
	\EndFor
	\State return $W_L$
	\EndProcedure
	\end{algorithmic}
\end{algorithm} 

\begin{algorithm}[h]
	\scriptsize
	\caption{Community Whitelist Algorithm}~\label{CWAlgo}
	\begin{algorithmic}[1]
	\Procedure{$CreateCommunityWhitelist$}{} \\
	\textbf{Input:} relaxed domain whitelists $(W^1_{RL}, \ldots, W^N_{RL})$ of $N$ users and $k$ the size of community whitelist\\
	\textbf{Output:} community whitelist $W_C$ of size $k$  
	\State $W_C = \{\}$, 
	\For {$j = 1 \ to \ N$}
		\For {$d_i \in W^j_{RL}$}
			\If {$d_i \notin W_C$}
				\State $W_C[d_i] = (N-1) \cdot 16$  /* $d_i$ is seen for the first time */
			\Else
				\State $W_C[d_i]  = W_C[d_i] - 16$ /* nullify the penalty */
			\EndIf	
			\State $W_C[d_i] $  += $W^j_{RL}[d_i]$ 		
		\EndFor	
	\EndFor
	\State sort domains in $W_C$ in the increasing order of the score
	\State return top $k$ entries from $W_C$
	\EndProcedure
	\end{algorithmic}
\end{algorithm} 

\begin{algorithm}[h]
	\scriptsize
	\caption{Whitelist Algorithm}~\label{SWAlgo}
	\begin{algorithmic}[1]
	\Procedure{$InitializeWhitelists$}{}\\
	\textbf{Input:} browsing history $H$ of the user\\
	\textbf{Output:} session $W_S$, local $W_L$, community $W_C$ and global $W_G$ whitelists
		\State $W_S$ = $\{\}$
		\State $H$ = $GetBrowsingHistory()$
		\If {$W_L$ is not created}
			\State $W_L$ = $CreateLocalWhitelist(H)$
		\Else
			\State $W^{prev}_S$ = $LoadPrevSessionWhitelist()$ 
			\State $W^{prev}_L$ = $LoadPrevLocalWhitelist()$ 
			\State $UpdateLocalWhitelist(H, W^{prev}_L, W^{prev}_S)$
		\EndIf
		\State $W_C$ $=$ $LoadCommunityWhitelist()$
		\State $W_G$ $=$ $LoadGlobalWhitelist()$
	\EndProcedure
	\\
	\Procedure{$CheckInWhitelist$}{} \\
	\textbf{Input:} domain $d$ requested by the user\\
	\textbf{Output:} boolean value indicating if the domain $d$ is present in one of the whitelists and updated session whitelist $W_S$
	\If {$d \in W_S$}
		\State return true	
	\ElsIf {$d \in W_L$}
		\State $W_S[d] = true$
		\State return true
	\ElsIf {$d \in W_C$}
		\State $W_S[d] = true$
		\State return true
	\ElsIf{$d \in W_G$}
		\State $W_S[d] = true$
		\State return true
	\ElsIf{$d$ appears in the top search engine results}
		\State $W_S[d] = true$
		\State return true
	\Else
		\State return false
	\EndIf
	\EndProcedure	
	\\/* domain is also added to the session whitelist if it is classified as safe by other components of PhishMatch */
	\end{algorithmic}
\end{algorithm} 

\begin{algorithm}[h]
	\scriptsize
	\caption{Algorithm to determine the transition type (context) of URL visit}~\label{ContextAlgo}
	\begin{algorithmic}[1]
	\Procedure{$DetectVisitContext$}{}\\
	\textbf{Input:} web request object $request$ corresponding to the current URL request, and $tabUrl$ map containing $tabId$ as key and URL as value\\
	\textbf{Output:} transition type $link$ or $typed$
	\State $transition = null$
	\If {$request.initiator$ is defined}
		\State $transition = link$
	\Else
		\If {$request.tabId \notin tabUrl$}
			\State $transition = link$
		\Else
			\State $transition = typed$
		\EndIf
	\EndIf
	\State return $transition$
	\EndProcedure
	\end{algorithmic}
\end{algorithm}

\begin{algorithm}[h]
	\caption{Algorithm to identify IDN attack}~\label{IDNAlgo}
	\scriptsize
	\begin{algorithmic}[1]
	\Procedure{$isPunycode$}{}\\
	\textbf{Input:} domain $d$ requested by the user\\
	\textbf{Output:} boolean value indicating if the domain is punycode or not
	\If {$d.startsWith($`xn-{}-'$)$}
		\State return true
	\Else
		\State return false
	\EndIf
	\EndProcedure
	\\
	\Procedure{$isIDNAttack$}{}\\
	\textbf{Input:} non-ASCII domain $d$ requested by the user\\
	\textbf{Output:} boolean value indicating if the non-ASCII domain resembles popular domain or not
	\State $d_{uni}$ = $punycode.toUnicode(d)$
	\State $d_{norm}$ = $d_{uni}.normalize(NFKD)$
	\State $d_{ascii}$ = ""
	\For {$l = 1$ to length($d_{norm}$)}
		\If {$isASCII(d_{norm}[l])$}
			\State $d_{ascii}$ += $d_{norm}[l]$
		\ElsIf {$confusables[d_{norm}[l]]$ is defined}
			\State $d_{ascii}$ += $confusables[d_{norm}[l]]$
		\Else
			\State continue /* skip non-ASCII character if not in confusables map*/
		\EndIf
	\EndFor
	\State $d_{ascii} = d_{ascii}.replace(/[^{\wedge}a$-$z \ 0$-$9$ . -$]/g,$ `'$)$
	\If {$CheckInWhitelist(d_{ascii})$}
		\State return true
	\Else
		\State return false
	\EndIf
	\EndProcedure
	\end{algorithmic}
\end{algorithm}

\begin{algorithm}[h]
	\scriptsize
	\caption{Algorithm to create Optimized Aho-Corasick Machine}~\label{AhoAlgo}
	\begin{algorithmic}[1]
	\Procedure{$CreateLexTrie$}{}\\
	\textbf{Input:} Set $Y = \{y_1, y_2, \ldots y_m\}$ of keywords\\
	\textbf{Output:} Lexicographic trie $T_L$
	\State $L_{trie}$ = create an empty trie consisting of start state 0
	\State $l_{max} = max(y_1.length, y_2.length, \ldots, y_m.length)$
	\State $Y_{sort}$ = sort keywords in $Y$ in lexicographic order
	\For {$l = 1 \ to \ l_{max}$}
		\For {$y_i \in Y_{sort}$}
			\If {$y_i.length >= l$}
				\State insert $y_i.substr(0,l)$ into trie $L_{trie}$  /*prefix $y_i[0, l-1]$*/
			\EndIf		
	\EndFor
	\EndFor
	\State return $L_{trie}$
	\EndProcedure
	\\
	\Procedure{$CreateOptimizedStates$}{}\\
	\textbf{Input:} Unoptimized Aho-Corasick Machine $A_{unopt}$ consisting of $goto$ ($lex \ trie$), $fail$ and $output$ functions\\
	\textbf{Output:} Optimized Aho-Corasick Machine $A_{opt}$ consisting of bitmap-lex and lex states
	\State $S =$ number of states in $A_{unopt}$ 
	\State $states = new \ Array(S)$
	\State $bitmap\_table = \{\}$
	\State $output\_map = \{\}$
	\For {$s \in A_{unopt}$}
		\State $states[s] = \{\}$
		\State $states[s].fail = A_{unopt}.fail[s]$
		\If {state $s$ has more than one transitions}
			\State $states[s].symbol = $ `$\#$'
			\State $states[s].smallest\_next = smallestNextState(s)$
			\State $bitmap\_table[s] = createBitmap(A_{unopt}.goto(s))$
		\ElsIf {state $s$ has one transition}
			\State $states[s].symbol$ =  $a \in \Sigma$ on which transition exists
			\State $states[s].smallest\_next = smallestNextState(s)$
		\Else    \quad /* state is a leaf node */
			\State  $states[s].symbol = $ `$\$$'
			\State $states[s].smallest\_next = null $
		\EndIf
		\If {state $s$ is an output state}
			\State $output\_map[s] = 0$
			\For {$o \in A_{unopt}.output(s)$}
				\State $output\_map[s]$ $=$ $max(o.length, output\_map[s])$
			\EndFor
		\EndIf
	\EndFor
	\EndProcedure
	\\
	\Procedure{$OptimizedAhoCorasick$}{}\\
	\textbf{Input:} Set $D = \{d_1, d_2, \ldots d_m\}$ of domains\\
	\textbf{Output:} Optimized Aho-Corasick Pattern Matching Machine $A_{opt}$
	\State $B$ = set of brands in $D$
	\State $T$ = set of TLDs in $D$
	\State $D_M$ = domains in $D$ containing multiple TLDs 
	\State $D_P$ = domains in $D$ whose brands are prefixes of other brands
	\State $B_{trie} = CreateLexTrie(B \cup D_M \cup D_P)$
	\For {$t_i \in T$}
		\State $T_{trie}[t_i] = CreateLexTrie(t_i)$	
	\EndFor
	\For {$d_i \in D \setminus (D_M \cup D_P)$}
		\State Split $d_i$ into brand $b_i$ and TLD $t_i$
		\State Concatenate end node of brand $b_i$ in $B_{trie}$ with the start node of $T_{trie}[t_i]$ trie on `.' transition
	\EndFor
	\State Set the failure pointer of start node in $B_{trie}$ to itself
	\State Set the failure pointer of nodes which end in `.' to start node of $B_{trie}$
	\For {$t_i \in T$}
		\State Set the failure pointer of start node in $T_{trie}[t_i]$ to start node of $B_{trie}$
	\EndFor
	\State Compute the failure pointer for the remaining states in $B_{trie}$ and $T_{trie}$
	\State Store the resulting pattern matching machine in $A_{unopt}$
	\State $CreateOptimizedStates(A_{unopt})$
	\EndProcedure
	\end{algorithmic}
\end{algorithm} 
	
\begin{algorithm}[h]
	\scriptsize
	\caption{Algorithm to determine matches using Optimized Aho-Corasick Machine}~\label{AhoMatch}
	\begin{algorithmic}[1]
	\Procedure{$Match$}{}\\
	\textbf{Input:} input text $x = a_1a_2\ldots a_n$ where $a_i \in \Sigma$ and optimized Aho-Corasick pattern matching machine $A_{opt}$\\
	\textbf{Output:} set of keywords found in $t$
	\State $s = 0$
	\State $i = 1$
	\State $matches = new \ Set()$
	\State $prevMatch = null$
	\While {$i \le n$}
		\State $a = x[i]$
		\If{$state[s].symbol \in \Sigma$}
			\If{$state[s].symbol == a$}
				\State $s = state[s].smallest\_next$
				\State $i$ += 1		
			\Else
				\State $f= state[s].fail$
			\EndIf
		\ElsIf{$state[s].symbol == $ `$\$$'}
			\State $f = state[s].fail$
		\ElsIf{$state[s].symbol == $ `$\#$'} 
			\State $bitmap = bitmap\_table[s]$
			\If {bit $b_{a}$ is set in $bitmap$}
				\State $next$ = $state[s].smallest\_next$ + $offset(bitmap, a)$
				\State $i$ += 1	
			\Else
				\State $f=  state[s].fail$	
			\EndIf	
		\EndIf	
		\If {$s \in output\_map$}
			\State $l = output\_map[s]$
			\State $match = t.substr(i-l+1, i+1)$
			\If {$match.startsWith($`.'$)$}
				\State $matches.add(prevMatch + match)$		
				\State $prevMatch = null$	
			\Else
				\State $matches.add(match)$
				\State $prevMatch = match$
			\EndIf				
		\EndIf
	\EndWhile
	\State return $matches$
	\EndProcedure
	\end{algorithmic}
\end{algorithm} 

\begin{algorithm}[h]
	\scriptsize
	\caption{Algorithm to determine different squatting techniques}~\label{MatchAlgo}
	\begin{algorithmic}[1]
	\Procedure{$SquattingCategory$}{}\\
	\textbf{Input:} URL $u$ requested by the user, optimized Aho-Corasick machine $A_{opt}$,
	and a filter function $R_{CB}$ which removes the occurance of common brands that appear
	as a substring of many other brands from a list of matches returned by $A_{opt}$\\
	\textbf{Output:} category of URL whether it is $benign$ or $phishing$ or $unknown$
	\State $d  = domain(u)$
	\State $b  = brand(u)$
	\State $s = subdomain(u)$
	\State $p = directory(u)$
	\If {$d \in A_{opt}.Match(d)$}
		\State return $benign$
	\ElsIf {$b \in A_{opt}.Match(d)$}
		\State return $wrongTLDsquatting$
	\ElsIf {$A_{opt}.Match(b).filter(R_{CB})$ is not empty}
		\State return $combosquatting$
	\ElsIf {$A_{opt}.Match(s).filter(R_{CB})$ is not empty}
		\State return $subdomain$-$spoofing$
	\ElsIf{$A_{opt}.Match(p).filter(R_{CB})$ is not empty}
		\State return $directory$-$spoofing$
	\Else
		\State return $unknown$
	\EndIf
	\EndProcedure
	\end{algorithmic}
\end{algorithm} 

\begin{algorithm}[h]
	\scriptsize
	\caption{Typosquatting Algorithm}~\label{TypoAlgo}
	\begin{algorithmic}[1]
	\Procedure{$Preprocessing$}{}\\
	\textbf{Input:} list of domains $D$\\
	\textbf{Output:} dictionary $L$ with length as key and subset of indices in $D$ as value, dictionary $T$ with trigram as key and subset of indices in $D$ as value, and dictionary $F$ with trigram as key and its frequency count in $D$ as value
	\State $T = \{\}$
	\State $L = \{\}$
	\State $F = \{\}$
	\For {$i = 0 \ to D.length-1$}
		\State $l = D[i].length$
		\If {$l \in L$}
			\State $L[l] = L[l].concat(i)$
		\Else
			\State $L[l] = [i]$ 
		\EndIf
		\If {$l < 6$ $or$ $l>20$}
			\State continue
		\EndIf
		\State  $d = sort(D[i])$
		\State $trigrams$ = $ExtractTrigrams(d)$
		\For {$t \in trigrams$}
			\If {$t \in T$}
				\State $T[t] = T[t].concat(i)$
				\State $F[t] += 1$
			\Else
				\State $T[t] = [i]$ 
				\State $F[t] = 1$
			\EndIf
		\EndFor	
	\EndFor
	\State return $T, L, F$
	\EndProcedure
	\\
	\Procedure{$Matching$}{} \\
	\textbf{Input:} input domain $d$, list of domains $D$, trigram dictionary $T$, frequency dictionary $F$ and length dictionary $L$\\
	\textbf{Output:} candidate popular domains similar to domain $d$  
	\State $l = d.length$
	\If {$l < 6$}
		\State $e = 1$
		\State return $BasicFilter(d, e, D, L)$	
	\ElsIf{$ l \le 10$}
		\State $e = 1$	
		\State $D_T = NgramFilter(d, e, D, T, F)$ 
		\State return $BasicFilter(d, e, D_T, L)$	
	\ElsIf {$l \le 20$}
		\State $e = 2$
		\State $D_T = NgramFilter(d, e, D, T, F)$ 
		\State return $BasicFilter(d, e, D_T, L)$	
	\Else
		\State $e = 2$
		\State return $BasicFilter(d, e, D, L)$	
	\EndIf
		
	\EndProcedure
	\\
	\Procedure{$NgramFilter$}{}\\
	\textbf{Input:} input domain $d$, error $e$, list of domains $D$, trigram dictionary $T$ and frequency dictionary $F$\\
	\textbf{Output:} candidate popular domains within distance $e$ of domain $d$  
	\State $d = sort(d)$
	\State Split domain $d$ into $e+1$ equal parts $d_1, \ldots d_{e+1}$
	\State $D_T = new \ Set()$
	\For{$q = 1 \ to \ e+1$} 
		\State $trigrams = ExtractTrigrams(d_q)$ 
		\State $s =$ use dict $F$ to sort $trigrams$ in increasing order of frequency
		\State $indices = T[s[0]] \cap T[s[1]] \cap \ldots T[s[s.length-1]]$
		\State $D_T = D_T \cup D[indices]$
	\EndFor	
	\State return $Array.from(D_T)$
	\EndProcedure
	\\
	\Procedure{$BasicFilter$}{}\\
	\textbf{Input:} input domain $d$, error $e$, list of domains $D$ and length dictionary $L$\\
	\textbf{Output:} candidate popular domains within distance $e$ of domain $d$  
		\State $l = d.length$
		\State /* filter $D$ by length */
		\State $D_L$ = retain domains in $D$ with length $l'$ such that $l - e \le l' \le l + e$
		\State /* filter $D_L$ using unigrams */
		\State $D_U$ = retain domains in $D_L$ having at most $e$ distinct characters from $d$
		\State Compute Damerau-Levenshtein distance between $d$ and domains in $D_U$
		\State Return domains in $D_U$ with distance at most $e$
	\EndProcedure
	\end{algorithmic}
\end{algorithm} 

\begin{figure}[h]
	\begin{subfigure}[b]{\linewidth}
	  \includegraphics[scale=0.27]{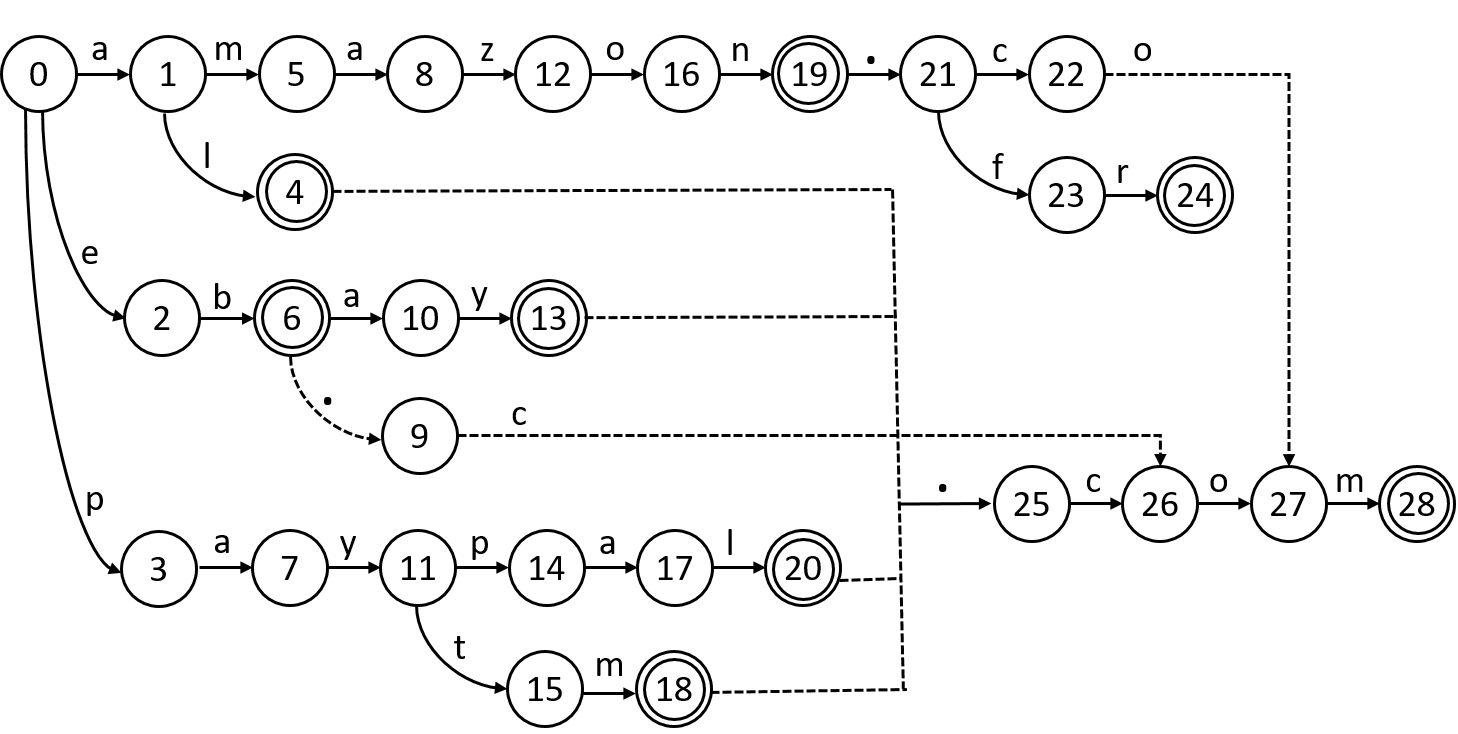}
	  \caption{Goto function}~\label{fig:trie_opt2}
	\end{subfigure}
	
	\begin{subfigure}[b]{\linewidth}
	  \centering
	  \includegraphics[scale=0.3]{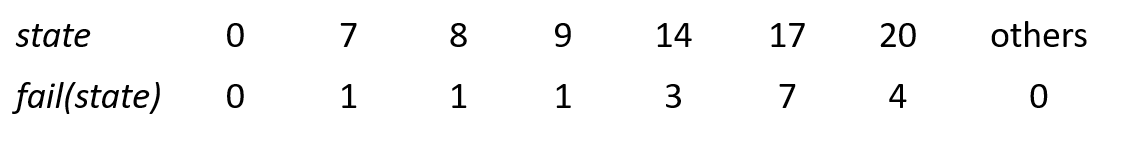}
	  \caption{Fail function}~\label{fig:fail_opt2}
	\end{subfigure}
	
	\begin{subfigure}[b]{\linewidth}
	  \centering
	  \includegraphics[scale=0.3]{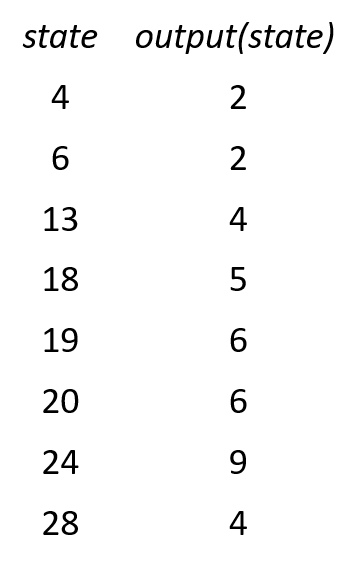}
	  \caption{Output function}~\label{fig:output_opt2}
	\end{subfigure}
	
	\begin{subfigure}[b]{\linewidth}
	  \centering
	  \includegraphics[scale=0.3]{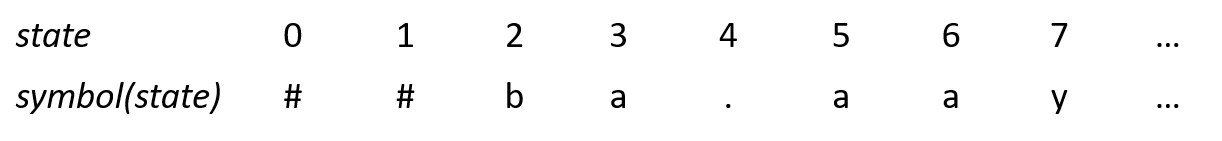}
	  \caption{Symbol array}~\label{fig:symbol_opt2}
	\end{subfigure}
	
	\begin{subfigure}[b]{\linewidth}
	  \centering
	  \includegraphics[scale=0.3]{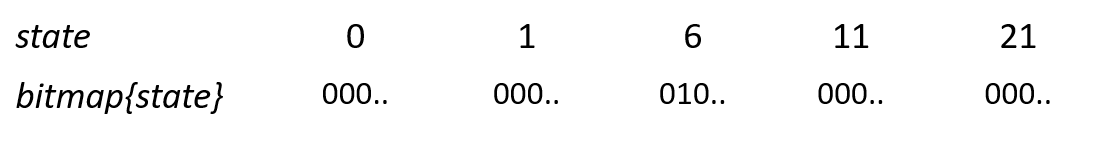}
	  \caption{Bitmap table}~\label{fig:bitmap_opt2}
	\end{subfigure}
	
	\begin{subfigure}[b]{\linewidth}
	  \centering
	  \includegraphics[scale=0.4]{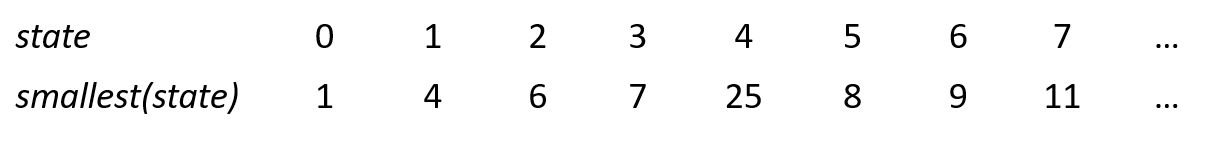}
	  \caption{Smallest next array}~\label{fig:small_opt2}
	\end{subfigure}
	
	\caption{Optimized pattern matching machine}~\label{fig:aho_opt2}
\end{figure}

\clearpage

\begin{table*}[h]
	\centering
	\scriptsize
	\caption{The browsing data of each user is split into two parts, training and testing. The table shows the number of unique domains in each training set, and the size of local whitelist $W_L$ and relaxed whitelist $W_{RL}$ (used for creating community whitelist $W_C$). It also shows the percentage of test URLs found in local $W_L$, community $W_C$ and global $W_G$ whitelists, and the percentage of URLs visited by each user from top Google Search results.}\label{tab:simwhitelist}
	\begin{tabular}{|c|c|c|c|c|c|c|c|c|c|c|}
	\hline
	\textbf{User} & \textbf{Train} & \textbf{Domains $|H|$} & \textbf{Local $|W_L|$} & \textbf{Relaxed $|W_{RL}|$} & \textbf{Test} & \textbf{$W_L$} & \textbf{$W_C$} & \textbf{$W_G$} & \textbf{User Initiated Google Search} & \textbf{Total}\\
	\hline
	User0  & 7990 & 43 & 9 & 17 & 5325 & 97.84 & 1.2 & 0.38 & 0.32 & 99.74\\
	User1  & 1107 & 69 & 7 & 34 & 1309 & 67 & 21.16 & 10.39 & 0.84 & 99.39\\
	User2  & 2340 & 94 & 10 & 30 & 1886 & 65.27 & 22.48 & 10.98 & 1.06 & 99.79\\
	User3  & 2570 & 41 & 3 & 18 & 5768 & 83.79 & 15.07 & 0.47 & 0.31 & 99.64\\
	User4  & 5774 & 187 & 7 & 26 & 3196 & 51.5 & 31.76 & 9.29 & 3.29 & 95.84\\
	User5  & 8652 & 219 & 24 & 47 & 3015 & 69.22 & 14.26 & 10.58 & 3.42 & 97.48\\
	User6  & 4441 & 240 & 15 & 43 & 3676 & 78.21 & 6.01 & 9.9 & 4.33 & 98.45\\
	User7  & 2174 & 92 & 2 & 30 & 1512 & 54.5 & 23.15 & 13.69 & 5.69 & 97.03\\
	User8  & 8684 & 162 & 18 & 41 & 5330 & 80.36 & 3.73 & 13.21 & 1.18 & 98.48\\
	User9  & 5828 & 100 & 15 & 40 & 3453 & 56.73 & 36.87 & 5.01 & 0.55 & 99.16\\
	User10  & 4778 & 75 & 12 & 26 & 5026 & 93.16 & 4.34 & 1.63 & 0.8 & 99.93\\
	User11  & 2485 & 42 & 5 & 14 & 770 & 83.9 & 7.79 & 5.58 & 2.21 & 99.48\\
	User12  & 7829 & 72 & 11 & 23 & 2434 & 78.76 & 4.07 & 16.64 & 0.12 & 99.59\\
	User13  & 8808 & 142 & 14 & 26 & 7323 & 89.7 & 6.28 & 1.83 & 1.98 & 99.79\\
	User14  & 8539 & 129 & 17 & 34 & 3401 & 89.06 & 4.06 & 4.62 & 0.88 & 98.62\\
	User15  & 9873 & 87 & 16 & 26 & 13830 & 96.44 & 2.52 & 0.81 & 0.21 & 99.98\\
	User16  & 8916 & 79 & 14 & 27 & 5864 & 51.4 & 42.05 & 5.46 & 0.77 & 99.68\\
	User17  & 2555 & 52 & 5 & 14 & 796 & 63.19 & 11.68 & 17.04 & 4.95 & 96.86\\
	User18  & 3007 & 136 & 13 & 34 & 1776 & 64.7 & 10.87 & 16.22 & 4.79 & 96.58\\
	User19  & 5226 & 124 & 18 & 28 & 1399 & 75.2 & 11.22 & 6.08 & 4.43 & 96.93\\
	User20  & 8573 & 158 & 26 & 43 & 4743 & 87.29 & 4.09 & 3.86 & 2.61 & 97.85\\
	User21  & 3564 & 103 & 10 & 37 & 1479 & 67.55 & 17.04 & 10.82 & 2.57 & 97.98\\
	User22  & 3583 & 110 & 8 & 32 & 2176 & 81.25 & 6.8 & 7.31 & 2.62 & 97.98\\
	User23  & 2864 & 128 & 9 & 33 & 1823 & 76.08 & 10.81 & 10.81 & 1.87 & 99.57\\
	User24  & 6113 & 192 & 18 & 48 & 14831 & 78.69 & 5.83 & 13.89 & 0.76 & 99.17\\
	
	\hline
	\end{tabular}%
\end{table*}

\end{document}